\documentclass{aa}  

\usepackage{graphicx}
\usepackage{txfonts}
\usepackage{color,hyperref}
\hypersetup{colorlinks,breaklinks,
            linkcolor=blue,urlcolor=blue,
            anchorcolor=blue,citecolor=blue}
\usepackage{caption}
\DeclareCaptionFormat{cont}{#1 (continued.)#2#3\par}

\begin{document}

\title{An APEX survey of outflow and infall toward the youngest protostars in Orion}


\titlerunning{APEX survey of the youngest protostars in Orion}
\authorrunning{}

   \author{Z. Nagy\inst{1,2,3}
          \and
          A. Menechella\inst{2}
          \and
          S. T. Megeath\inst{2}
          \and         
          J. J. Tobin\inst{4}
          \and
          J. J. Booker\inst{2}
          \and          
          W. J. Fischer\inst{5}
          \and          
          P. Manoj\inst{6}
          \and
          T. Stanke\inst{7}
          \and
          A. Stutz\inst{8,9}
          \and          
          F. Wyrowski\inst{10}
          }

\institute{
Center for Astrochemical Studies, Max-Planck-Institut f\"ur extraterrestrische Physik, Giessenbachstrasse 1, 85748 Garching, Germany
\and
Ritter Astrophysical Research Center, Department of Physics and Astronomy, University of Toledo, 2801 West Bancroft Street, Toledo, OH 43606, USA
\and
Konkoly Observatory, Research Centre for Astronomy and Earth Sciences, Konkoly Thege Mikl\'os \'ut 15-17, H-1121 Budapest, Hungary \\
\email{nagy.zsofia@csfk.mta.hu}
\and
National Radio Astronomy Observatory, 520 Edgemont Rd., Charlottesville, VA 22903, USA
\and
Space Telescope Science Institute, 3700 San Martin Drive, Baltimore, MD 21218, USA
\and
Department of Astronomy and Astrophysics, Tata Institute of Fundamental Research, Colaba, Mumbai 400005, India
\and
European Southern Observatory, 85748 Garching bei M\"unchen, Germany
\and
Departmento de Astronom\'ia, Facultad Ciencias F\'isicas y Matem\'aticas, Universidad de Concepci\'on, Av. Esteban Iturra s/n Barro Universitario, Casilla 160-C, Concepci\'on, Chile
\and
Max-Planck-Institut f\"ur Astronomie, D-69117 Heidelberg, Germany
\and
Max-Planck-Institut fur Radioastronomie, Auf dem H\"{u}gel 69, 53121, Bonn, Germany
}

\date{}


\abstract
{}
{We aim to characterize the outflow properties of a sample of early Class 0 phase low-mass protostars in Orion, which were first identified by the \textit{Herschel} Space Observatory. We also look for signatures of infall in key molecular lines.}
{Maps of CO $J$=3-2 and $J$=4-3 toward 16 very young Class 0 protostars were obtained using the Atacama Pathfinder EXperiment (APEX) telescope. We searched the data for line wings indicative of outflows and calculated masses, velocities, and dynamical times for the outflows. We used additional HCO$^+$, H$^{13}$CO$^+$, and NH$_3$ lines to look for infall signatures toward the protostars.}
{We estimate the outflow masses, forces, and mass-loss rates based on the CO $J$=3-2 and $J$=4-3 line intensities for eight sources with detected outflows. We derive upper limits for the outflow masses and forces of sources without clear outflow detections. 
The total outflow masses for the sources with clear outflow detections are in the range between 0.03 and 0.16 $M_\odot$ for CO $J$=3-2 and between 0.02 and 0.10 $M_\odot$ for CO $J$=4-3. 
The outflow forces are in the range between $1.57\times10^{-4}$ and $1.16\times10^{-3}$ $M_\odot$ km s$^{-1}$ yr$^{-1}$ for CO $J$=3-2 and between $1.14\times10^{-4}$ and $6.92\times10^{-4}$ $M_\odot$ km s$^{-1}$ yr$^{-1}$ for CO $J$=4-3. Nine protostars in our sample show asymmetric line profiles indicative of infall in HCO$^+$, compared to H$^{13}$CO$^+$ or NH$_3$.}
{The outflow forces of the protostars in our sample show no correlation with the bolometric luminosity, unlike those found by some earlier studies for other Class 0 protostars. The derived outflow forces for the sources with detected outflows are similar to those found for other, more evolved, Class 0 protostars, suggesting that outflows develop quickly in the Class 0 phase.}


\maketitle

\section{Introduction}
\label{sec:intro}

Surveys of molecular clouds with the \textit{Spitzer} and \textit{Herschel} space telescopes have now identified a large sample of protostars in nearby molecular clouds (\citealp{evans2009}, \citealp{dunham2013}, \citealp{manoj2013}, \citealp{stutz2013}, \citealp{megeath2016}, \citealp{furlan2016}). These surveys provide the sensitivity, wavelength coverage, and sample size needed to identify short-lived phases in the formation of stars, and they refine our picture of protostellar evolution.  \textit{Herschel} observations of the Orion molecular clouds discovered 19 protostars with bright 70~$\mu$m and 160~$\mu$m fluxes and faint 24~$\mu$m fluxes, also known as PACS Bright Red sources (PBRs; \citealp{stutz2013}, \citealp{tobin2015}, \citealp{tobin2016}). Eleven of these were not previously identified as protostars and four were not detected in 3-24 $\mu$m \textit{Spitzer} images \citep{megeath2012,megeath2016}. 

On the basis of their spectral energy distributions (SEDs), protostars are broadly classified as Class 0, Class I, or flat spectrum protostars, which approximately correspond to different stages in the formation of protostars (e.g., \citealp{andre1993}, \citealp{lada1987}, \citealp{wilking1989}, \citealp{whitney2003}; \citealp{furlan2016}).  
\citet{stutz2013} argue that these new Orion protostars are very young Class 0 protostars, with $T_{\rm{bol}} < 50$~K (compared to $< 70$~K for all Class 0 objects) and $L_{\rm{submm}}/L_{\rm{bol}}$ of 0.6\% to 6\% (compared to $>$0.5\% for all Class 0 objects). Their bolometric luminosities range from 0.6 to 30 $L_{\rm{bol}}$, confirming that they are true protostars in a rapid mass infall phase (\citealp{stutz2013}, \citealp{furlan2016}). Subsequent Combined Array for Research in Millimeter-wave Astronomy (CARMA) 3 mm continuum observations supported this interpretation by showing that many of the PBRs also have higher $L_{\rm{3mm}}/L_{\rm{bol}}$ than other, previously observed, Class 0 protostars \citep{tobin2015}. This high ratio suggests that they have more massive envelopes for a given luminosity than most Class 0 sources. The rarity of the young Orion sources implies ages less than 25,000 years \citep{stutz2013}; this is 1/6 the duration of the Class 0 phase \citep{dunham2014b}. 
Although the high luminosities suggest that most of the PBRs are not first hydrostatic cores (FHSCs), recent  Atacama Large Millimeter/submillimeter Array (ALMA) and Karl G. Jansky Very Large Array (VLA) imaging suggests that one PBR may contain a FHSC while another may not contain any \citep{karnath2020}.

To further study these young protostars, we carried out a molecular line survey using the 12-m submillimeter Atacama Pathfinder EXperiment (APEX; \citealp{guesten2008}) at Llano de Chajnantor in Chile designed to search for the signatures of infall and to measure the properties of their outflows.
These observations build upon previous surveys with single-dish telescopes that show changes in infall and outflow signatures with evolutionary class. This is the first systematic survey for infall toward PBRs and the second survey for outflow (after the work by \citealp{tobin2016}).

Single-dish studies of infall have used asymmetry in self-absorbed line profiles as an indicator of infall, with red-shifted self-absorption as a sign of infall \citep{evans2003}. Although these types of profiles may also be produced by outflow and rotation, an excess number of protostars showing redshifted self-absorption in a sample of protostars is evidence for infall. \citet{mardones1997} used the difference in the velocities of the peaks of optically thick and thin lines to search for infall asymmetry; they found that Class 0 protostars show an excess of red-shifted self-absorption, which is expected from infall. In contrast, Class I sources show equal numbers in red- and blue-shifted line peaks. Hence, a higher incidence of red-shifted self-absorption profiles is a characteristic of Class 0 protostars.  

Single-dish studies have also shown the evolution of protostellar outflows with evolutionary class \citep{bontemps1996}. These measurements primarily trace gas from the envelope and parental cloud entrained in the outflow by the transfer of momentum from an accretion driven jet and wind \citep{arce2007}. 
One of the most important physical parameters of outflows is the force, which measures the rate at which momentum is injected into the outflow; from conservation of momentum, this is a direct constraint on the momentum flow rate of jet and wind powering the outflow. 
Existing observations show that outflow force of low mass protostars increases with their bolometric luminosity, but decreases as the protostars evolve from the Class 0 to Class I phase (\citealp{bontemps1996}, \citealp{takahashi2008}, \citealp{curtis2010}). 
Similarly, based on \textit{Herschel} data of 50 protostars, \citet{manoj2016} found a correlation between the far-infrared CO luminosity, which is proportional to the mass-loss rate, and the bolometric luminosity.
However, the initial development of outflows during the early part of the Class 0 phase is still not well understood and lacks observations that can be used to test models of the evolution of outflows (e.g., \citealp{machidahosokawa2013}). 
In this paper, we aim to characterize the earliest state of outflows from low-mass protostars by analyzing observations of some of the youngest protostars. These observations complement those of \citet{tobin2016}, who observed 14 PBRs with CARMA in CO $J$=1-0 at angular resolutions of 3$''$ to 6$''$ and detected outflows toward eight of the 14 sources. 
Outflows from four of the PBRs analyzed in this paper (HOPS 400, 401, 403, and 404) have also been detected with ALMA at an angular resolution of $0.25''\times0.24''$ \citep{karnath2020}.

In this paper, we present results from CO $J$=3-2 and $J$=4-3 maps observed at angular resolutions of 14-19$''$ toward 16 PBRs. We also present HCO$^+$ $J$=3-2, H$^{13}$CO$^+$ $J$=3-2, and NH$_3$ (1,1) spectra to search for infall toward 12 PBRs. 
The paper is organized as follows. In Section \ref{sec:obs} we present the observations used in this study. Section \ref{sec:outflowparam} discusses the properties of the sample of protostars analyzed in this paper. We derive physical properties for the outflows of the protostars in Sect. \ref{sec:physical_properties}. We discuss infall signatures detected toward the protostars in Sect. \ref{sec:infall}. We compare our results to previous works in Sect. \ref{sec:discussion}.

\section{Observations}
\label{sec:obs}

The CO $J$=3-2 (345.795989 GHz) and CO $J$=4-3 (461.040768 GHz) maps were observed in April 2013 using the First Light APEX Submillimeter Heterodyne receiver (FLASH, \citealp{heyminck2006}) instrument of APEX (project number m-091.f-0032-2013). Approximately 1.5$\times$1.5 arcminute maps were observed toward 16 PBRs.
The data were converted from an antenna temperature to main-beam brightness temperature scale using main beam efficiencies of 0.73 for CO $J$=3-2 and 0.60 for CO $J$=4-3. 
The beam sizes of the CO $J$=3-2 and CO $J$=4-3 maps are 19.2$''$ and 14.4$''$, respectively.
The reduction of the data was done using standard routines in the CLASS package of GILDAS\footnote{http://www.iram.fr/IRAMFR/GILDAS}.

We used additional single pointing observations of HCO$^+$ $J$=3-2 (267.557626 GHz) and H$^{13}$CO$^+$ $J$=3-2 (260.255339 GHz) data observed with the Swedish Heterodyne Facility Instrument (SHeFI) receiver of APEX in March 2013. HCO$^+$ data are available for 12 of the PBRs, and were used as an optically thick tracer to check for infall motions. As an optically thin tracer we used the H$^{13}$CO$^+$ $J$=3-2 data for some of the sources. The main beam efficiency for the HCO$^+$ $J$=3-2 and H$^{13}$CO$^+$ $J$=3-2 data is 0.74, and the beam size is $\sim$25$''$. The H$^{13}$CO$^+$ data were smoothed from their original velocity resolution of $\sim$0.09 km s$^{-1}$ to $\sim$0.18 km s$^{-1}$ to get a better signal-to-noise ratio (S/N).

For eight sources we used NH$_3$ (1,1) data as an optically thin tracer of infall motions, which were observed as single pointings using frequency switching in April 2012 with the National  Radio Astronomy Observatory (NRAO) \textit{Robert C. Byrd} Green Bank Telescope (GBT). 
We used the K-band Focal Plane Array receiver (KFPA) with the GBT autocorrelation spectrometer as our backend. The NH$_3$ (1,1) transition was observed in a 50 MHz spectral window divided into 4096 12.207 kHz channels ($\sim$0.154 km s$^{-1}$).
The 7 beams are laid out in a hexagonal pattern with a central beam and 6 outer beams separated by 96$''$.
Each beam simultaneously observes in both left and right circular polarization. The beam size of the GBT at this frequency is 32$''$ ($\sim$13500 AU). 
For the analysis presented here we only used the central KFPA beam, centered on the protostar position.
The observations were carried out in excellent conditions, the 23 GHz opacity at zenith was $\sim$0.04 and system temperatures were typically $\sim$45 K. 
Each source was observed for at least 2.5 minutes, some sources were revisited to achieve a higher S/N with up to 10 minutes of time on source. The typical rms noise was 0.035 K ($T_{\rm{mb}}$).
The raw data were reduced using the GBTIDL package\footnote{\texttt{http://gbtidl.nrao.edu/}}. For each beam, the two polarizations were averaged and the data were put on the $T_{\rm{mb}}$ scale by correcting for atmospheric opacity and the main beam efficiency of 0.89. The data were then baseline subtracted using a 7th order polynomial over line free regions outside of and between the hyperfine emission lines. 

\section{Identification of outflows}
\label{sec:outflowparam}

The sample of 16 PBRs which the maps were centered on is summarized in Table \ref{tab:source_properties}, which includes the bolometric luminosities and temperatures of the protostars, based on \citet{furlan2016}. We also show the parameters for the Class 0 source HOPS 402, which is covered by the HOPS 401 map, the Class I source HOPS 223, which is close to HOPS 397, and the Class 0 source HOPS 340, which is close to HOPS 341.
Figure \ref{CO32_spec} shows the average CO $J$=3-2 and $J$=4-3 line profiles observed toward the maps. For HOPS 404 there is only CO $J$=3-2 data available, all other sources were mapped in both transitions. Toward nine of the mapped PBRs there is evidence of broad, high velocity wings in both transitions relative to the systemic velocities of the sources: HOPS 341, 354, 358, 372, 373, 394, 397, 399, and 401. The outflows detected in CO $J$=3-2 are always detected in CO $J$=4-3 as well.
The remainder of the protostars that show no clear detections of line wings may be confused with other sources.

\begin{table*}[ht]  
\centering    
\caption{The protostars analyzed in this paper and their observed properties based on \citet{furlan2016}. The inclinations and the CARMA CO $J$=1-0 detections are based on \citet{tobin2016}. Most sources listed here are the sources which the maps are centered on, except for HOPS 223, 340, and HOPS 402. Additional sources which are (partially) covered by the maps are shown in Figures \ref{fig:int_co32} and \ref{fig:int_co43}.
Notes: $^1$Based on \citet{furlan2016}, $^2$Based on \citet{stutz2013}, $^3$Tentative blended outflow detection. $^4$The red-shifted outflow emission detected toward the HOPS 401 map is probably related to HOPS 316/358. $^5$Tentative detection, low S/N.
}
\label{tab:source_properties}
\begin{tabular}{l c c c c c c c c}
\hline
HOPS$^{1}$&	PBR$^2$& RA (deg)& Dec (deg)& $L_{\rm{bol}}$& $T_{\rm{bol}}$& $i$& \multicolumn{2}{c}{Outflow detection}\\
& & (J2000)& (J2000)& ($L_\odot$)& (K)& (deg)& CO $J$=3-2, $J$=4-3& CARMA CO $J$=1-0\\
\hline
340&           &  86.7554& +0.4393&  1.85&    40.6&               &       yes&             \\

341&    	       &  86.7541& +0.4395&	2.07&    39.4&               &   blended&             \\

354&	           &  88.6011& +1.7387&	6.57&    34.8& nearly edge-on&       yes&             \\

358&	           &  86.5301& -0.2250& 24.96&    41.7&               &       yes&             \\

359&	           &  86.8534& +0.3500& 10.00&    36.7&               &        no&             \\

372&	           &  85.3598& -2.3056&   4.80&	 37.3&               & confusion& extended$^3$\\

373&	     093003&  86.6279& -0.0431&   5.32&	 36.9&       $\sim$50&       yes&     extended\\

394&	     019003&  83.8497& -5.1315&   6.56&	 45.5&       $\sim$30& tentative&     extended\\

397&	     061012&  85.7036& -8.2696&   1.66&	 46.1&   intermediate& confusion&    tentative\\

223&           &  85.7019& -8.2762&  19.25&  247.5&   intermediate&       yes&     extended\\

398&     082005&  85.3725& -2.3547&   1.01&	 23.0&               &        no&           no\\

399&	     082012&  85.3539& -2.3024&   6.34&	 31.1&       $\sim$50&       yes&     extended\\

400&     090003&  85.6885& -1.2706&   2.94&	 35.0&       $\sim$30&        no&      compact\\

401&     091015&  86.5319& -0.2058&   0.61&	 26.0&               & tentative$^4$&       no\\

402&           &  86.5415& -0.2047&   0.55&   24.2&               &        no&           no\\

403&     093005&  86.6156& -0.0149&   4.14&	 43.9&       $\sim$30&        no&      compact\\ 

404&     097002&  87.0323& +0.5641&   0.95&	 26.1&               &        no&           no\\ 

405&     119019&  85.2436& -8.0934&   1.60&   35.0&        edge-on&        no&      extended\\ 

407&     302002&  86.6177& +0.3242&   0.71&	 26.8&       $\sim$80&        no&  extended$^5$\\ 
\hline
\end{tabular}
\end{table*}

\begin{figure*}[ht]
\centering
\includegraphics[width=5.5cm]{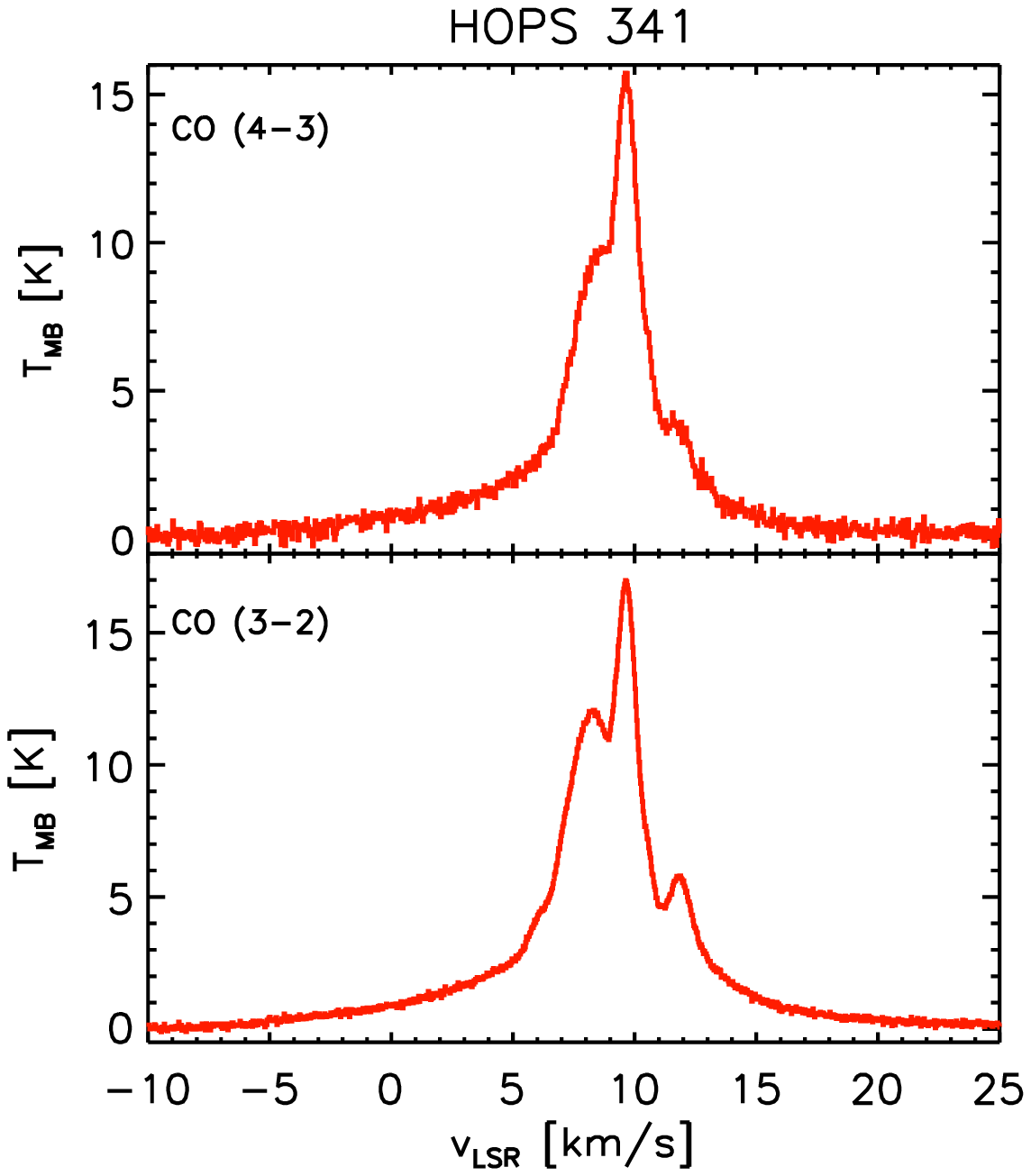}
\includegraphics[width=5.5cm]{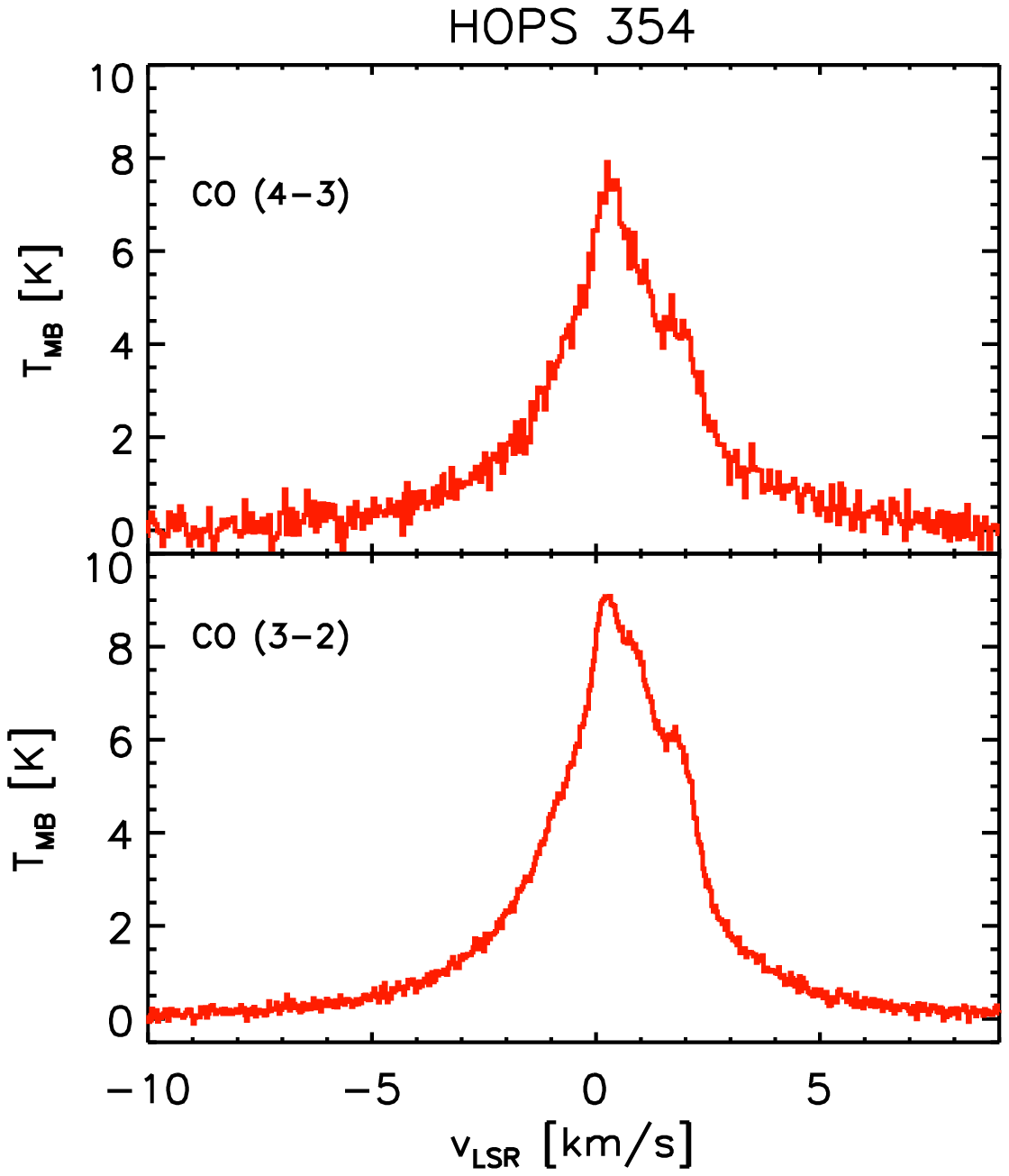}
\includegraphics[width=5.5cm]{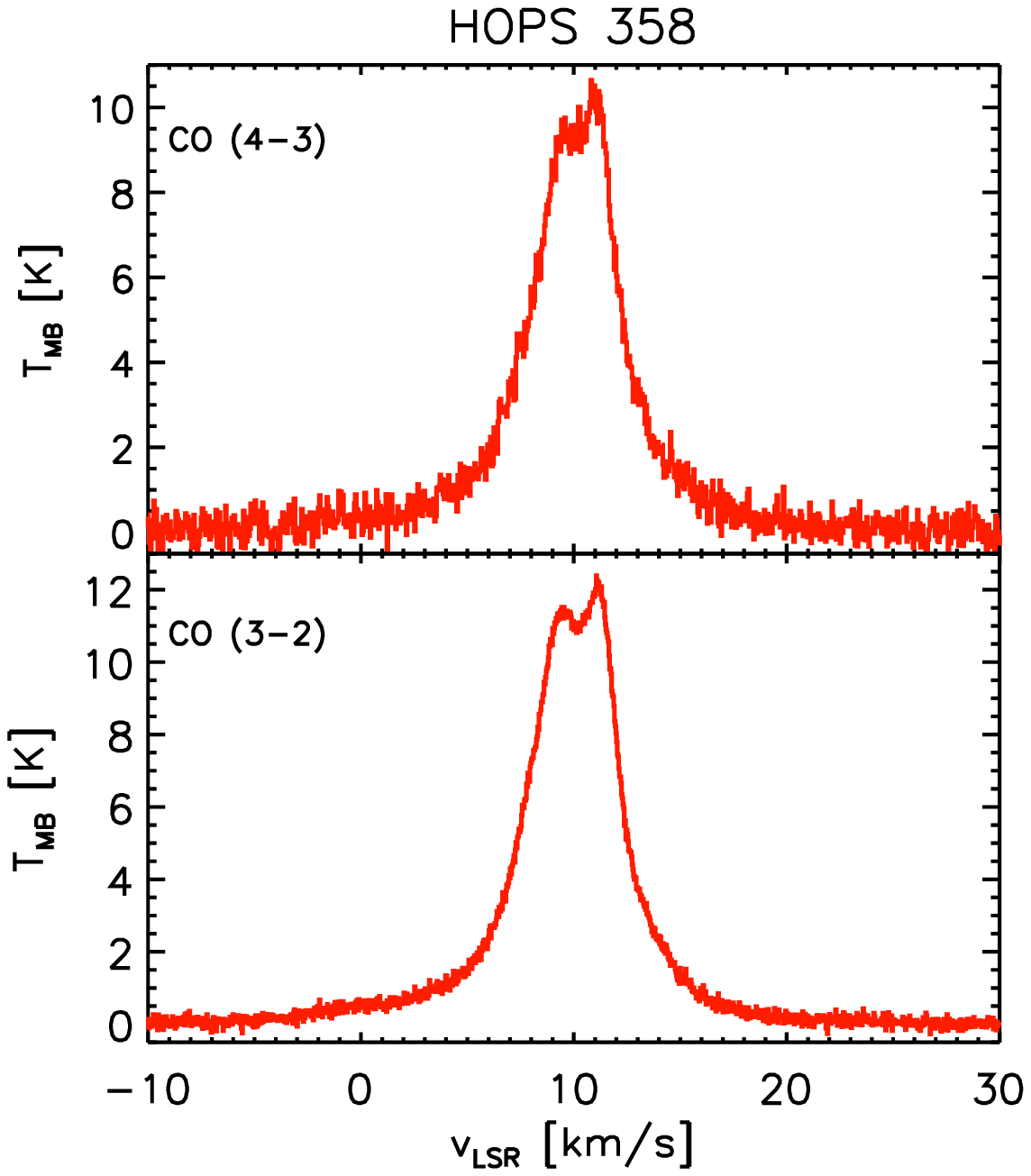}
\vskip+0.5cm
\includegraphics[width=5.5cm]{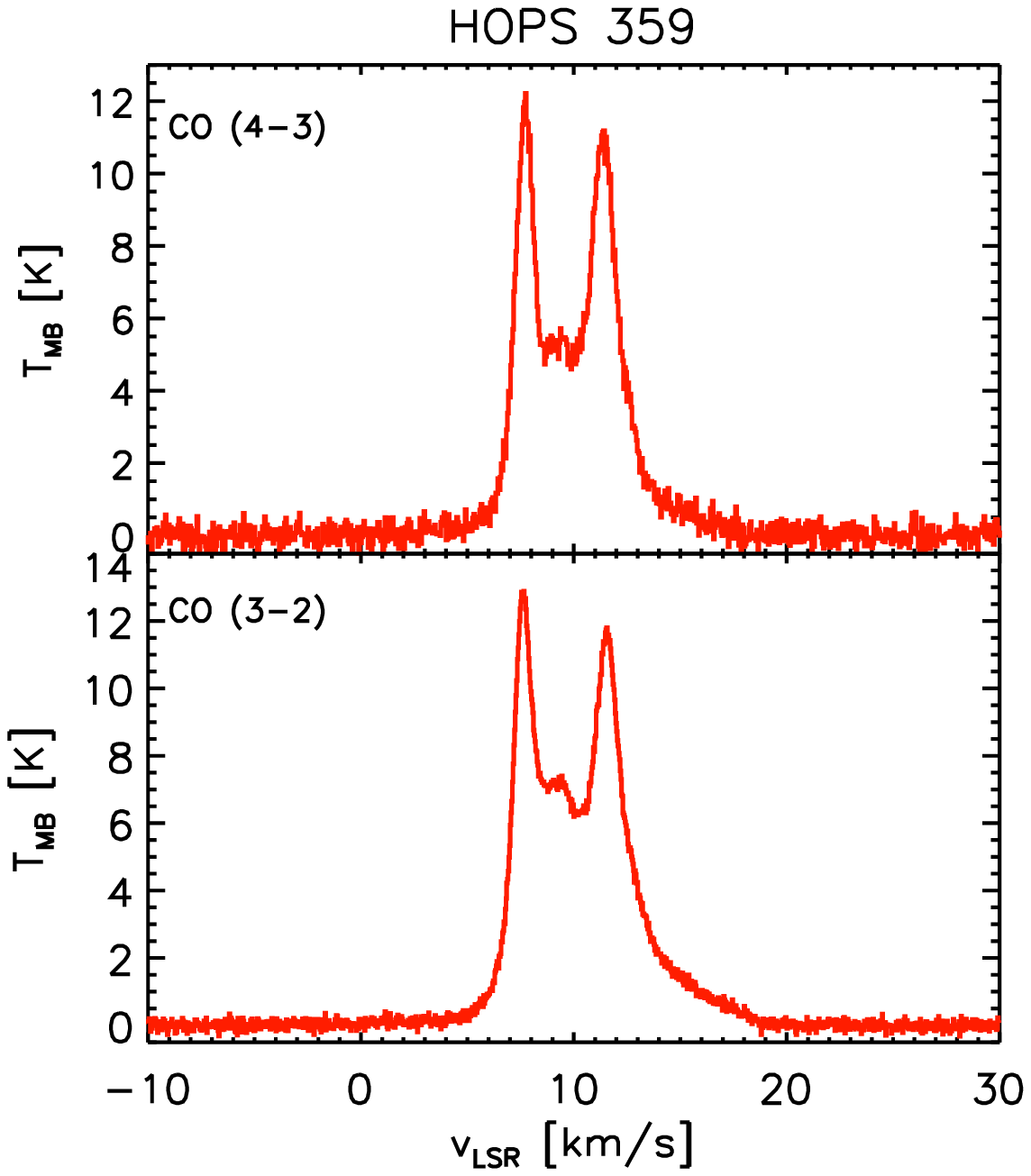}
\includegraphics[width=5.5cm]{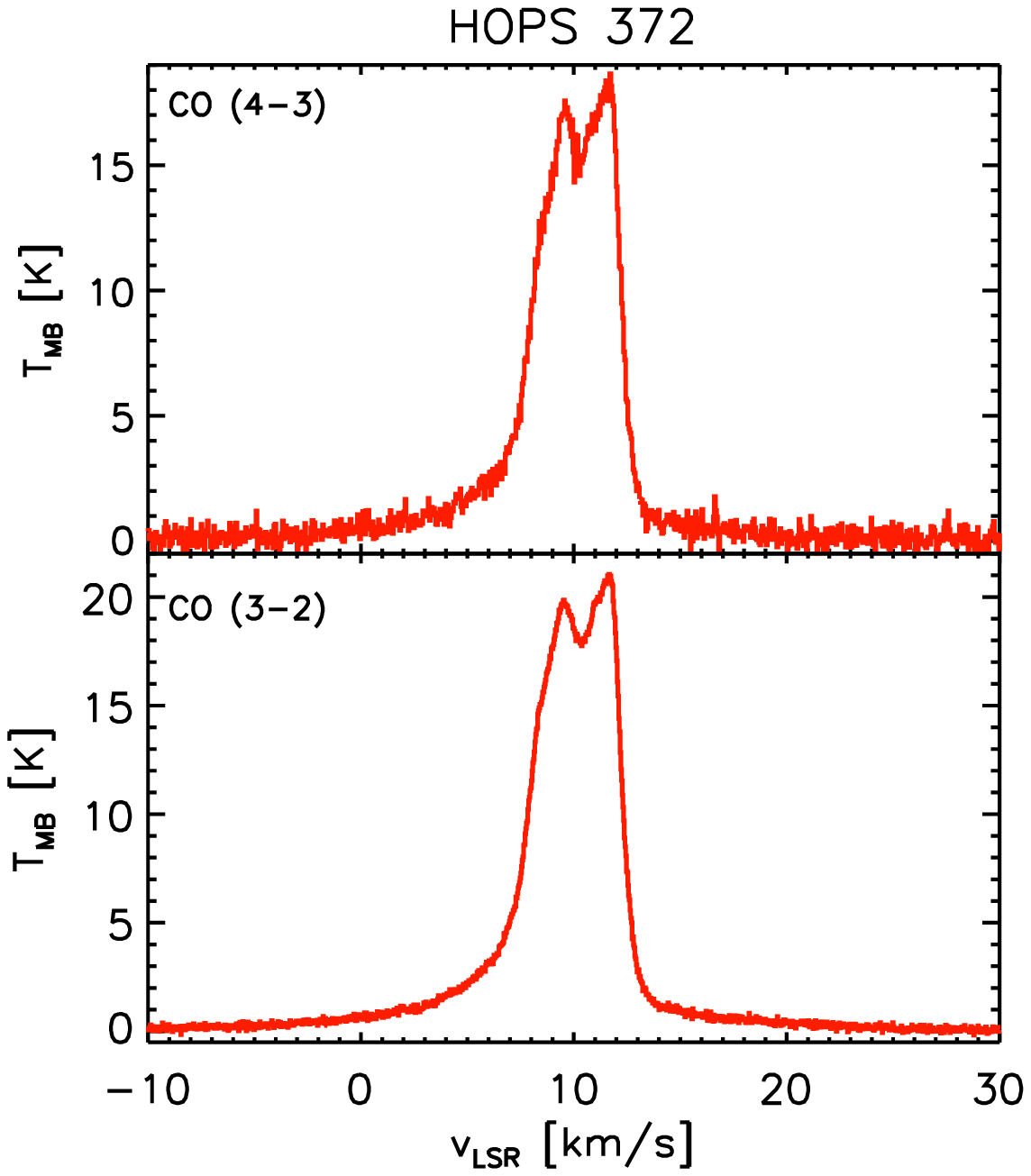}
\includegraphics[width=5.5cm]{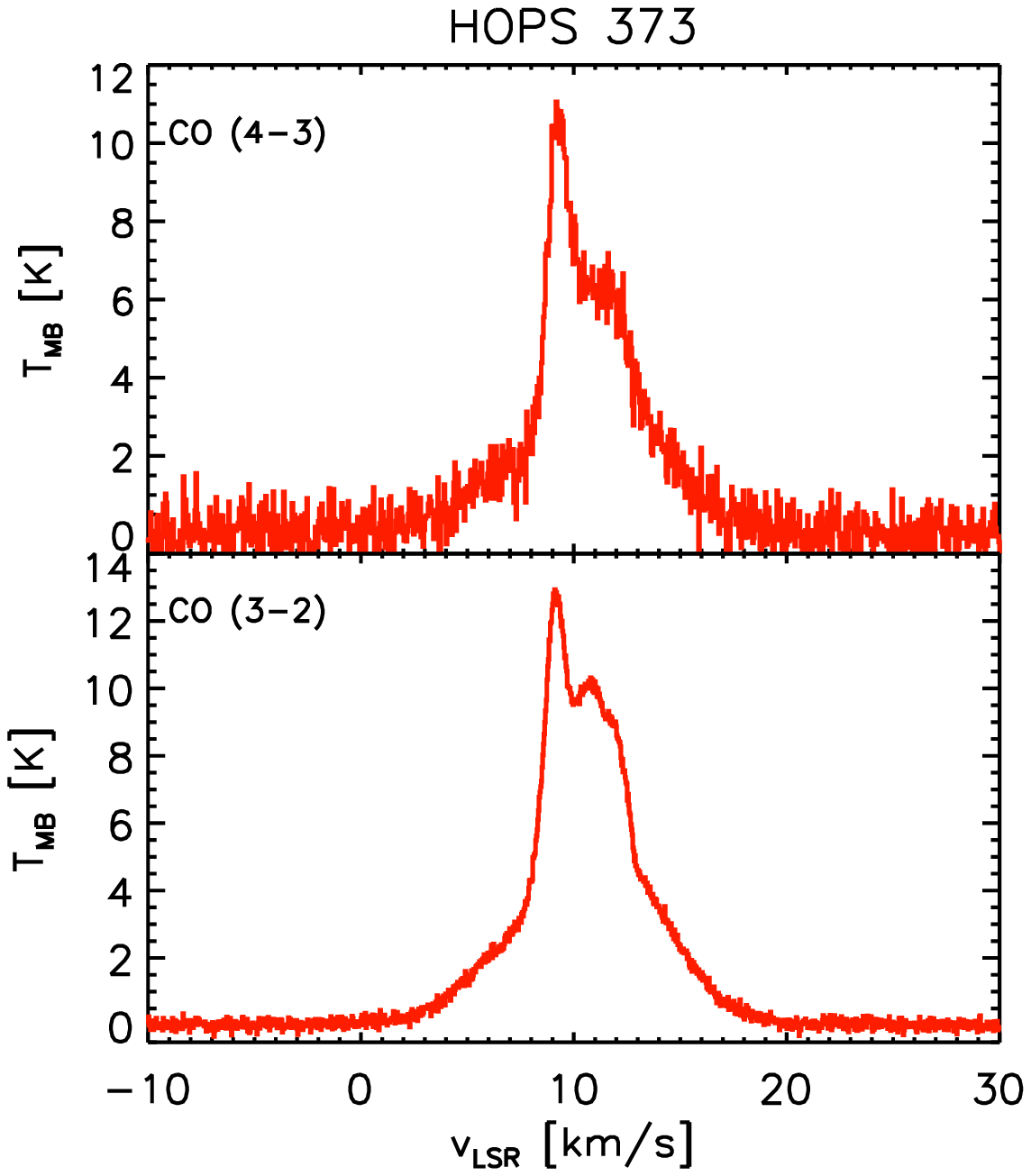}
\vskip+0.5cm
\includegraphics[width=5.5cm]{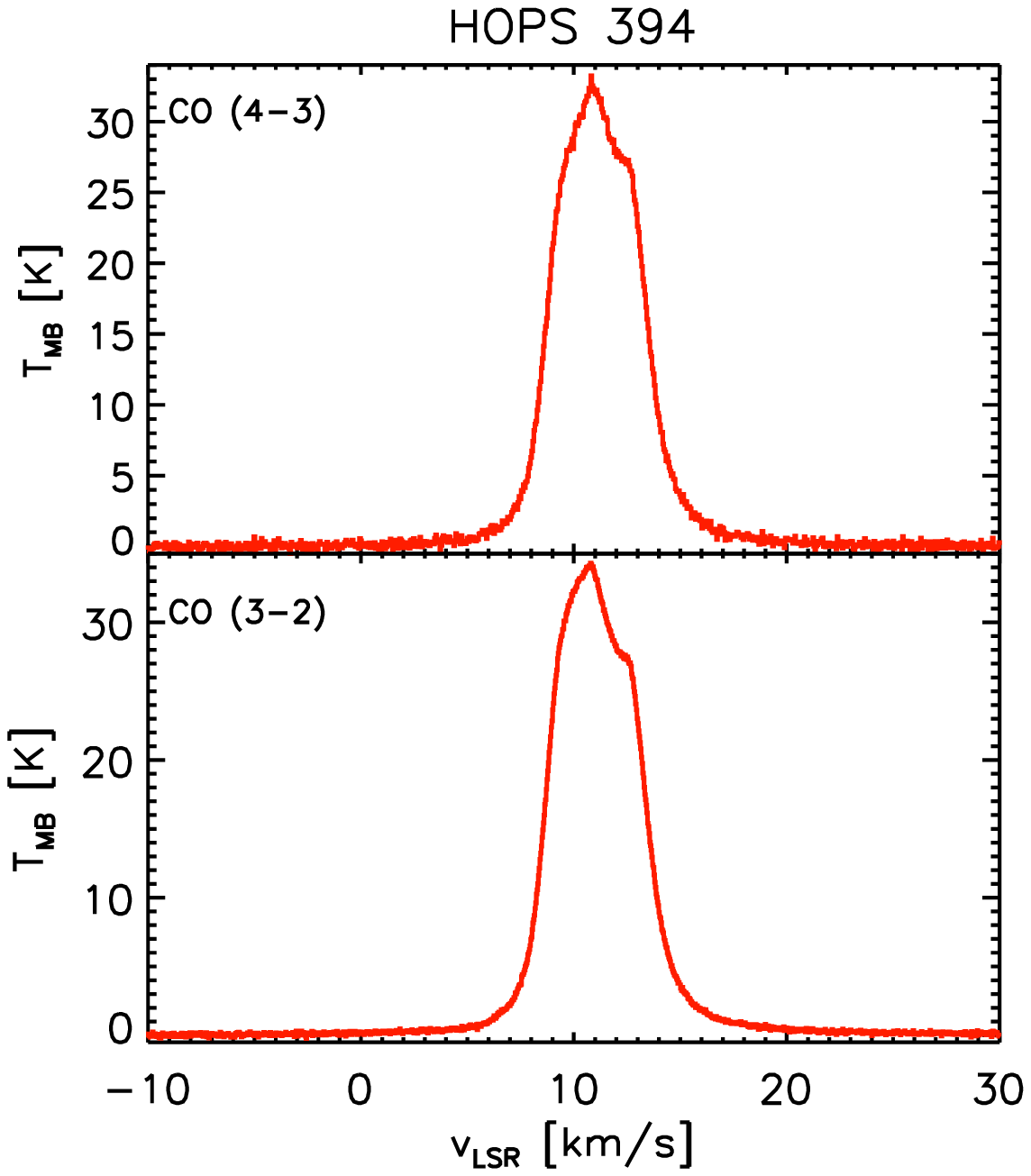}
\includegraphics[width=5.5cm]{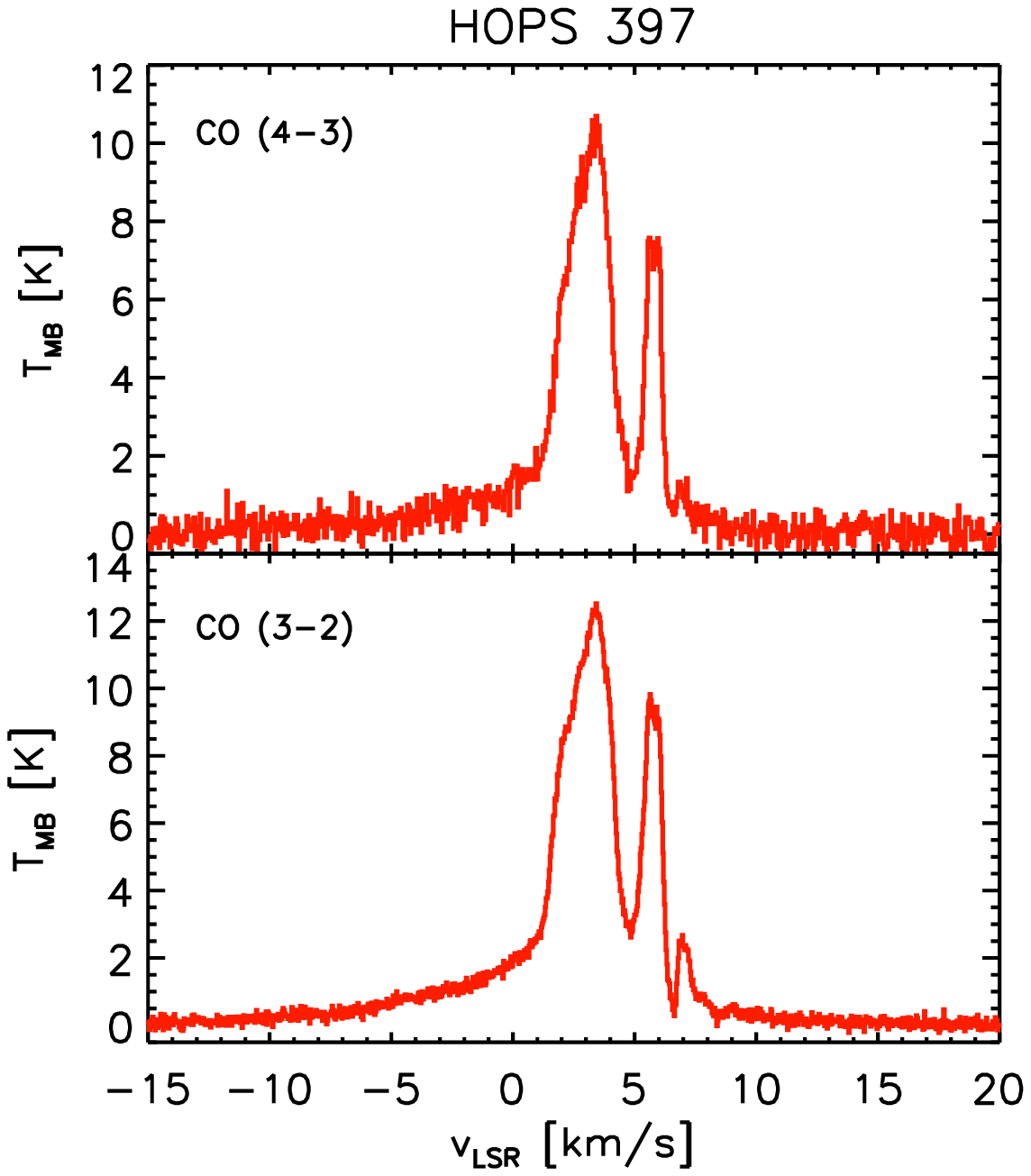}
\includegraphics[width=5.5cm]{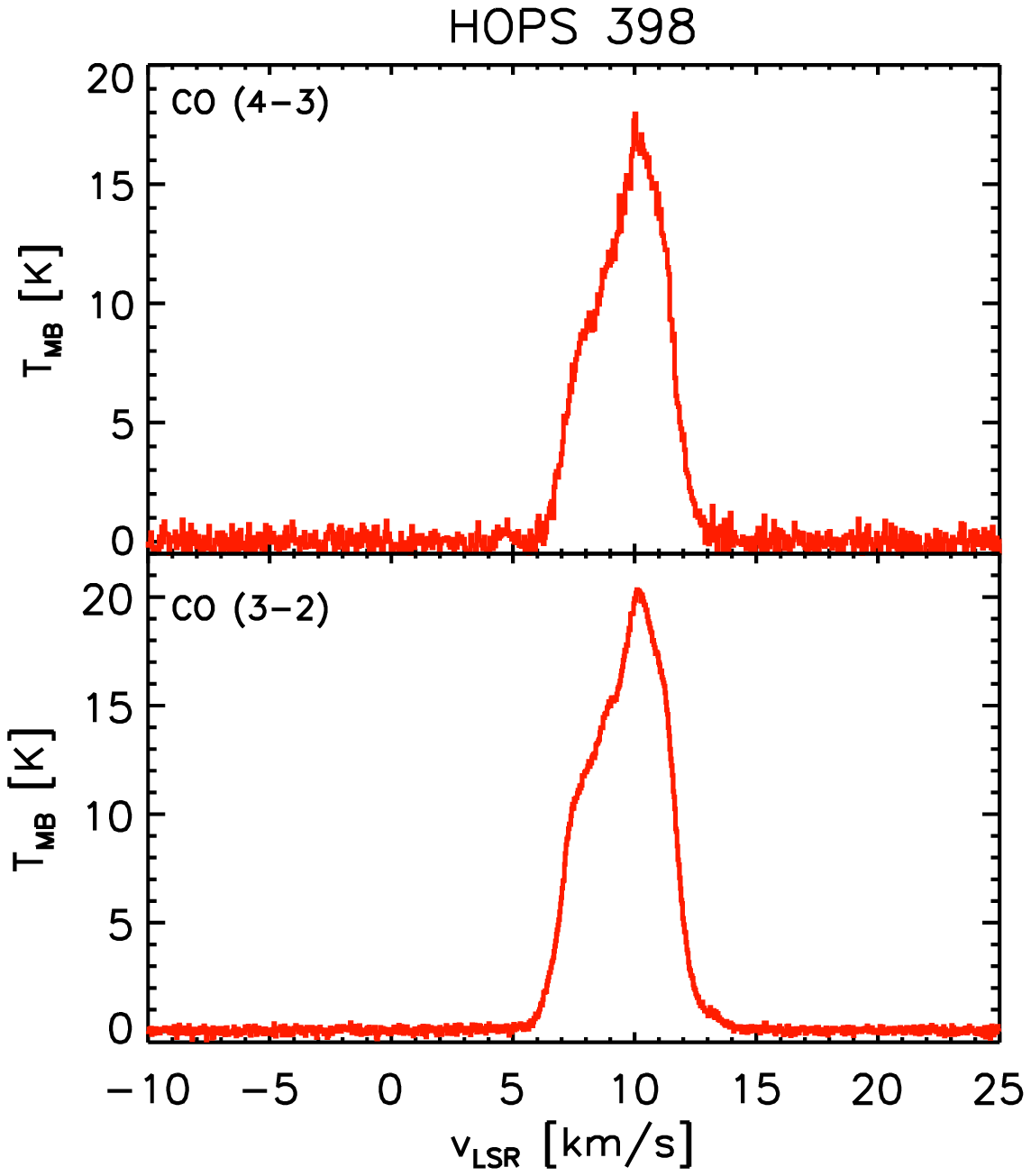}
\caption{
CO $J$=3-2 and CO $J$=4-3 line profiles averaged over the maps. 
The source mentioned for each figure is the center of the maps. The outflow emission detected toward HOPS 341 likely corresponds to the Class 0 protostar HOPS 340. The line wings detected toward HOPS 372 and HOPS 399 are from the same outflow, attributed to HOPS 399. The outflow toward the HOPS 397 map is dominated by the Class I protostar HOPS 223.
}
\label{CO32_spec}
\end{figure*}

\begin{figure*}[ht]
\centering
\ContinuedFloat
\captionsetup{list=off,format=cont}
\includegraphics[width=5.5cm]{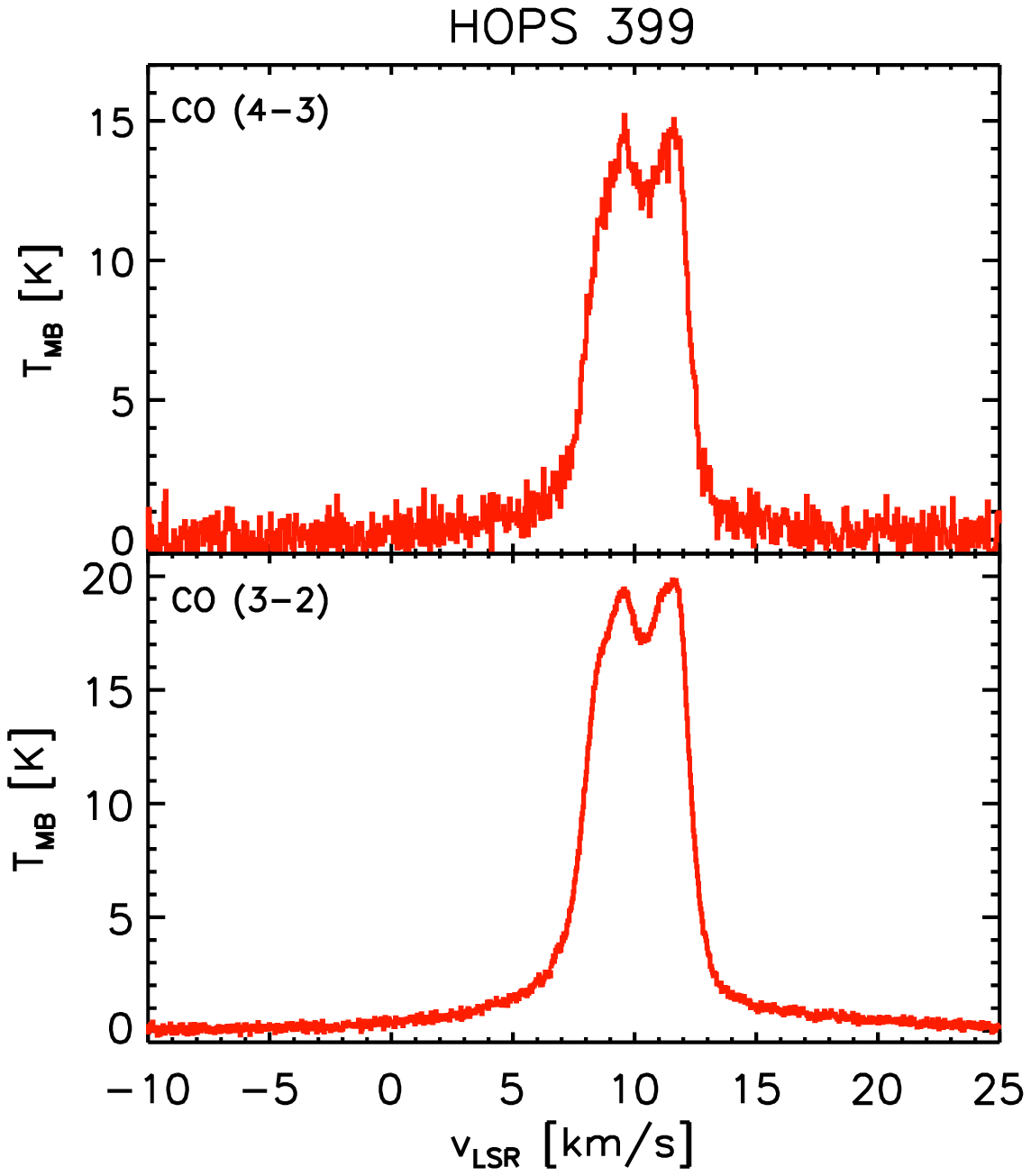}
\includegraphics[width=5.5cm]{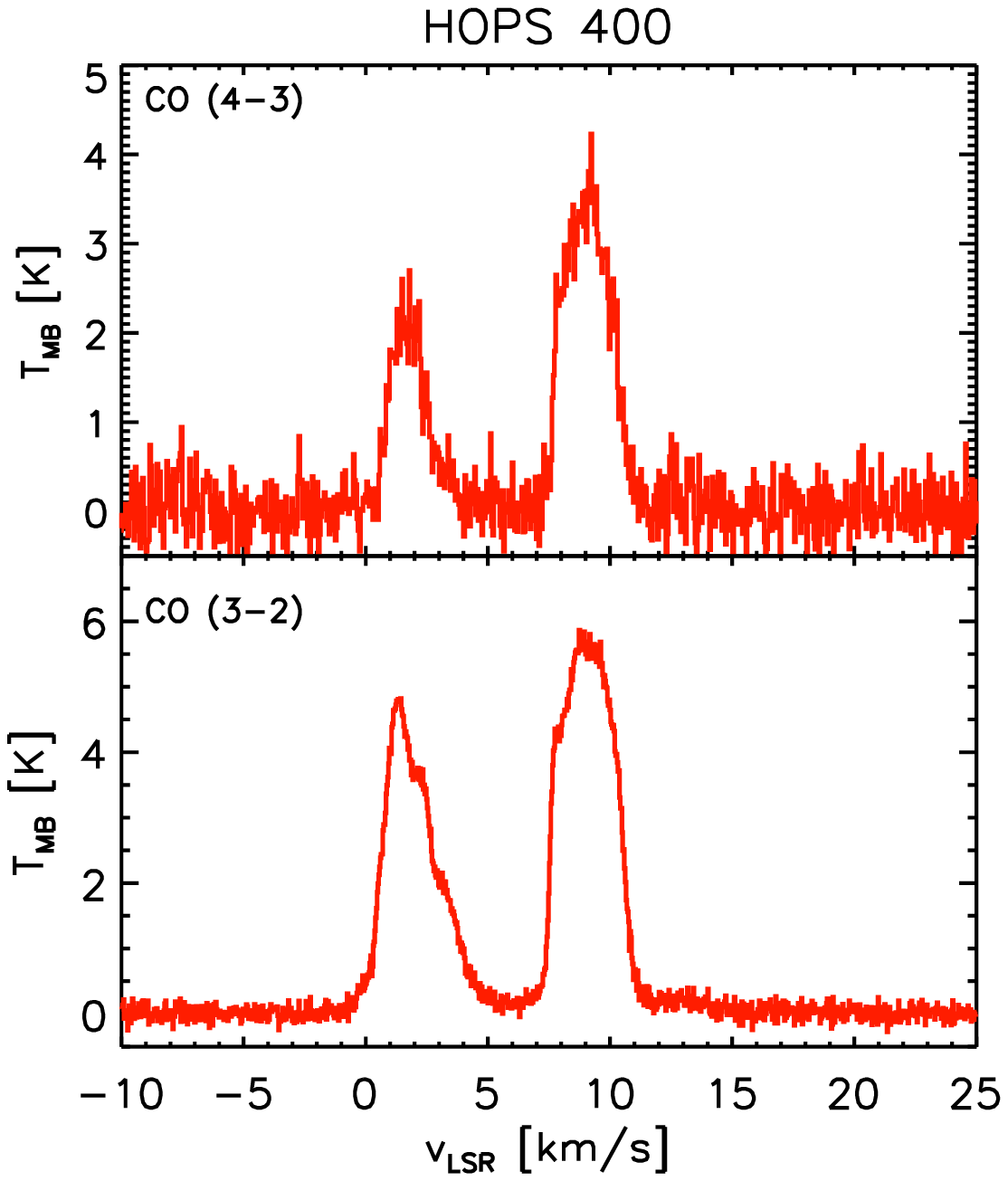}
\includegraphics[width=5.5cm]{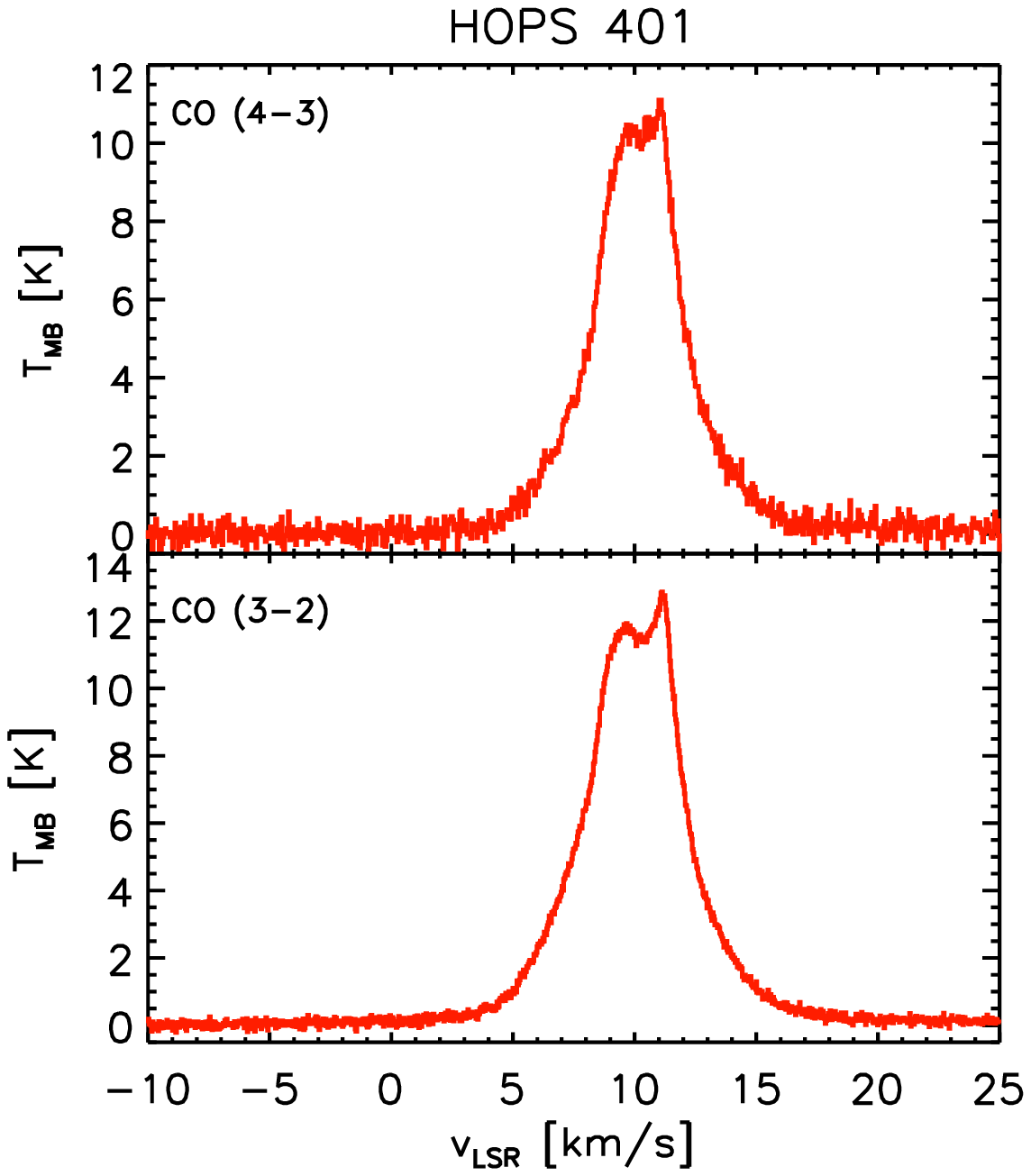}
\vskip+0.5cm
\includegraphics[width=5.5cm]{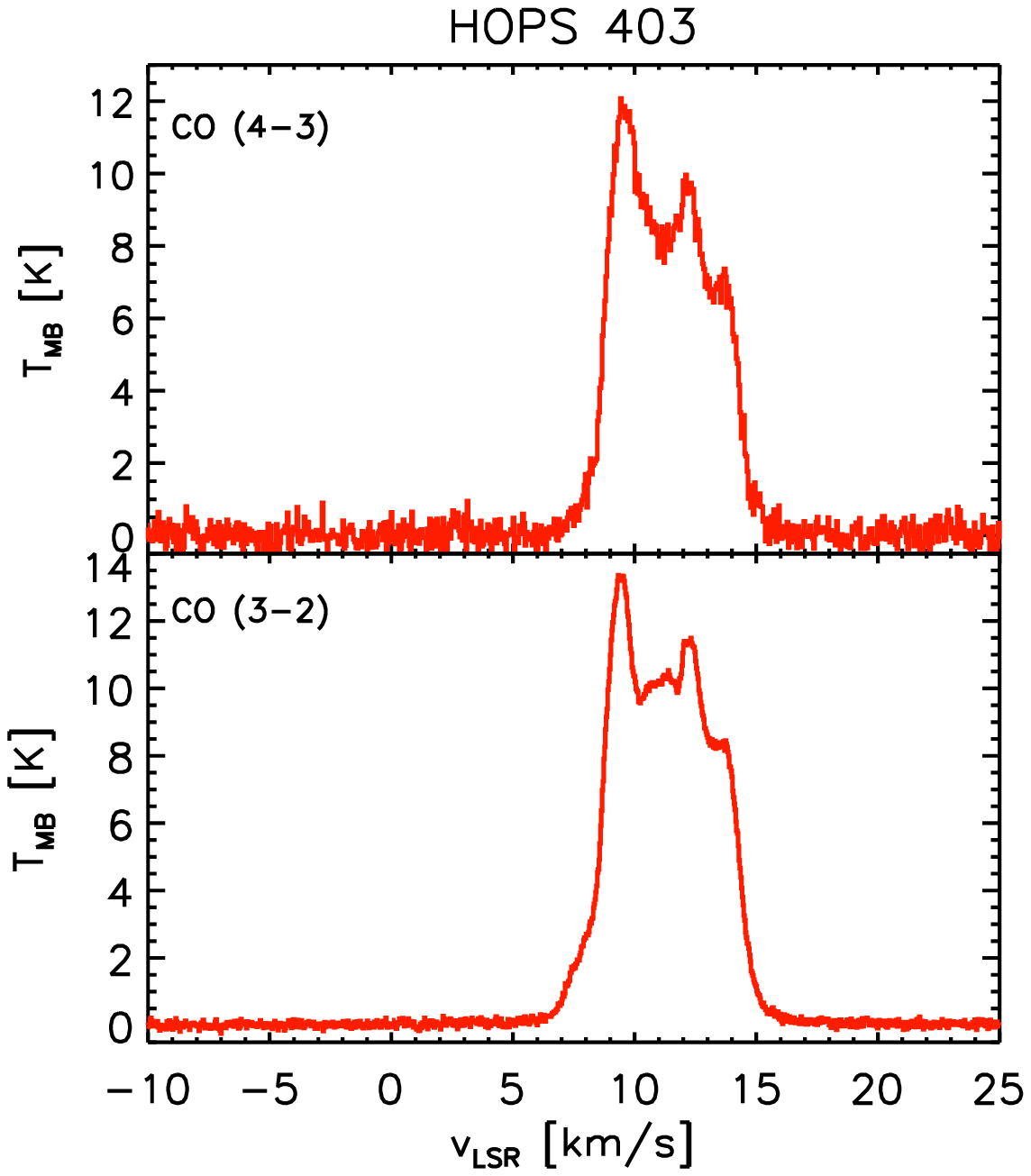}
\includegraphics[width=5.5cm]{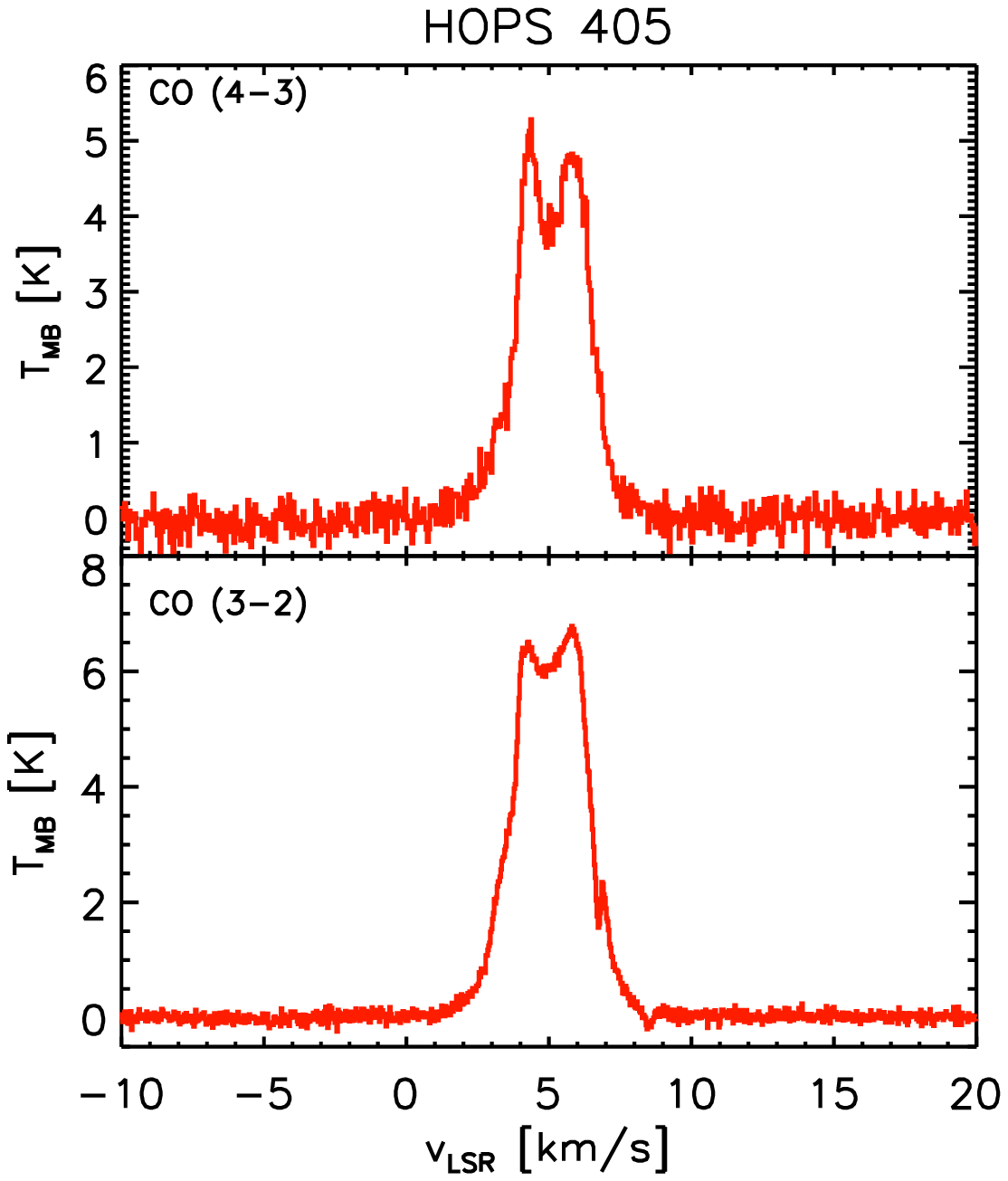}
\includegraphics[width=5.5cm]{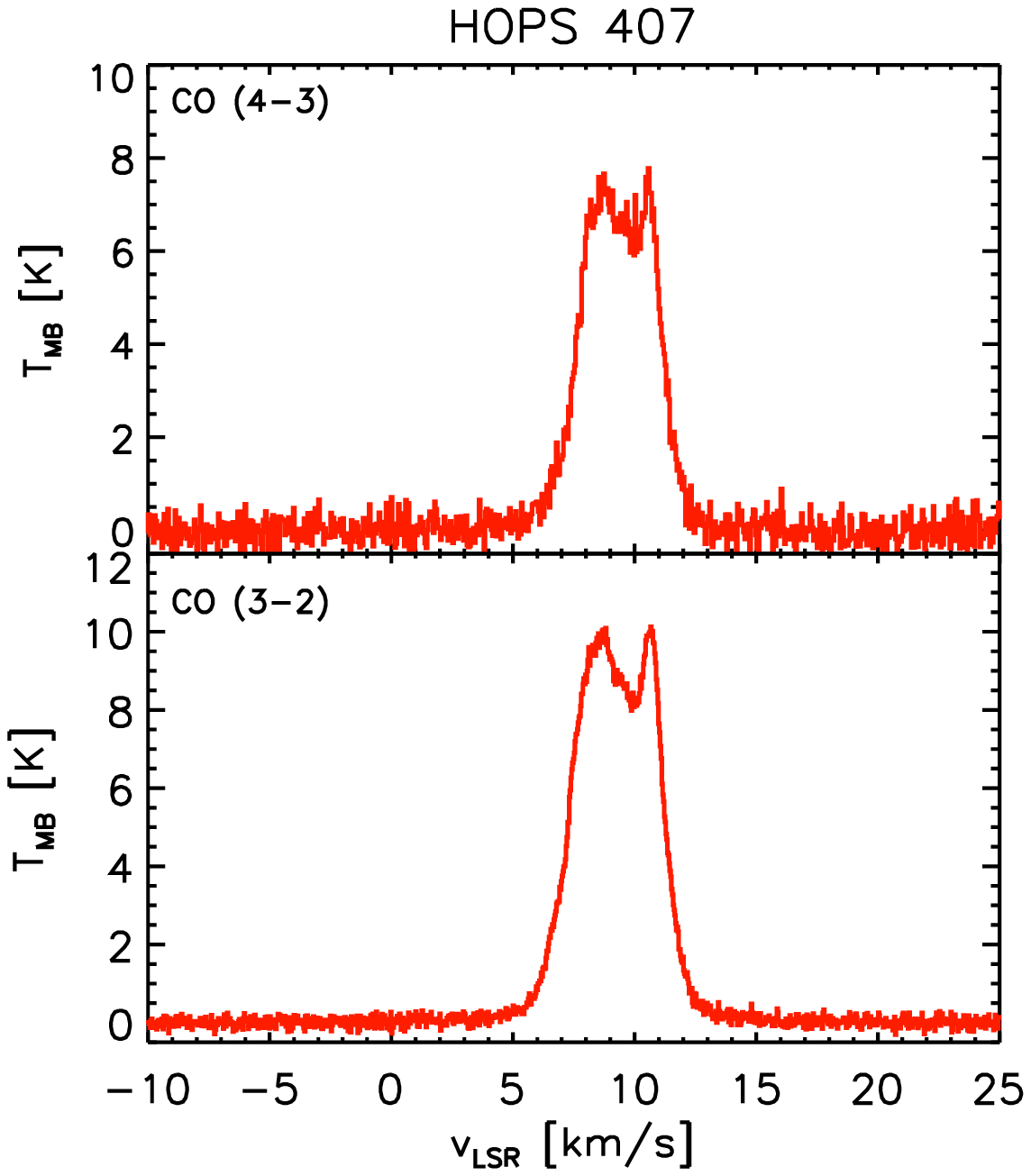}
\includegraphics[width=5.5cm]{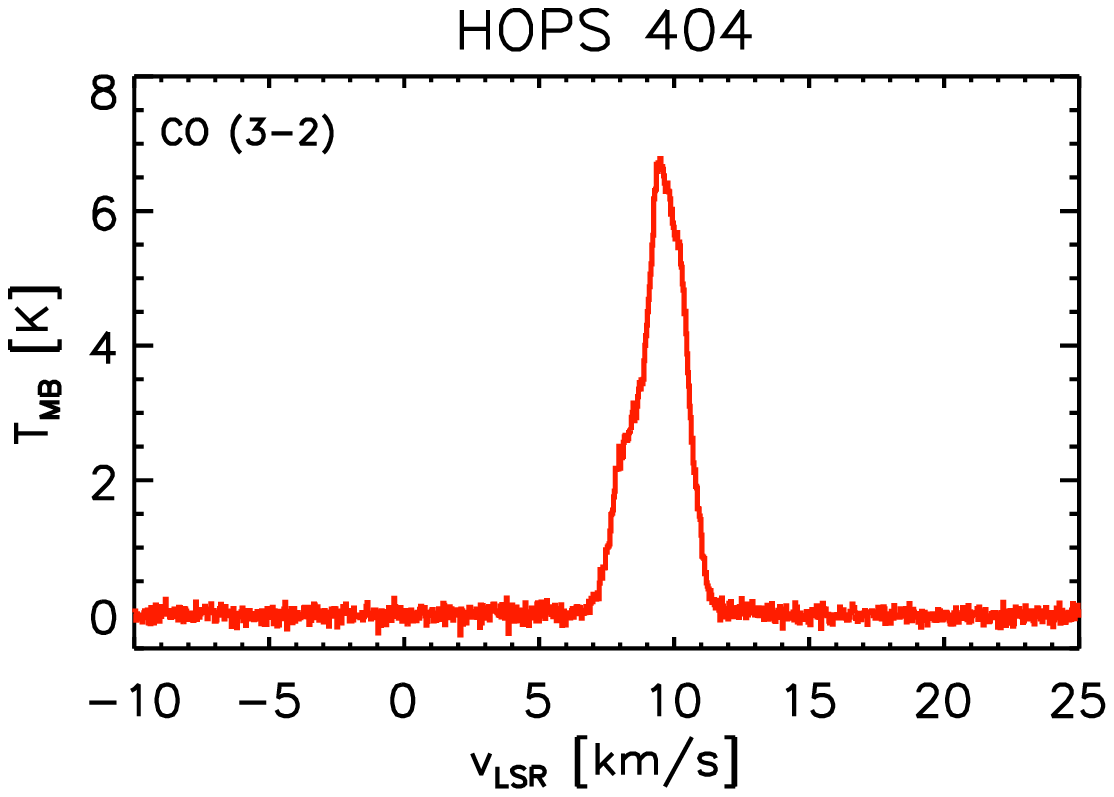}
\caption{}
\label{CO43_spec}
\end{figure*}

Intensity maps integrated over 2-5 km s$^{-1}$ velocity intervals are shown in Sect. \ref{sect_channelmaps}, while maps of the emission integrated over the velocities of the line wings are shown in Sect. \ref{sec:int_maps}. These maps show the structure of the high velocity gas for the sources where outlfow wings were detected. Outflows centered on the protostars with bipolar morphologies are apparent for HOPS 354, 358, 373, and 399. 

HOPS 341 shows two nearly perpendicular outflows centered on the source and the nearby ($\sim5''$) HOPS 340; these outflows are also apparent in 4.5~$\mu$m imaging of the NGC 2071 region \citep{megeath2012}. In our analyzis of the APEX data, we treat this as a single outflow. Based on the 4.5~$\mu$m data, the outflow in the north-south direction is related to the Class 0 source HOPS 340. As seen in the intensity maps shown in Sect. \ref{sect_channelmaps}, the north-south outflow dominates the outflow emission detected toward HOPS 341 / 340, indicating that the outflow emission toward this source is dominated by HOPS 340.
As HOPS 372 and 399 are located $\sim$20$''$ from each other, the outflow parameters calculated for these sources correspond to the same outflow (Figures \ref{fig:int_co32} and \ref{fig:int_co43}). Based on the CARMA interferometer data \citep{tobin2016}, the high-velocity outflow emission toward HOPS 372 / 399 is attributed to HOPS 399. The outflow of HOPS 372 might contribute at low velocities, but we attribute the entire outflow to HOPS 399 in our analyzis. We label HOPS 372 as ``confusion'' in Table \ref{tab:source_properties} to suggest that it is not detected due to the HOPS 399 outflow.
HOPS 394 only shows red-shifted outflow emission, which may include a substantial contribution from outflows originating from HOPS 68/69.
HOPS 401 and 402 are around 19$''$ offset from each other. The red-shifted outflow emission toward the map centered on HOPS 401 is related to HOPS 358, but not HOPS 401 and 402. The blue-shifted outflow emission in this map may include a contribution from HOPS 401 and/or 402. Based on the results of \citet{karnath2020}, there is no indication of an outflow toward HOPS 402 (which appears more luminous). There is a hint of an outflow toward HOPS 401, but not a clear detection. We consider this as a tentative detection of an outflow related to HOPS 401. 
HOPS 397 is found on the edge of an outflow from HOPS 223. We consider the line wings toward HOPS 397 as part of the HOPS 223 outflow. In their higher resolution interferometer data, \citet{tobin2016} find tentative evidence of an outflow at this position.
Similar to the case of HOPS 372, we label this outflow as confusion in Table \ref{tab:source_properties}.
HOPS 354, one of our four protostars with isolated bipolar outflow detections, is found to have an outflow aligned with the plane of the sky with both blue- and red-shifted features on either side of the source. The outflow of HOPS 405 is also found to be in the plane of the sky in the interferometer data of \citet{tobin2016}. The outflow is also apparent in our intensity maps at velocities bracketing the systemic velocity (Sect. \ref{sect_channelmaps}); however, since the spectra of this source do not show distinct line wings, we exclude HOPS 405 from our analysis. 
\citet{tobin2016} and \citet{karnath2020} also detect compact outflows toward HOPS 400, 403, and 404 at higher angular resolution in CARMA and ALMA data, respectively. These outflows are not identified in our lower angular resolution data.

To define the velocity range over which the outflows are observed, we used the line emission as a function of velocity, as shown in Sect. \ref{sect_channelmaps}.
At velocity channels close to the systemic velocity the channel maps show large morphological variations due to changes in the line profile related to the superposition of different physical components and due to the opacity. The channel maps at higher velocities (relative to the systemic velocity) trace only the outflow component, and therefore do not show large spatial variations from channel to channel. By visual inspection of emission from channel to channel, we defined the velocity ranges corresponding to the blue- and the red-shifted outflow wings. The velocity intervals and the lengths of each outflow are listed in Table \ref{table:outflowmass}. Figures \ref{fig:int_co32} and \ref{fig:int_co43} show the integrated intensities corresponding to the blue-shifted and red-shifted outflow lobes integrated over the velocities shown in Table \ref{table:outflowmass}.

\section{Physical properties of the outflows}
\label{sec:physical_properties}

\subsection{Outflow masses}
\label{sec:mass}

We derived outflow masses in the LTE approximation, using the method described in \citet{yildiz2015}.
To calculate the mass of the blue- and red-shifted outflow lobes, we used the intensities integrated over the velocity intervals listed in Table \ref{table:outflowmass}.
We calculated the upper level column density ($N_{\rm{u}}$) for each pixel toward which we detected high-velocity wings corresponding to the outflow as
\begin{equation}
\centering
N_u=\frac{\beta f^2 \int T_{\rm{mb}} {\rm{d}} V}{A_{\rm{ul}}},
\end{equation}
where $\beta$ is a constant that equals to 8$\pi k/hc^3$ or 1937 cm$^{-2}$ GHz$^{-2}$ K$^{-1}$ km$^{-1}$, $f$ is the frequency in GHz, the integral of $\int T_{\rm{mb}} {\rm{d}}V$ is the integrated intensity for the determined velocity interval, and $A_{\rm{ul}}$ is the Einstein $A$ coefficient.

The total column density corresponding to a single pixel then can be derived as
\begin{equation}
N_{\rm{total}} = \frac{N_{\rm{u}}}{g_{\rm{u}}} Q(T) \exp\frac{E_{\rm{u}}}{k T_{\rm{ex}}},
\end{equation}
where $g_{\rm{u}}$ is (2$J$+1), $Q(T)$ is the partition function, $E_{\rm{u}}$ is the energy of the upper energy level measured from ground state, $k$ is the Boltzmann constant, and $T_{\rm{ex}}$ is the excitation temperature assumed to be 75 K, similar to \citet{yildiz2015}.

The total outflow mass then can be derived using a sum over all pixels showing high velocity emission
\begin{equation}
M_{\rm{outflow}} = \mu_{\rm{H}_2} m_{\rm{H}} A \left[ \frac{{\rm{H_2}}}{{\rm{^{12}CO}}} \right] \sum_j N_{{\rm{total}},j},
\end{equation}
where $\mu_{\rm{H}_2}$ is 2.8, $m_{\rm{H}}$ is the mass of hydrogen, and $A$ is the projected area of an individual pixel at the adopted distance of 420 pc \citep{menten2007}. For the [H$_2$/$^{12}$CO] abundance ratio, we used $1.2\times10^4$ (e.g., \citealp{yildiz2015}). Table \ref{table:outflowmass} shows the outflow masses for each source. 
The mass estimate for HOPS 394 was made by taking all the outflow emission seen toward the source into account, but is considered to be an upper limit, given that some of the emission might correspond to other sources, such as HOPS 68. To estimate a lower limit for the outflow mass of the red-shifted lobe of HOPS 394, we assume that only the northern one of the two blobs seen in the outflow maps (\ref{fig:int_co32} and \ref{fig:int_co43}) corresponds to HOPS 394. This gives a mass estimate of $\sim$0.02 $M_\odot$ for both CO transitions, about half of the value found by taking all the red-shifted emission into account for the mass estimate.

\begin{table*}[ht]
\begin{minipage}[!h]{\linewidth}
\centering
\caption{Outflow masses, the largest spatial extent of the outflow lobes ($R$), and the velocity intervals for which the physical parameters of the blue- and red-shifted outflow lobes for the CO $J$=3-2 and $J$=4-3 transitions were calculated.
The uncertainties of the outflow masses are about 40\%, consisting of 20\% calibration error and 20\% error resulting from the uncertainty in the assumed excitation temperature.
The system temperature ($T_{\rm{sys}}$) is given to compare the sensitivity of observations.
The outflow detected toward the Class 0 protostar HOPS 340 might have a contamination from the PBR HOPS 341, and the outflow detected toward the Class I protostar HOPS 223 might have a contamination from the PBR HOPS 397.
We do not show the parameters of the outflow detected toward HOPS 372, as it is attributed to HOPS 399.
}
\label{table:outflowmass}
\begin{tabular}{l c c c c c c c c c c c c c}

\hline

Source& Line& $T_{\rm{sys}}$& $M_{\rm{blue}}$& $M_{\rm{red}}$& $R_{\rm{blue}}$& $R_{\rm{red}}$& 
$V_{\rm{blue}}$& $V_{\rm{red}}$\\
& & (K)& ($M_\odot$)& ($M_\odot$)& (AU)& (AU)& (km s$^{-1}$)& (km s$^{-1}$)\\
\hline

HOPS 340& CO $J$=3-2& 229& 
0.11& 0.05&
2.9$\times$10$^{4}$& 2.9$\times$10$^{4}$& 
(-5, 6.1)& (13.6, 25)\\ 

& CO $J$=4-3& 758&
0.07& 0.03&
2.9$\times$10$^{4}$& 2.9$\times$10$^{4}$& 
(-5, 6.4)& (13.1, 25)\\

HOPS 354& CO $J$=3-2& 280&
0.02& 0.06&
2.9$\times$10$^{4}$& 2.2$\times$10$^{4}$& 
(-15, -2.8)& (2.5, 15)\\

& CO $J$=4-3& 1001&
0.02& 0.02&
2.9$\times$10$^{4}$& 2.2$\times$10$^{4}$& 
(-15, -2.0)& (2.7, 15) \\

HOPS 358& CO $J$=3-2& 226&
0.07& 0.05&
3.3$\times$10$^{4}$& 3.8$\times$10$^{4}$& 
(-5, 6.7)& (13.6, 25)\\ 

& CO $J$=4-3& 701&
0.04& 0.05&
3.3$\times$10$^{4}$& 3.8$\times$10$^{4}$& 
(-5, 6.8)& (12.7, 25)\\

HOPS 373& CO $J$=3-2& 318& 
0.06& 0.09&
3.2$\times$10$^{4}$& 3.3$\times$10$^{4}$& 
(-5, 8.0)& (12.7, 25)\\

& CO $J$=4-3& 1671&
0.03& 0.04&
3.2$\times$10$^{4}$& 3.3$\times$10$^{4}$& 
(-5, 7.7)& (13.3, 25)\\

HOPS 394& CO $J$=3-2& 219&
--& 0.05&
--& 2.9$\times$10$^{4}$& 
--& (14.6, 25)\\

& CO $J$=4-3& 618&
--& 0.04&
--& 2.9$\times$10$^{4}$& 
--& (14.3, 25)\\

HOPS 223& CO $J$=3-2& 234&
0.07& 0.02&
3.3$\times$10$^{4}$& 3.3$\times$10$^{4}$& 
(-10, 0.8)& (7.0, 20)\\

& CO $J$=4-3& 716&
0.03& 0.01&
3.3$\times$10$^{4}$& 3.3$\times$10$^{4}$& 
(-10, 1.0)& (6.5, 20)\\

HOPS 399& CO $J$=3-2& 331&
0.06& 0.05&
3.6$\times$10$^{4}$& 3.3$\times$10$^{4}$& 
(-5, 6.8)& (13, 25)\\

& CO $J$=4-3& 1785&
0.03& 0.03&
3.6$\times$10$^{4}$& 3.3$\times$10$^{4}$& 
(-5, 7.0)& (12.8, 25)\\

HOPS 401& CO $J$=3-2& 225&
0.03& --&
3.3$\times$10$^{4}$& --& 
(-5, 6.1)& --\\

& CO $J$=4-3& 676&
0.02& --&
3.3$\times$10$^{4}$& --& 
(-5, 6.8)& --\\

\hline
\end{tabular}
\end{minipage}
\end{table*}

\subsection{Outflow forces}
\label{sec:force}

We calculated the outflow force $F_{\rm{CO}}$ using a method referred to as ``the v$_{\rm{max}}$ method'' by \citet{vandermarel2013}. This is one of the most commonly used methods to estimate the outflow force (\citealp{vandermarel2013} and references therein) and as such, it gives a good comparison to other studies. The force is given by
\begin{equation}
\label{eqn_force}
F_{\rm{CO}} = c_i \frac{M_{\rm{outflow}} V_{\rm{max}}^2}{R_{\rm{CO}}},
\end{equation}
where $V_{\rm{max}}$ is defined as $|V_{\rm{out}} - V_{\rm{LSR}}|$, with $V_{\rm{LSR}}$ as the systemic velocity of the source and $V_{\rm{out}}$ as the lowest (for the blue lobe) and highest (for the red lobe) velocity at which the outflow wing is detected. $V_{\rm{out}}$ for each source is shown in Table \ref{table:outflowmass} as a part of the velocity ranges defined for the outflows, $V_{\rm{blue}}$ and $V_{\rm{red}}$. 
The $V_{\rm{LSR}}$ values are around 10 km s$^{-1}$ for all the sources except HOPS 354 and HOPS 223, for which it is around 0 km s$^{-1}$ and 5 km s$^{-1}$, respectively, as seen in Figures \ref{CO43_spec} and \ref{HCOp_HNC_NH3_spec}. Based on these values and those for $V_{\rm{blue}}$ and $V_{\rm{red}}$ from Table \ref{table:outflowmass}, $V_{\rm{max}}$ is around 15 km s$^{-1}$ for all sources in both transitions.
The value of $c_i$ is a correction factor for the inclination taken from \citet{cabritbertout1990} for the inclinations shown in Table \ref{tab:source_properties}. For sources with intermediate inclination, and for those with no derived inclination we assumed an intermediate value of 50$^\circ$.
$R_{\rm{CO}}$ is the projected size of the outflow lobes as seen in the intensity maps corresponding to the blue- and red-shifted outflow lobes (Figures \ref{fig:int_co32} and \ref{fig:int_co43}).
Table \ref{tab:forces} shows the outflow forces calculated for the blue- and red-shifted outflow lobes for the CO $J$=3-2 and the CO $J$=4-3 transitions.
As mentioned in Sect. \ref{sec:mass}, the southern emission blob seen toward HOPS 394 (as seen in Figs. \ref{fig:int_co32} and \ref{fig:int_co43}) likely corresponds to other sources. For the calculation of the value in Table \ref{tab:forces}, we assumed that all the detected outflow emission corresponds to HOPS 394, therefore that value is an upper limit of the outflow force of the red-shifted outflow lobe of HOPS 394. When assuming that only the northern emission blob corresponds to HOPS 394, the outflow force for both CO transitions is $\sim$1.7$\times$10$^{-4}$ $M_\odot$ km s$^{-1}$ yr$^{-1}$, using an $R_{\rm{CO}}$ value of 1.7$\times$10$^4$ AU.

\begin{figure}[h!]
\centering
\includegraphics[width=9.0cm]{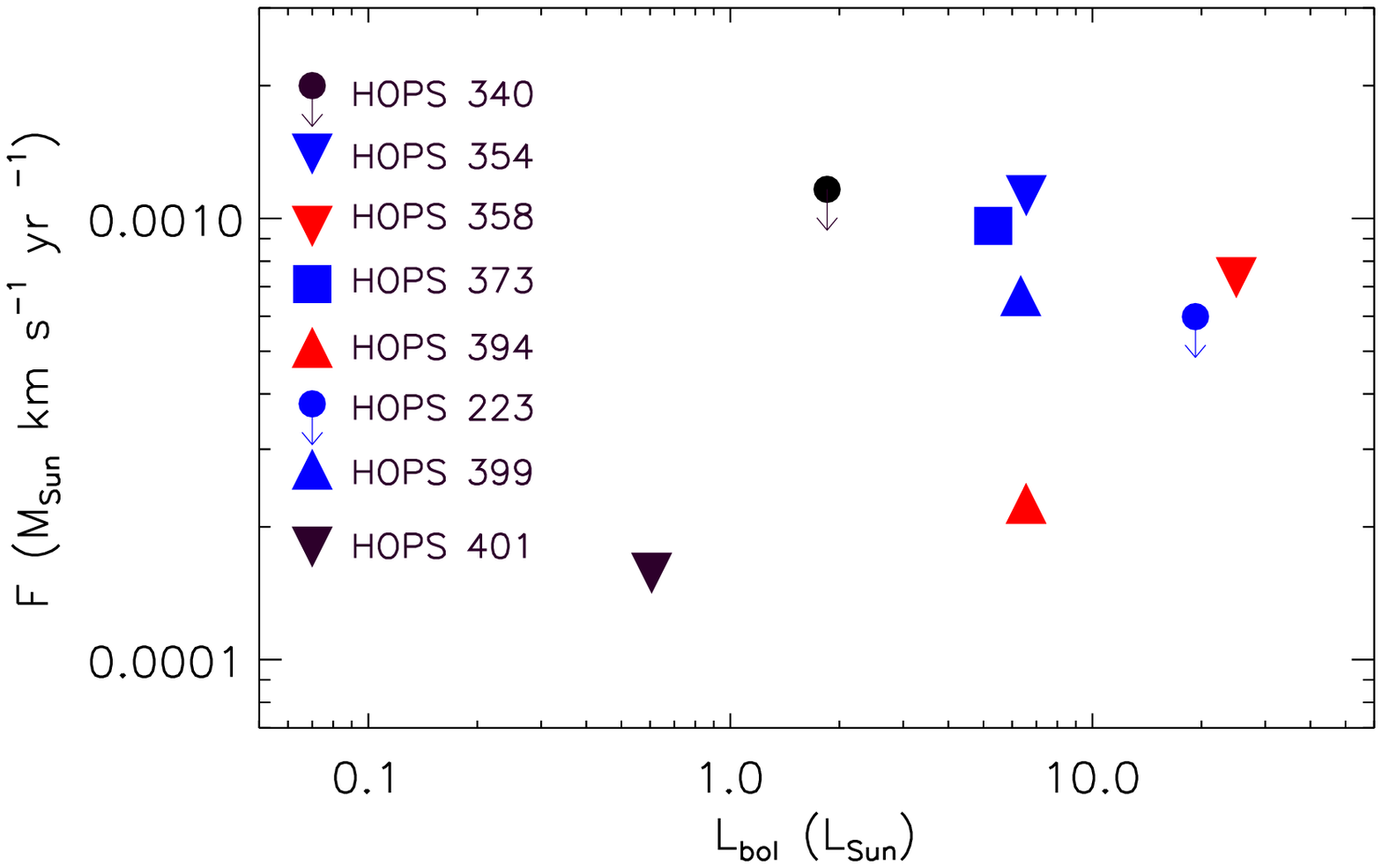}
\caption{
Outflow forces (for the combined blue and red outflow lobes) ($M_\odot$ km s$^{-1}$ yr$^{-1}$) versus $L_{\rm{bol}}$ ($L_\odot$) calculated using the CO $J$=3-2 transition for the protostars analyzed in this paper. The upper limit toward the Class 0 protostar HOPS 340 is due to possible contamination from the PBR HOPS 341, and the upper limit toward the Class I source HOPS 223 is due to possible contamination from the PBR HOPS 397. The detection of the HOPS 401 outflow is tentative.}
\label{ForceComparison}
\end{figure}

\subsection{Dynamical age, mass-loss rate, and kinetic luminosity}

Based on the parameters derived above, the dynamical age of the outflow can be calculated as
\begin{equation}
t_{\rm{dyn}}=\frac{R_{\rm{CO}}}{V_{\rm{max}}},
\end{equation}
which assumes the velocity to be constant over the extent of the outflow. The outflow mass-loss rate is then determined by
\begin{equation}
\dot{M}_{\rm{outflow}}=\frac{M_{\rm{outflow}}}{t_{\rm{dyn}}}.
\end{equation}
Finally the kinetic luminosity is calculated using
\begin{equation}
L_{\rm{kin}}=\frac{1}{2} F_{\rm{CO}} V_{\rm{max}}.
\end{equation}
Table \ref{tab:forces} includes the kinetic luminosities, and Table \ref{tab:tdyn_massloss_rates} the dynamical ages and mass-loss rates of the sources with outflow detections.
For HOPS 394, when assuming that only the northern blob seen in Figures \ref{fig:int_co32} and \ref{fig:int_co43} corresponds to its red-shifted outflow emission, its dynamical age is $\sim$5$\times$10$^3$ year, its outflow mass-loss rate is $\sim$3.7$\times$10$^{-6}$ $M_\odot$ yr$^{-1}$, and its kinetic luminosity is $\sim$0.2 $L_\odot$.
As the outflow forces, mass-loss rates, and kinetic luminosities are calculated using the same $R_{\rm{CO}}$ values derived for each source, the fact that some of the outflows may be more extended than the maps does not affect the derived values of $F_{\rm{CO}}$, $\dot{M}_{\rm{outflow}}$, and $L_{\rm{kin}}$.

\subsection{Sources with no outflow detection}

The CO observations showed high-velocity wings toward nine of the 16 observed maps. For the rest of the maps (centered on seven PBRs), no outflow wings could be identified.
For these sources we derive upper limits of their outflow masses and forces. For this estimate, we assume the outflow to be detected toward 25\% of pixels across the map, similar to the sources with outflow detections. We also assume the outflow wings to be detected over a range of velocities equal to the average value of the velocity ranges over which the sources with clear outflow detections show outflow wings: $\sim$12 km s$^{-1}$ for both the blue- and red-shifted outflow wings. Then we derive the integrated intensity corresponding to the outflows based on the 1-$\sigma$ rms noise level (RMS) measured around the CO $J$=3-2 and CO $J$=4-3 lines as ${\rm{RMS}}~\Delta V N_{\rm{chan}} N_{\rm{pix}}$, where $\Delta V$ is the channel width, $N_{\rm{chan}}$ is the number of channels that cover the adopted velocity range, and $N_{\rm{pix}}$ is the number of pixels toward which the outflow is expected to be detected. The rms values used are the average rms levels measured toward the sources. In each case, we assumed both a red- and a blue-shifted outflow wing.
Based on the integrated intensity upper limits, we derived the corresponding outflow masses and forces using the methods described above in Sect. \ref{sec:mass} and Sect. \ref{sec:force}. The results are shown in Table \ref{tab:upperlimits}.

\begin{table*}[ht]
\begin{minipage}{\linewidth}
\centering
\caption{Outflow forces and kinetic luminosities of the outflows.}
\label{tab:forces}
\begin{tabular}{lcccccccc}
\hline
& \multicolumn{4}{c}{$F_{\rm{CO}}$ ($M_\odot$ km s$^{-1}$ yr$^{-1}$)}& 
\multicolumn{4}{c}{$L_{\rm{kin}}$ ($L_\odot)$}\\
Source& Blue $J$=3-2& Red $J$=3-2& Blue $J$=4-3& Red $J$=4-3&  
Blue $J$=3-2& Red $J$=3-2& Blue $J$=4-3& Red $J$=4-3\\
\hline

HOPS 340& 
8.22$\times$10$^{-4}$& 3.42$\times$10$^{-4}$& 4.99$\times$10$^{-4}$& 1.93$\times$10$^{-4}$& 
1.02& 0.42& 0.62& 0.24\\

HOPS 354& 
2.18$\times$10$^{-4}$& 9.08$\times$10$^{-4}$& 1.86$\times$10$^{-4}$& 3.37$\times$10$^{-4}$&
0.18& 0.65& 0.15& 0.24\\

HOPS 358& 
4.68$\times$10$^{-4}$& 2.66$\times$10$^{-4}$& 2.53$\times$10$^{-4}$& 2.44$\times$10$^{-4}$&
0.58& 0.33& 0.31& 0.30\\

HOPS 373& 
3.86$\times$10$^{-4}$& 5.76$\times$10$^{-4}$& 1.72$\times$10$^{-4}$& 2.27$\times$10$^{-4}$&
0.45& 0.76& 0.20& 0.30\\

HOPS 394& 
--& 2.26$\times$10$^{-4}$& --& 1.85$\times$10$^{-4}$&
--& 0.28& --& 0.23\\

HOPS 223& 
4.47$\times$10$^{-4}$& 1.52$\times$10$^{-4}$& 2.00$\times$10$^{-4}$& 7.24$\times$10$^{-5}$&
0.48& 0.21& 0.21& 0.10\\

HOPS 399& 
3.34$\times$10$^{-4}$& 3.34$\times$10$^{-4}$& 1.90$\times$10$^{-4}$& 1.71$\times$10$^{-4}$&
0.41& 0.41& 0.23& 0.21\\

HOPS 401& 
1.57$\times$10$^{-4}$& --& 1.14$\times$10$^{-4}$& --&
0.19& --& 0.14& --\\

\hline
\end{tabular}
\end{minipage}
\end{table*}

\begin{table*}[ht]
\begin{minipage}{\linewidth}
\centering
\caption{Dynamical timescales and mass-loss rates of the outflows.}
\label{tab:tdyn_massloss_rates}
\begin{tabular}{lccccccccc}
\hline
& \multicolumn{4}{c}{$t_{\rm{dyn}}$ (10$^3$ yr)}& \multicolumn{4}{c}{$\dot{M}$ (10$^{-6}$ $M_\odot$ yr$^{-1}$)}\\
Source& Blue $J$=3-2& Red $J$=3-2& Blue $J$=4-3& Red $J$=4-3& Blue $J$=3-2& Red $J$=3-2& Blue $J$=4-3& Red $J$=4-3\\
\hline

HOPS 340& 9.2& 9.2& 9.2& 9.2& 
12.0& 5.5& 7.6& 3.3\\

HOPS 354& 9.2& 7.0& 9.2& 7.0& 
2.2& 8.6& 2.2& 2.9\\

HOPS 358& 10.4& 12.0& 10.4& 12.0& 
6.7& 4.2& 3.8& 4.2\\

HOPS 373& 10.1& 10.4& 10.1& 10.4&
5.9& 8.6& 3.0& 3.8\\

HOPS 394& --& 9.2& --& 9.2&
--& 5.5& --& 4.4\\

HOPS 223& 10.4& 10.4& 10.4& 10.4&
6.7& 1.9& 2.9& 1.0\\

HOPS 399& 11.4& 10.4& 11.4& 10.4&
5.3& 4.8& 2.6& 2.9\\

HOPS 401& 10.4& --& 10.4& --&
2.9& --& 1.9& --\\

\hline
\end{tabular}
\end{minipage}
\end{table*}

\begin{table*}[h!]
\centering
\caption{1-$\sigma$ upper limits on the outflow masses and forces of sources with no outflow detections. The system temperature ($T_{\rm{sys}}$) is given to compare the sensitivity of observations.}
\begin{tabular}{lcccccc}
\hline
& \multicolumn{2}{c}{$T_{\rm{sys}}$ (K)}& \multicolumn{2}{c}{$M$ ($M_\odot$)}& \multicolumn{2}{c}{$F$ (10$^{-4}$ $M_\odot$ km s$^{-1}$ yr$^{-1}$)}\\
Source&	CO $J$=3-2& CO $J$=4-3& CO $J$=3-2& CO $J$=4-3& CO $J$=3-2& CO $J$=4-3\\

\hline

HOPS 359& 221&  620& 0.03&  0.04&  1.81&  2.48\\
HOPS 398& 275& 1179& 0.03&  0.07&  1.93&  4.15\\
HOPS 400& 243&  831& 0.05&  0.08&  1.87&  3.29\\
HOPS 403& 267& 1061& 0.04&  0.06&  1.73&  2.45\\
HOPS 404& 271&    -& 0.03&     -&  1.73&     -\\
HOPS 405& 224&  650& 0.02&  0.03&  2.49&  3.47\\
HOPS 407& 228&  704& 0.03&  0.04&  2.86&  4.30\\
		
\hline
\end{tabular}
\label{tab:upperlimits}
\end{table*}

\section{Infall signatures}
\label{sec:infall}

We used the HCO$^+$, H$^{13}$CO$^+$, and NH$_3$ spectra to search for infall signatures toward the PBRs. 
The typical kinematic signature of infall is red-shifted self-absorption in an optically thick line (e.g., \citealp{evans2003}). This can lead to a double-peaked, self-absorbed line with a stronger blue-shifted than red-shifted peak, or an asymmetric line which peaks at a blue-shifted velocity \citep{myers1996}. An optically thin tracer is used to establish the systemic velocity of the source.
We used HCO$^+$ $J$=3-2 as an optically thick tracer, and H$^{13}$CO$^+$ $J$=3-2 or NH$_3$ (1,1) as optically thin tracers to probe the infall asymmetry. The HCO$^+$, H$^{13}$CO$^+$, and NH$_3$ line profiles observed toward 12 PBRs are shown in Figure \ref{HCOp_HNC_NH3_spec}. 
As the noise levels for the NH$_3$ (1,1) lines are lower than those for the H$^{13}$CO$^+$ lines, we used the NH$_3$ (1,1) lines as an optically thin tracer whenever the NH$_3$ (1,1) spectra were observed (for 8 sources). For the rest of the sources we used the H$^{13}$CO$^+$ line. To avoid the effect of opacity on the results, we used one of the satellite lines of NH$_3$ (1,1): the $F_1$=0-1, $F$=1/2-3/2 transition at a frequency of 23696.0406 MHz. To estimate the opacity of NH$_3$ (1,1), we fit its hyperfine structure using Gildas/CLASS. The results are shown in Table \ref{tab:nh3_parameters}. The opacities derived correspond to the main group of the hyperfine components. The opacity of the component used in this paper for the infall analyzis is about 1/5 of that of the main line \citep{hotownes1983}.

\begin{figure}[h!]
\centering
\includegraphics[width=9.0cm, angle=0, trim=1cm 0cm 0cm 0cm,clip=true]{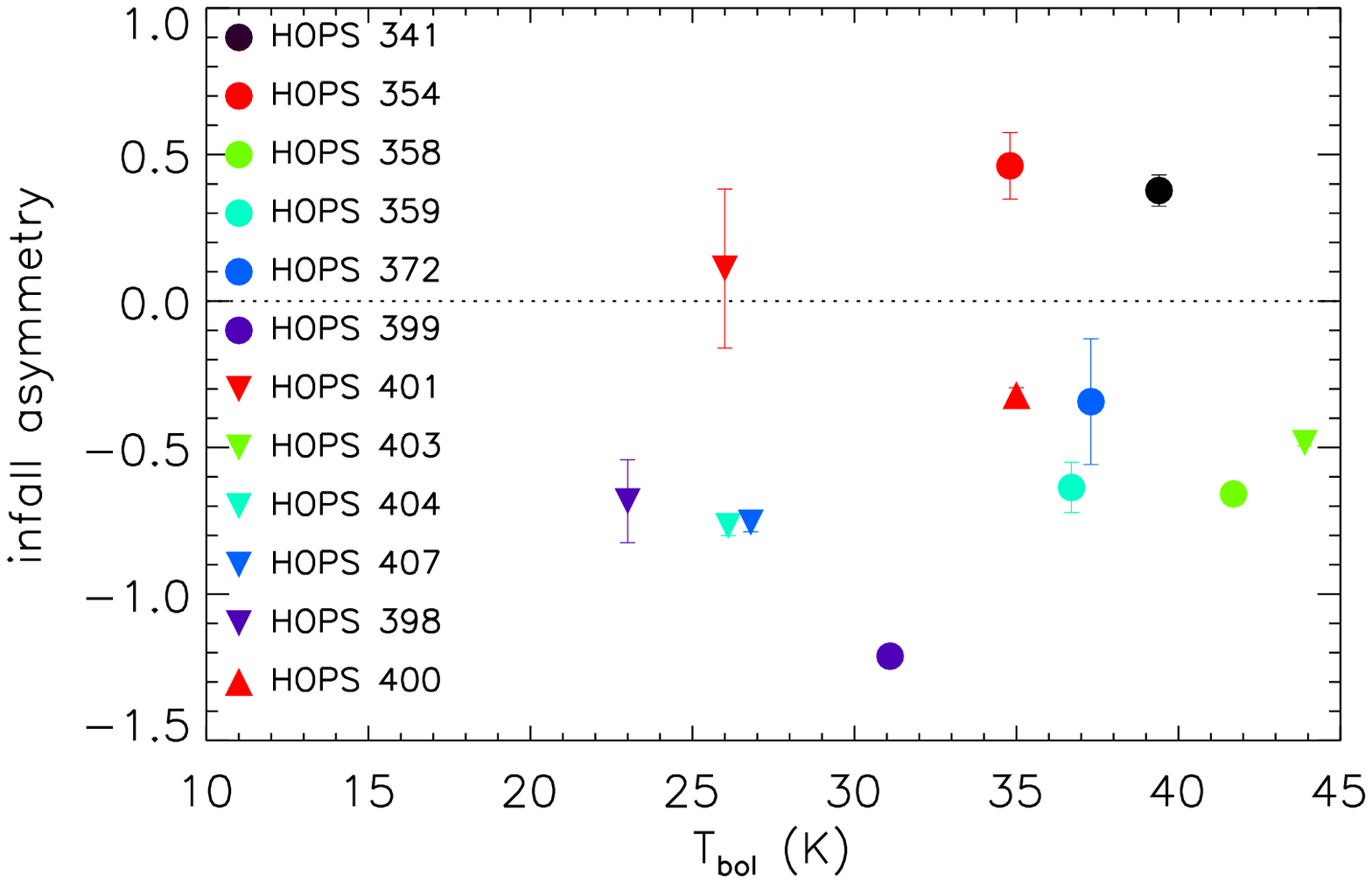}
\caption{Infall signatures based on the HCO$^+$, H$^{13}$CO$^+$, and NH$_3$ data shown in Figure \ref{HCOp_HNC_NH3_spec}. The infall asymmetry is defined as the velocity of the optically thick line (HCO$^+$) minus the velocity of the optically thin line (H$^{13}$CO$^+$ or NH$_3$) divided by the FWHM line width of the optically thin line as defined by \citet{mardones1997}.}
\label{fig:infall_asymmetry}
\end{figure}

We used the approach of \citet{mardones1997} to probe the infall asymmetries using the nondimensional parameter defined as 
\begin{equation}
\label{infall_eqn}
\delta V = (V_{\rm{thick}}-V_{\rm{thin}})/\Delta V_{\rm{thin}},
\end{equation}   
where $V_{\rm{thick}}$ is defined as the velocity of the strongest peak of the optically thick line (HCO$^+$), and $V_{\rm{thin}}$ and $\Delta V_{\rm{thin}}$ are defined as the velocity and full width at half maximum (FWHM) of the optically thin line (H$^{13}$CO$^+$ or NH$_3$). 
To measure the $V_{\rm{thin}}$ and $\Delta V_{\rm{thin}}$ parameters, we fit Gaussians to the H$^{13}$CO$^+$ and NH$_3$ (1,1) $F_1$=0-1, $F$=1/2-3/2 line profiles. The parameters of the fits are summarized in Table \ref{tab:nh3_parameters}. For $V_{\rm{thick}}$ we took the velocity corresponding to the peak intensity of the HCO$^+$ line profile for each source. For HOPS 372 and HOPS 398 there are two velocity components in the optically thin tracer. For the calculation of the infall parameters we consider the brighter velocity component in both cases.
The derived infall asymmetries (i.e., $\delta V$ defined in Eqn. \ref{infall_eqn}) are shown in Fig. \ref{fig:infall_asymmetry}. Based on this simple analysis nine of the twelve PBRs observed in HCO$^+$ and H$^{13}$CO$^+$ or NH$_3$ show infall asymmetry. The three PBRs that do not show an infall asymmetry have a detected outflow (HOPS 341 and 354) or a potential outflow (HOPS 401). However, considering the error of the infall parameter of HOPS 401, an infall asymmetry cannot be ruled out. The larger error is due to the lower S/N data for this source.

The infall signatures toward some of the sources may correspond to more than one protostar, in case of nearby sources like HOPS 341 and 340, which are located $\sim$6$''$ from each other.
HOPS 358 has a nearby source HOPS 316, which is 13$''$ away. HOPS 316 is a weak sub-mm source \citep{furlan2016}, which was not detected in the VLA/ALMA Nascent Disk and Multiplicity (VANDAM) Survey \citep{tobin2020}. Therefore, even if it is less than a beam size away, it is unlikely to contaminate the line profiles toward HOPS 358.
HOPS 399 and HOPS 372 are located $\sim$25$''$ from each other, which is about the beam size of the observations, therefore, contamination of the two sources in the analyzis of infall is not significant. However, given that the H$^{13}$CO$^+$ line profile toward HOPS 372 shows two peaks, confusion of the emission corresponding to HOPS 372 and HOPS 399 cannot be ruled out at the angular resolution of the observations. HOPS 372 has a lower luminosity than HOPS 399, and is a less deeply embedded source with weaker HCO$^+$ lines. Therefore, the infall detection toward HOPS 372 may be due to confusion with HOPS 399.
There is another source which shows two peaks in the optically thin tracer used for the infall analysis: HOPS 398. However, since it is an isolated source, with the nearest source located 2 arcmin south \citep{stutz2013}, there is no contamination of the emission from another protostar.
The case of HOPS 401 and 402 is similar to HOPS 399 and HOPS 372: even if they are nearby sources, they are separated by $\sim$35$''$ and the line parameters of HOPS 401 are not affected by HOPS 402.
HOPS 407 is about 21.4$''$ from HOPS 331, which is a flat spectrum source with a low-mass envelope \citep{furlan2016}, and therefore the infall parameter derived for HOPS 407 is unlikely to be strongly contaminated by HOPS 331.
HOPS 400 is binary ($<$2$''$ separation, \citealp{karnath2020}), which is effectively isolated on larger scales. The infall signature toward this source therefore suggests infall onto the binary.
HOPS 403 may also be a binary on a small scale ($<$2$''$ separation, \citealp{karnath2020}), but isolated on large scales, therefore, infall toward this source may also be infall onto a binary.

\begin{table*}[ht]
\centering
\caption{Parameters of the NH$_3$ (1,1) and H$^{13}$CO$^+$ $J$=3-2 lines used in this paper. The integrated intensity, LSR velocity, FWHM line width, and peak intensity were obtained from the Gaussian fitting of the $F_1=0-1$, $F=1/2-3/2$ transition of NH$_3$ (1,1) at 23696.0406 MHz. The opacity of the main group is obtained from fitting the hyperfine structure of NH$_3$ (1,1). 
}
\begin{tabular}{cccrcccrr}
\hline
Source& Line& Intensity& $V_{\rm{thin}}$& $\Delta V_{\rm{thin}}$& $T_{\rm{MB}}$& $\tau_{\rm{main}}$& $V_{\rm{thick}}$& $\delta V$\\	
      & & (K km s$^{-1}$)& (km s$^{-1}$)& (km s$^{-1}$)&           (K)& & (km s$^{-1}$)&                    \\
\hline

HOPS 341& NH$_3$ (1,1)& 0.61$\pm$0.02&  9.83$\pm$0.01& 0.53$\pm$0.02& 1.07$\pm$0.03& 1.47$\pm$0.07& 
10.03$\pm$0.04& 0.38$\pm$0.05\\

HOPS 358& NH$_3$ (1,1)& 0.99$\pm$0.02& 10.39$\pm$0.01& 1.20$\pm$0.03& 0.78$\pm$0.03& 1.05$\pm$0.01& 
9.60$\pm$0.04& -0.66$\pm$0.04\\

HOPS 398& NH$_3$ (1,1)& 0.26$\pm$0.03& 10.46$\pm$0.02& 0.60$\pm$0.06& 0.41$\pm$0.03& 0.86$\pm$0.15& 
10.05$\pm$0.04& -0.68$\pm$0.14\\

HOPS 399& NH$_3$ (1,1)&   1.09$\pm$0.02&  9.39$\pm$0.01& 0.66$\pm$0.01& 1.56$\pm$0.02& 1.86$\pm$0.07& 
8.59$\pm$0.04&  -1.21$\pm$0.02\\

HOPS 400& NH$_3$ (1,1)&   1.07$\pm$0.02&  8.51$\pm$0.01& 0.56$\pm$0.01& 1.79$\pm$0.03& 2.77$\pm$0.02&
8.33$\pm$0.04&  -0.32$\pm$0.03\\

HOPS 403& NH$_3$ (1,1)&   0.74$\pm$0.03& 10.46$\pm$0.03& 1.59$\pm$0.01& 0.43$\pm$0.04& 0.44$\pm$0.06&
9.69$\pm$0.04&  -0.48$\pm$0.01\\

HOPS 404& NH$_3$ (1,1)&   0.67$\pm$0.02&  9.63$\pm$0.01& 0.43$\pm$0.01& 1.48$\pm$0.04& 2.41$\pm$0.09&
9.30$\pm$0.04&  -0.77$\pm$0.03\\

HOPS 407& NH$_3$ (1,1)&   0.91$\pm$0.02&  9.61$\pm$0.01& 0.45$\pm$0.01& 1.91$\pm$0.04& 4.32$\pm$0.09&
9.27$\pm$0.04&  -0.76$\pm$0.03\\

HOPS 354& H$^{13}$CO$^+$& 1.49$\pm$0.09&  1.02$\pm$0.05& 1.56$\pm$0.11& 0.90$\pm$0.11& --&
1.74$\pm$0.04&  0.46$\pm$0.11\\

HOPS 359& H$^{13}$CO$^+$& 1.54$\pm$0.08&  9.30$\pm$0.03& 0.99$\pm$0.06& 1.46$\pm$0.13& --&
8.67$\pm$0.04&  -0.64$\pm$0.09\\

HOPS 372& H$^{13}$CO$^+$& 0.76$\pm$0.09&  9.18$\pm$0.05& 0.99$\pm$0.15& 0.73$\pm$0.13& --&
8.84$\pm$0.04&  -0.34$\pm$0.21\\
                
HOPS 401& H$^{13}$CO$^+$& 0.53$\pm$0.08& 10.19$\pm$0.08& 0.99$\pm$0.19& 0.51$\pm$0.11& --&
10.30$\pm$0.04&  0.11$\pm$0.27\\

\hline
\end{tabular}
\label{tab:nh3_parameters}
\end{table*}

Although rotation and outflow may produce asymmetric profiles with either positive or negative values of $\delta V$ for randomly oriented protostars, \citet{mardones1997} argue that the preponderance of negative (i.e., blue-shifted) values of $\delta V$ are indicative of infall. 
They find an excess of blue-shifted values of $\delta V$ (with $\delta V < -0.25$) in a single dish survey of Class 0 protostars, with 50\% of the sources showing blue-shifted velocities.
A similar excess of blue-shifted velocity was not found toward Class I protostars; although these protostars almost certainly also have infall. \citet{mardones1997} suggest that the difference is due to different physical conditions in the envelopes surrounding Class 0 and I protostars. The PBRs also show an excess of blue-shifted velocities, similar to the Class 0 protostars shown by \citet{mardones1997}. This is the first direct observational evidence for infall motions toward PBRs. This is also further evidence that the PBRs are Class 0 protostars.

\section{Discussion}
\label{sec:discussion}

We observed CO $J$=3-2 and $J$=4-3 maps toward 16 very young Class 0 protostars, classified as PBRs. 
Of the 16 protostars, only four do not show even tentative evidence of high velocity outflow emission in the APEX, CARMA or recent ALMA imaging: HOPS 359, 398, 402, HOPS 404 (Table \ref{tab:source_properties}).
We estimated the masses and forces of the outflows adopting the LTE approximation.
The total outflow masses (including the contribution from both the red- and blue-shifted outflow lobes) for the sources with clear outflow detection are in the range between 0.03 and 0.16 $M_\odot$ for CO $J$=3-2, and in the range between 0.02 and 0.10 $M_\odot$ for CO $J$=4-3, assuming an excitation temperature of 75~K.
The outflow forces are in the range between $1.57\times10^{-4}$ and $1.16\times10^{-3}$ $M_\odot$ km s$^{-1}$ yr$^{-1}$ for CO $J$=3-2 and in the range between $1.14\times10^{-4}$ and $6.92\times10^{-4}$ $M_\odot$ km s$^{-1}$ yr$^{-1}$ for CO $J$=4-3.
Most sources of our sample show an infall asymmetry based on the observed HCO$^+$, H$^{13}$CO$^+$, and NH$_3$ lines.

Some of the PBRs show no outflow emission in the CO $J$=3-2 and $J$=4-3 APEX maps but have outflow detections in CO $J$=1-0 at higher angular resolution (3-6$''$) with CARMA \citep{tobin2016}. These sources are HOPS 400, 403, 405, and 407. \citet{karnath2020} detected compact outflows at even higher angular resolution ($0.25''\times0.24''$) in CO $J$=3-2 with ALMA toward HOPS 400, 403, 404, and possibly toward HOPS 401. HOPS 404 has a very compact and low velocity outflow that may be driven by a FHSC.
These sources stress the importance of high angular resolution when separating contributions of outflows related to different sources and when detecting compact outflows such as those observed with ALMA and CARMA toward HOPS 400, 403, and 404.
It is not established that the high velocity emission detected near HOPS 401 and 402 in our APEX data is related to outflows from these sources, as a clear bipolar morphology is not apparent (Fig. \ref{fig:int_co32} and Fig. \ref{fig:int_co43}). 
HOPS 401 and 402 were were not detected in the CO $J$=1-0 CARMA maps \citep{tobin2016}. There is a hint of an outflow detected with ALMA toward HOPS 401 \citep{karnath2020}, therefore, the high velocity emission detected with APEX may at least partially be related to HOPS 401. 
As these sources were not detected at 24 $\mu$m and show high opacities at 870 $\mu$m, they might represent very early stages of protostellar evolution \citep{karnath2020}.

Three sources are clearly detected in both the high-resolution CARMA CO $J$=1-0 data and in the APEX CO $J$=3-2 and $J$=4-3 maps: HOPS 373, HOPS 394, and HOPS 399. Due to the very different angular resolutions, the outflow parameters are not directly comparable, however, the APEX CO $J$=3-2 and CARMA data of HOPS 399 show very similar forces, $(6-7) \times 10^{-4}$ $M_\odot$ km s$^{-1}$ yr$^{-1}$; this is the highest force measured in the \citet{tobin2016} sample.
Two of these sources (HOPS 373 and HOPS 399) show a clear jet morphology even at the angular resolution of the APEX data. The more evolved protostar HOPS 223 is also detected by both APEX and CARMA with a jet morphology.

\begin{figure}[h!]
\centering
\includegraphics[width=8.5cm]{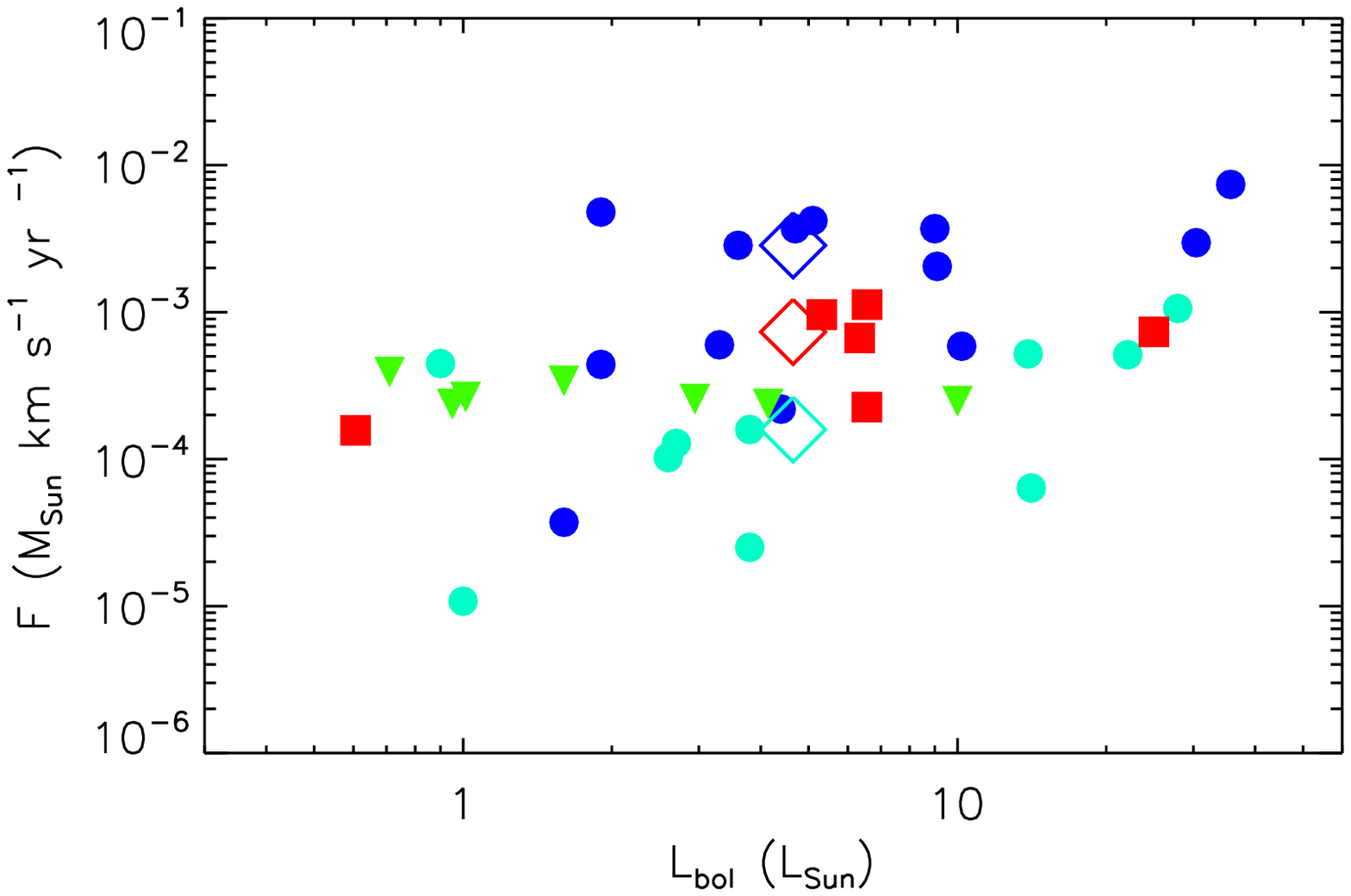}
\includegraphics[width=8.5cm]{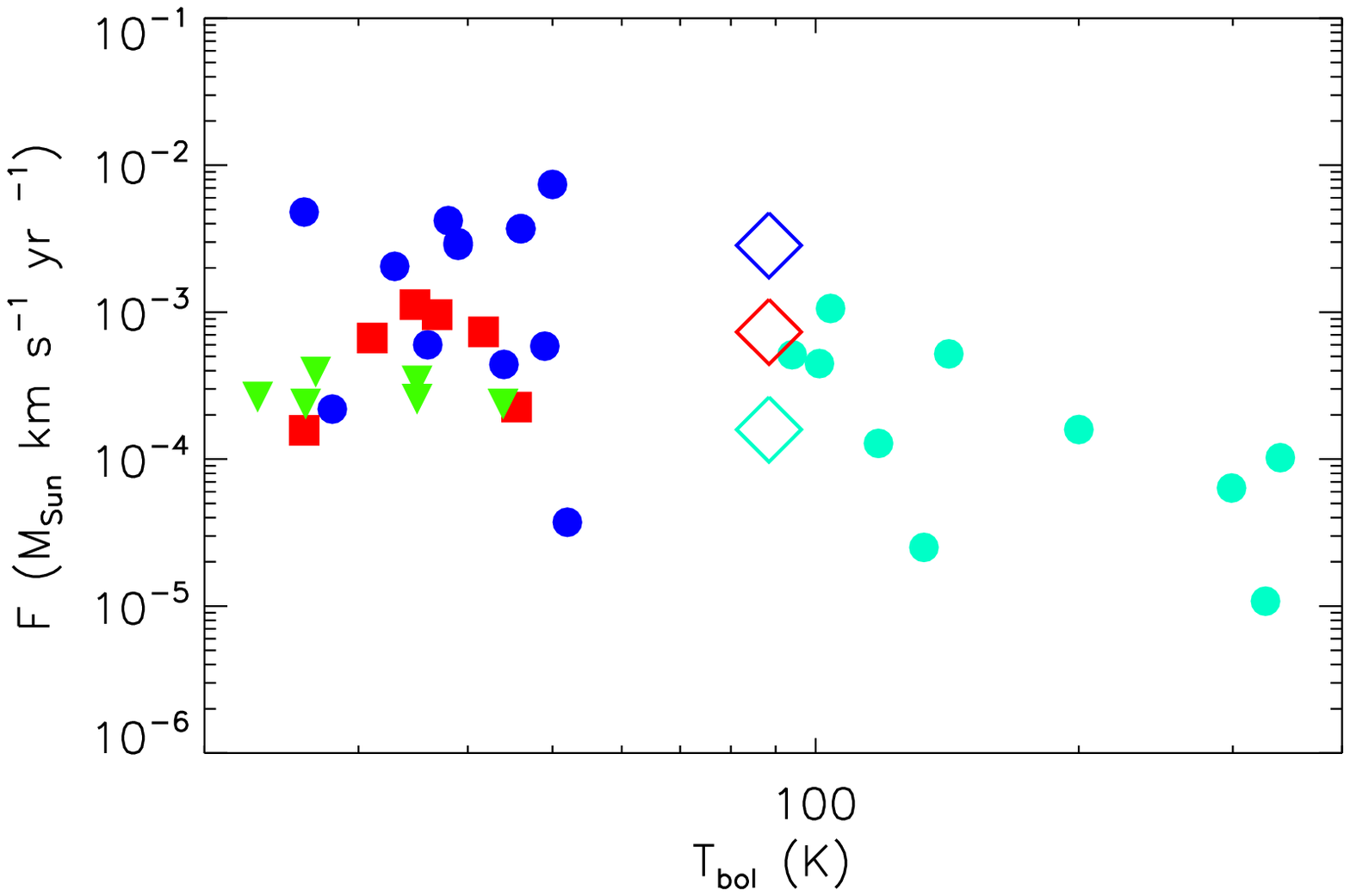}
\caption{
Outflow forces (for the combined blue and red outflow lobes) ($M_\odot$ km s$^{-1}$ yr$^{-1}$) versus $L_{\rm{bol}}$ ($L_\odot$) (top panel) and $T_{\rm{bol}}$ (K) (bottom panel) calculated using the CO $J$=3-2 transition and assuming an excitation temperature of 75~K for the PBRs analyzed in this paper (red squares). The Class 0 source HOPS 340 and Class I source HOPS 223 are not shown. The outflow forces calculated using the same transition and excitation temperature by \citet{yildiz2015} for a sample of Class 0 protostars (blue circles) and Class I protostars (cyan circles). The green triangles are the 1-$\sigma$ upper limits for the PBRs with no detected outflow emission. The diamonds represent the median values of each sample: Class 0 protostars from \citet{yildiz2015} (blue), Class I protostars from \citet{yildiz2015} (cyan), and the protostars studied in this paper (red).}
\label{fig:forces_comparison}
\end{figure}

The outflow forces derived for the PBRs with detected outflows are in the same range as those measured for other Class 0 protostars (Fig. \ref{fig:forces_comparison}). 
HOPS 400 and HOPS 403, which are upper limits in this diagram, show similar forces in ALMA CO $J$=3-2 imaging \citep{karnath2020}.
The forces for the Class 0 sources from the sample of \citet{yildiz2015} plotted in Fig. \ref{fig:forces_comparison} were also derived from CO $J$=3-2 APEX data in an LTE approximation, assuming an excitation temperature of 75~K, and therefore, are directly comparable to our PBRs results. 
The small fraction of protostars that are PBRs indicates that they have ages $\leq$ 25,000 year, 1/6 of the Class 0 phase \citep{stutz2013}.
The result that the outflow forces measured for the PBRs are similar to those derived for more evolved Class 0 protostars suggests that outflows develop quickly during the Class 0 phase.
ALMA imaging of the PBRs HOPS 402 and HOPS 404 shows no outflow or, in the case of HOPS 404, an unusually low-velocity outflow. These may show protostars before a high-velocity outflow is established \citep{karnath2020}.

Furthermore, the correlation between the outflow force and the bolometric luminosity suggested by previous studies (i.e., \citealp{bontemps1996}, \citealp{curtis2010}, \citealp{vandermarel2013}) is not seen for the sources with outflow detections. This may be due to the small number of sources which cover a limited range of bolometric luminosities, with 5 of the 8 sources in the range between an $L_{\rm{bol}}$ of $\sim$1.7 $L_\odot$ and $\sim$10 $L_\odot$.
Also, as seen in Figures \ref{fig:int_co32} and \ref{fig:int_co43}, some of the outflows detected toward the PBRs may be confused with outflow emission from other nearby protostars. Separating the contribution of CO emission related to the different sources is not possible at the angular resolution of APEX (and other single dish telescopes). Another difficulty may be that the CO $J$=3-2 transition was suggested to be subthermally excited in outflows (\citealp{ginsburg2011}, \citealp{dunham2014}), and therefore underestimates the outflow masses.

The large scatter in outflow properties and the lack of a significant dependence with $T_{\rm{bol}}$ or other SED properties has also been found for more evolved prorototars.
Habel et al. (in prep.) found that outflow properties do not vary in a predictable way with SED class. They measured the outflow opening angles of protostars from 1.6 $\mu$m images obtained with the \textit{Hubble} Space Telescope. They found no evidence of a correlation between the outflow opening angles and evolutionary tracers derived from SED fitting, particularly across the Class I phase \citep{furlan2016}.

\section{Summary}

We presented CO $J$=3-2 and $J$=4-3 maps for 16 very young Class 0 protostars observed with APEX at angular resolutions of 14-19$''$. Outflows were detected toward six of the 16 PBRs, and toward an additional Class 0 and a Class I protostar.
Only four protostars do not show evidence of high velocity outflows in recent APEX, CARMA, and ALMA imaging. 
We derived physical properties for the outflows of these sources including masses, forces, and upper limits for the masses and forces of the PBRs without outflow detections. We found that the outflow forces are in the same range as those derived for Class 0 protostars. This result suggests that outflows develop quickly in the Class 0 phase.
There is no evidence for a correlation between the outflow force and bolometric luminosity for the protostars with outflow detections.

Using HCO$^+$, H$^{13}$CO$^+$, and NH$_3$ (1,1) data measured toward 12 of the PBRs, we found infall asymmetry toward nine of them. An excess of sources with infall asymmetry was also found earlier for Class 0 protostars, which is further evidence that the PBRs are primarily Class 0 protostars.

\begin{acknowledgements}
We thank the referee for comments which helped to improve our paper.
Support for Z. N. was provided by NASA Origin of the Solar System program 13-OSS13-0094 (Megeath PI). A. S. acknowledges funding through Fondecyt regular (project code 1180350) and Chilean Centro de Excelencia en Astrof\'isica y Tecnolog\'ias Afines (CATA) BASAL grant AFB-170002.
J.J.T. acknowledges support from NSF AST-1814762. The National Radio Astronomy
Observatory is a facility of the National Science Foundation operated under cooperative agreement by Associated Universities, Inc. The Green Bank Observatory is a facility of the National Science Foundation operated under cooperative agreement by Associated Universities, Inc.
This publication is based on data acquired with the Atacama Pathfinder Experiment (APEX). APEX is a collaboration between the Max-Planck-Institut fur Radioastronomie, the European Southern Observatory, and the Onsala Space Observatory.
\end{acknowledgements}

%
%

\onecolumn

\begin{appendix}

\section{Line profiles observed toward the PBRs}

\begin{figure*}[h!]
\begin{center}
\includegraphics[width=4.7cm]{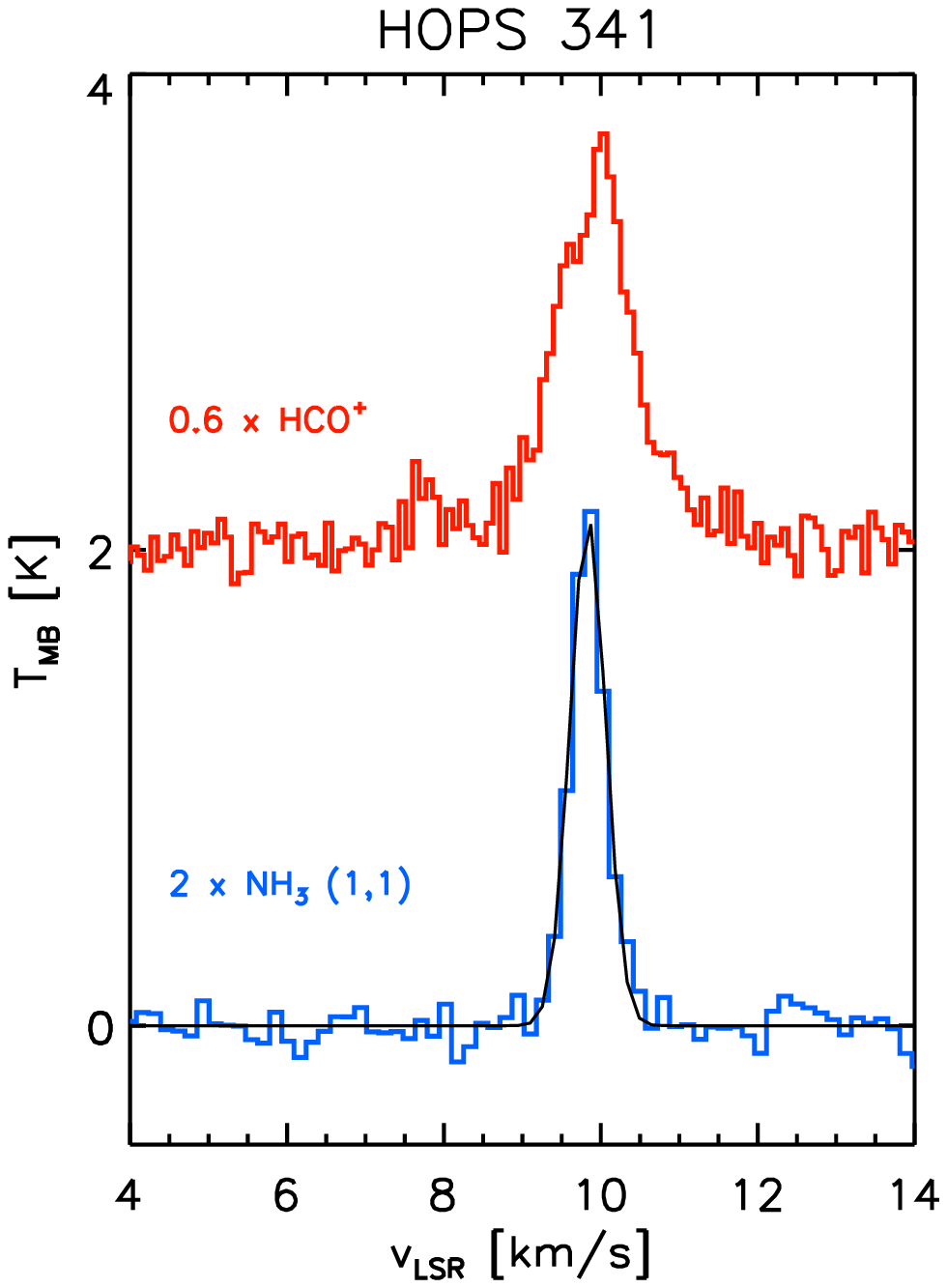}
\includegraphics[width=4.7cm]{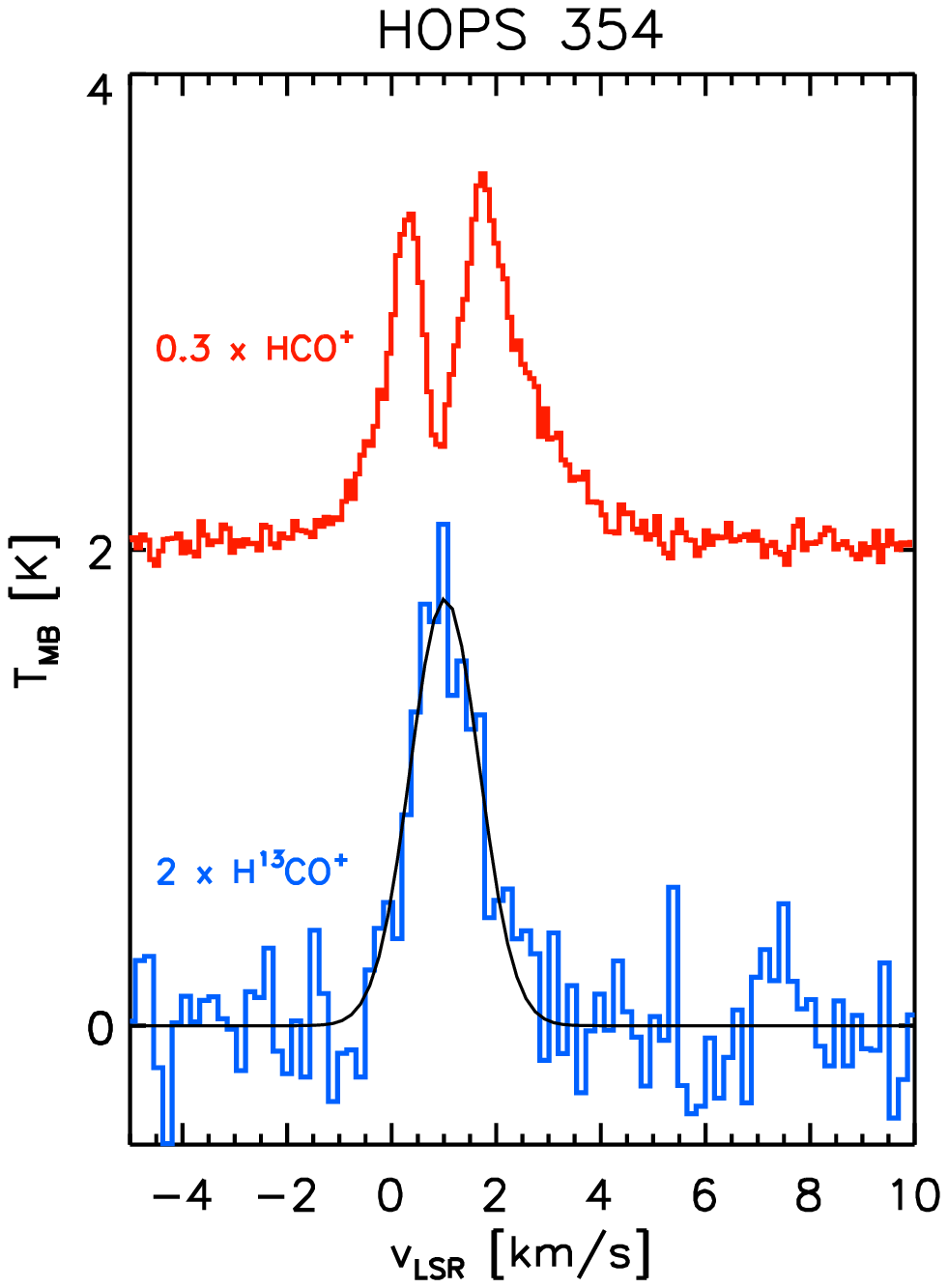}
\includegraphics[width=4.7cm]{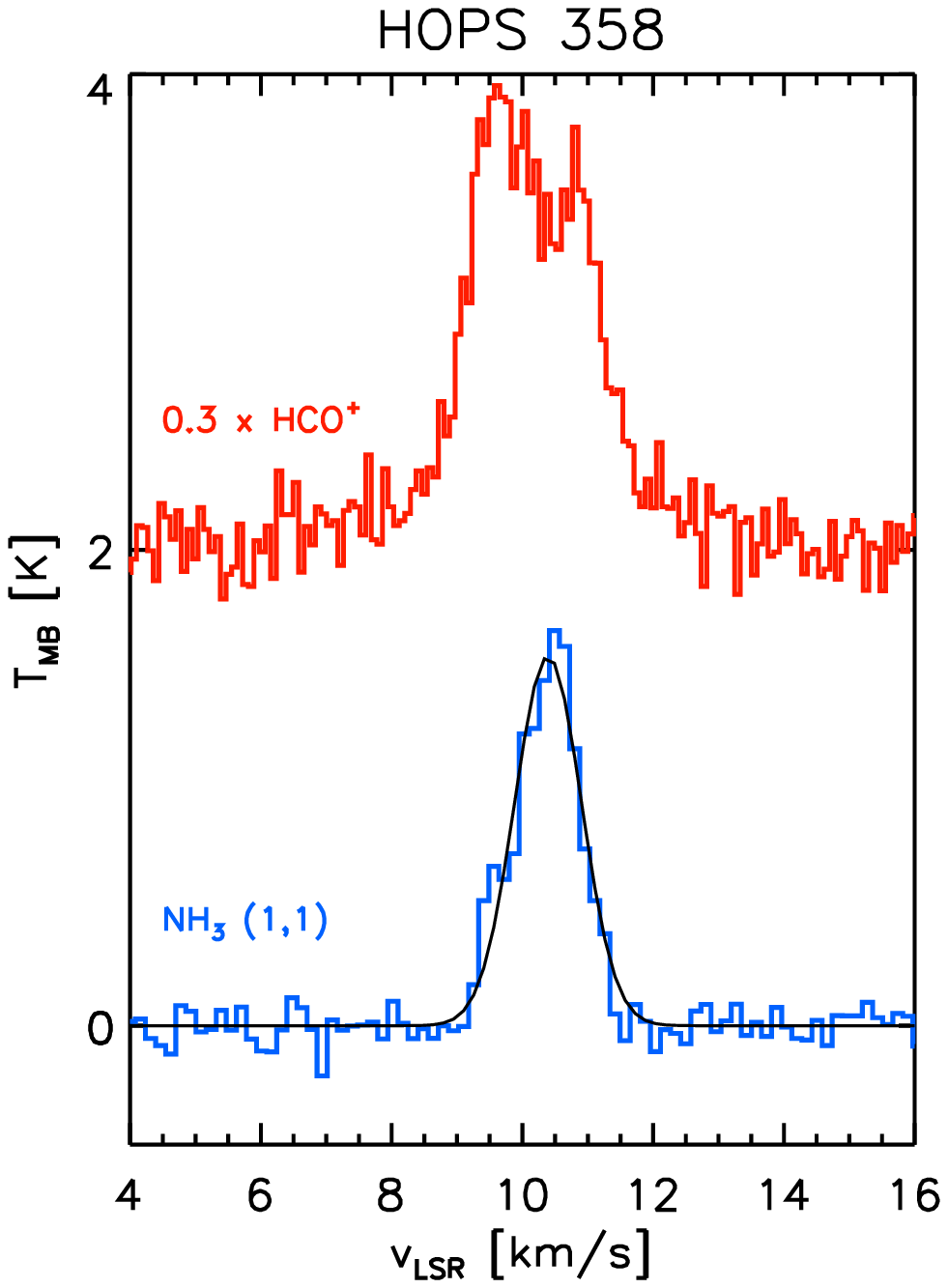}
\includegraphics[width=4.7cm]{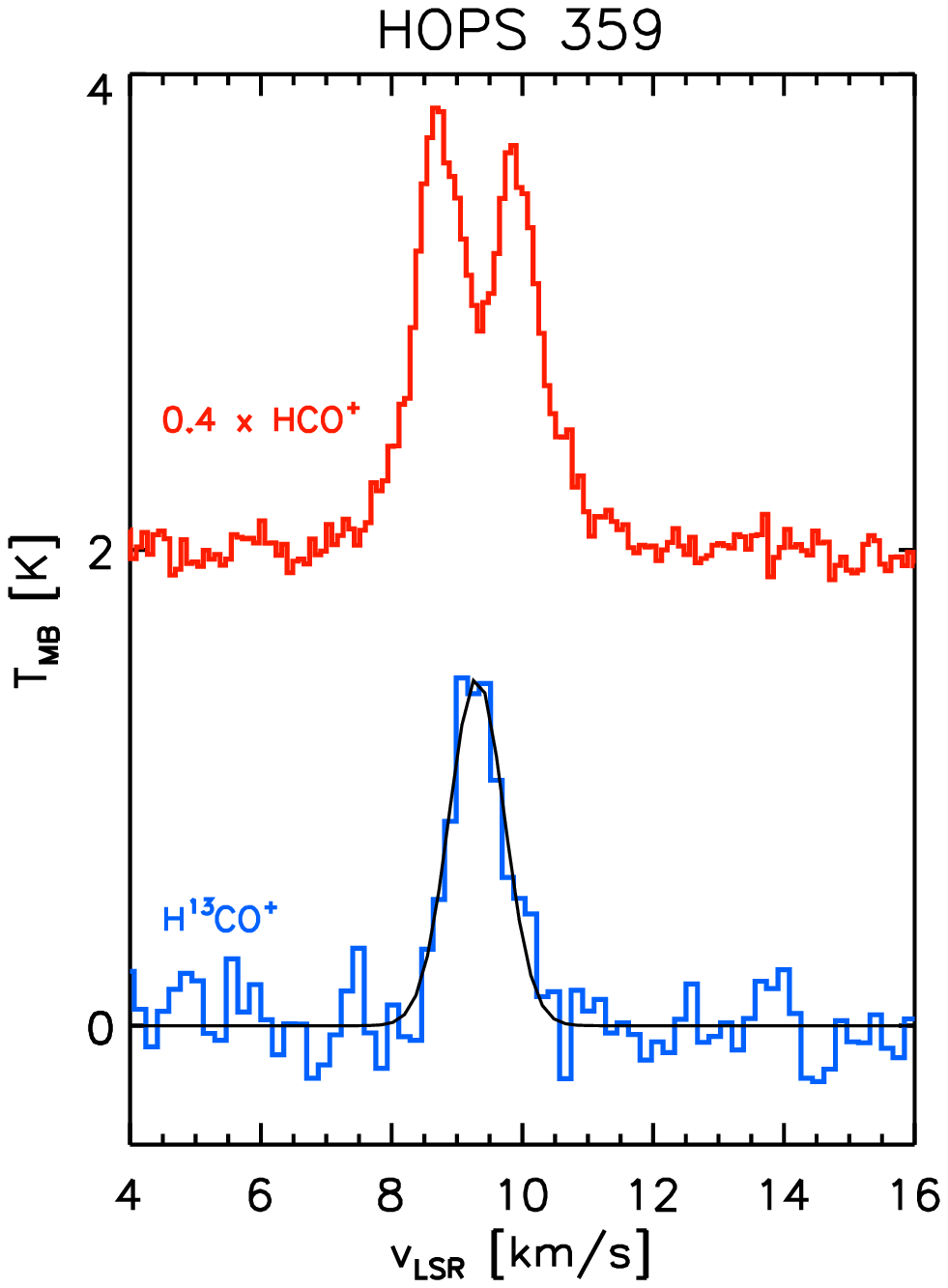}
\includegraphics[width=4.7cm]{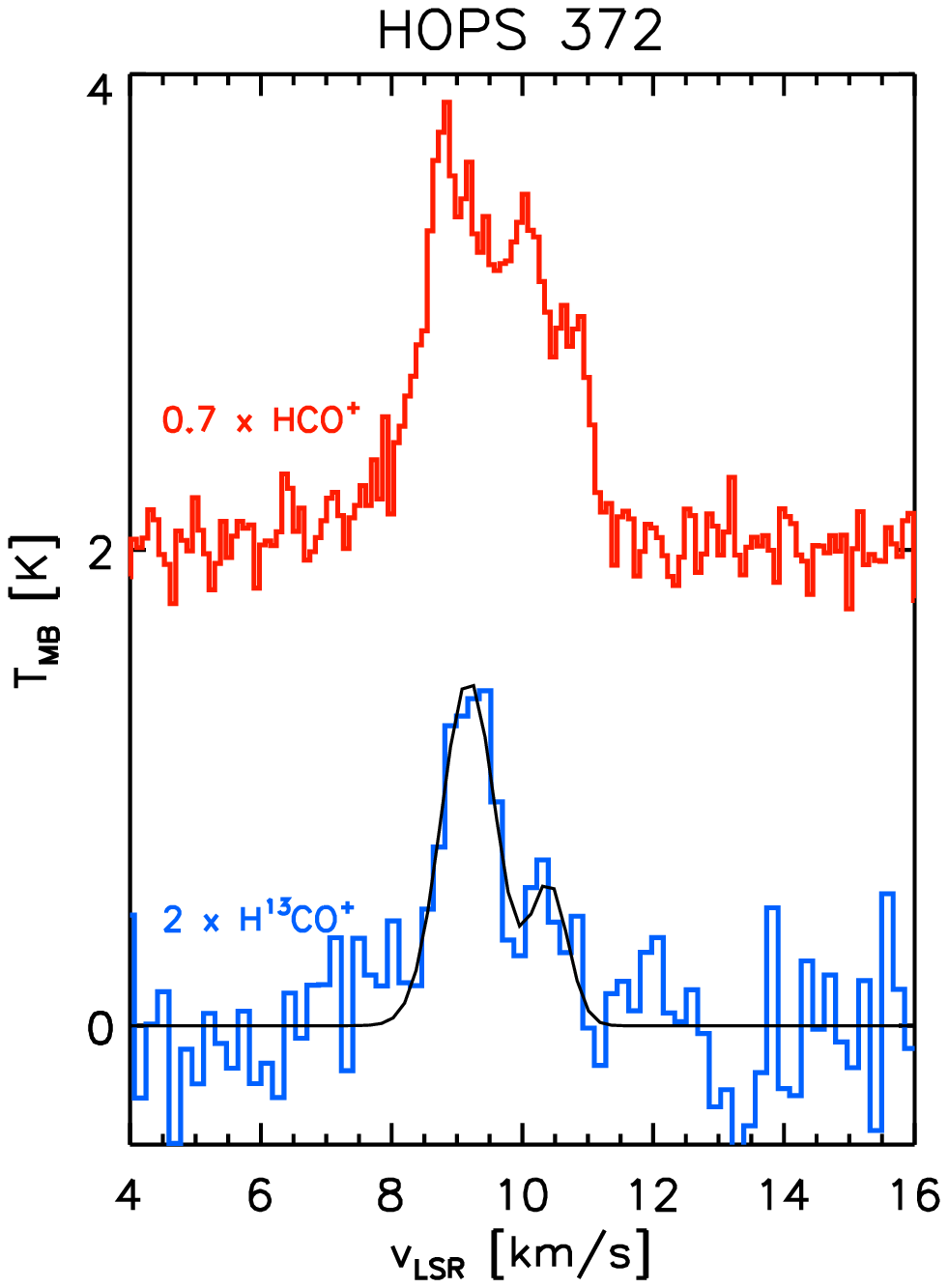}
\includegraphics[width=4.7cm]{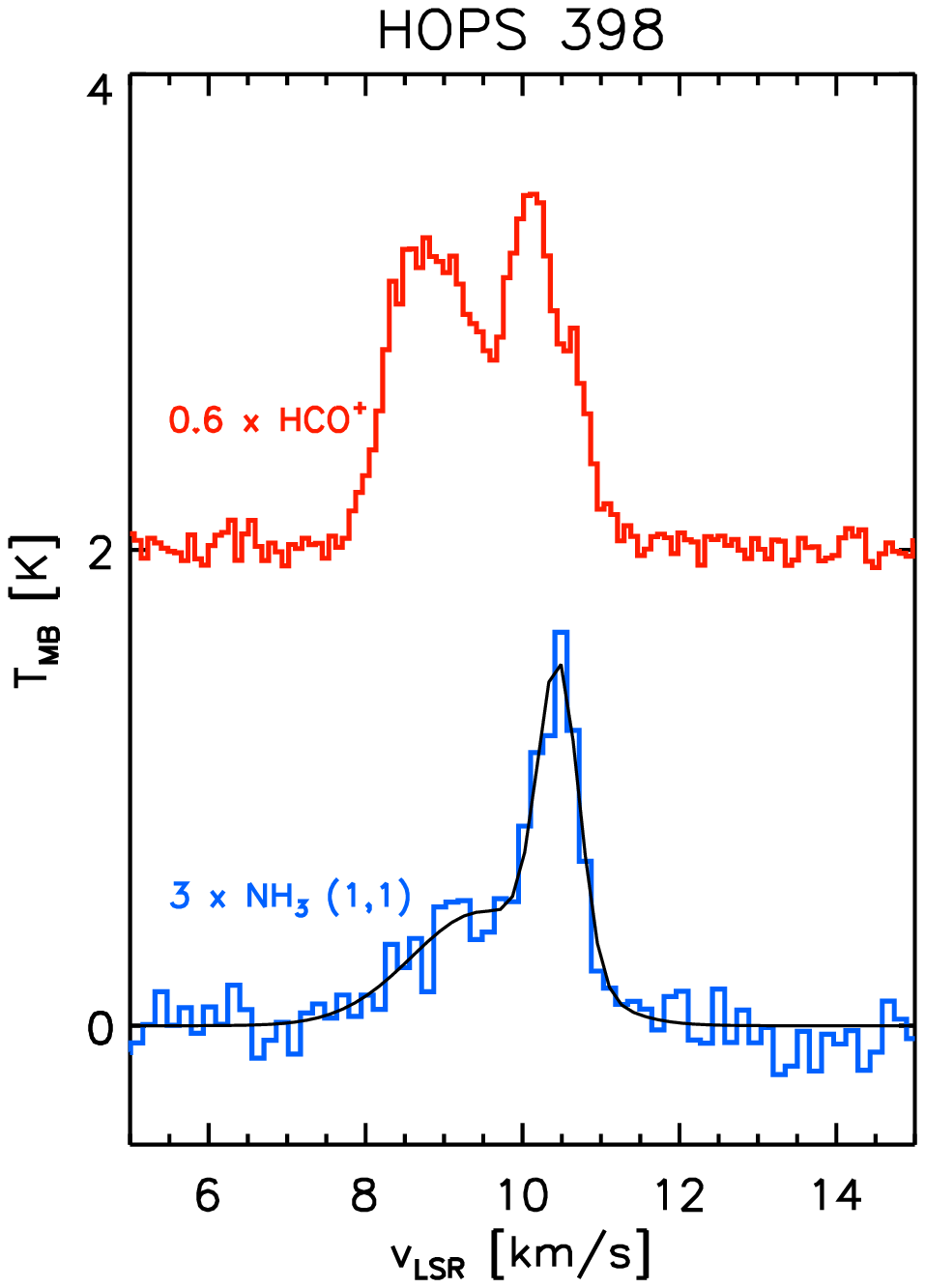}
\includegraphics[width=4.7cm]{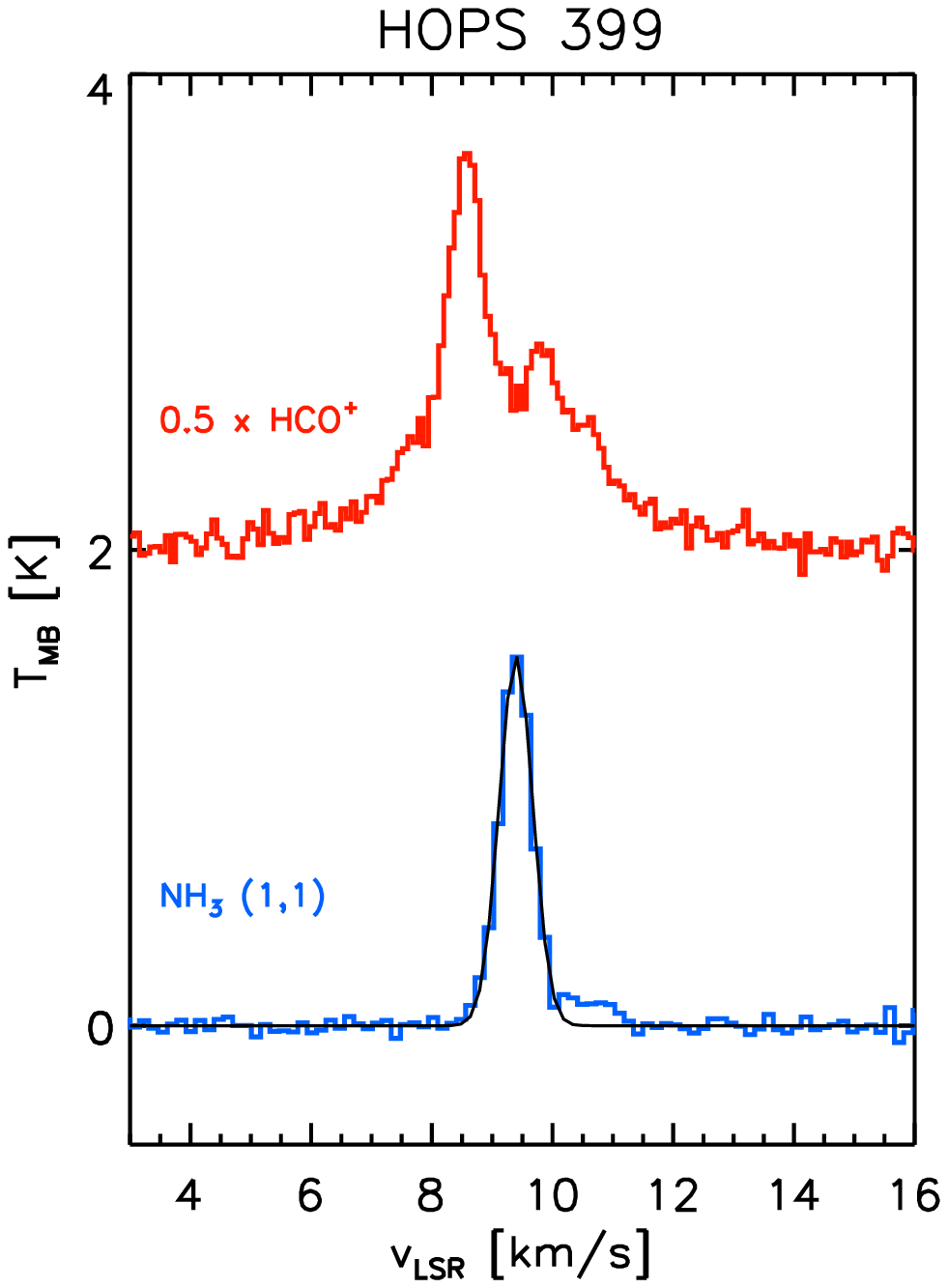}
\includegraphics[width=4.7cm]{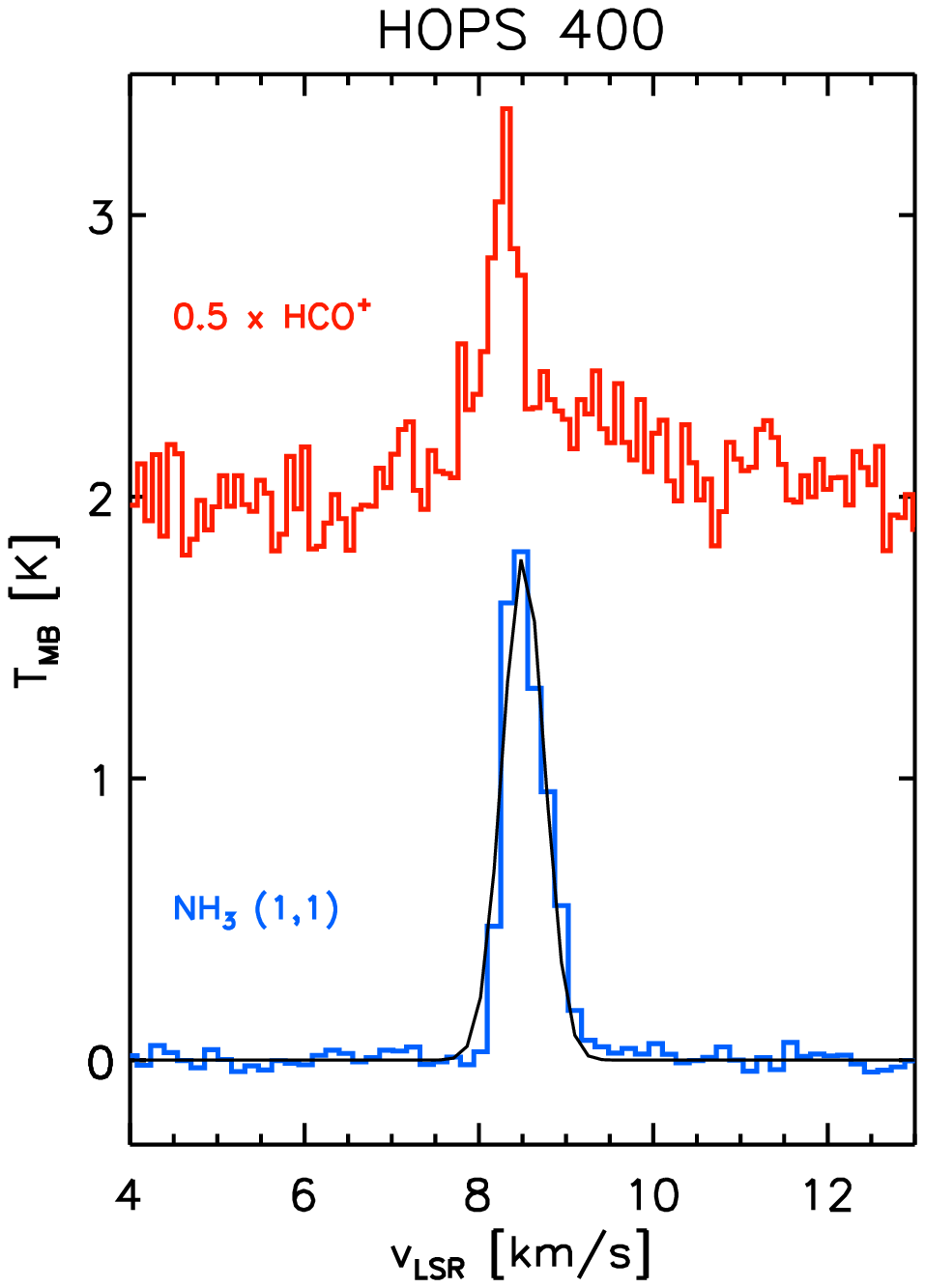}
\includegraphics[width=4.7cm]{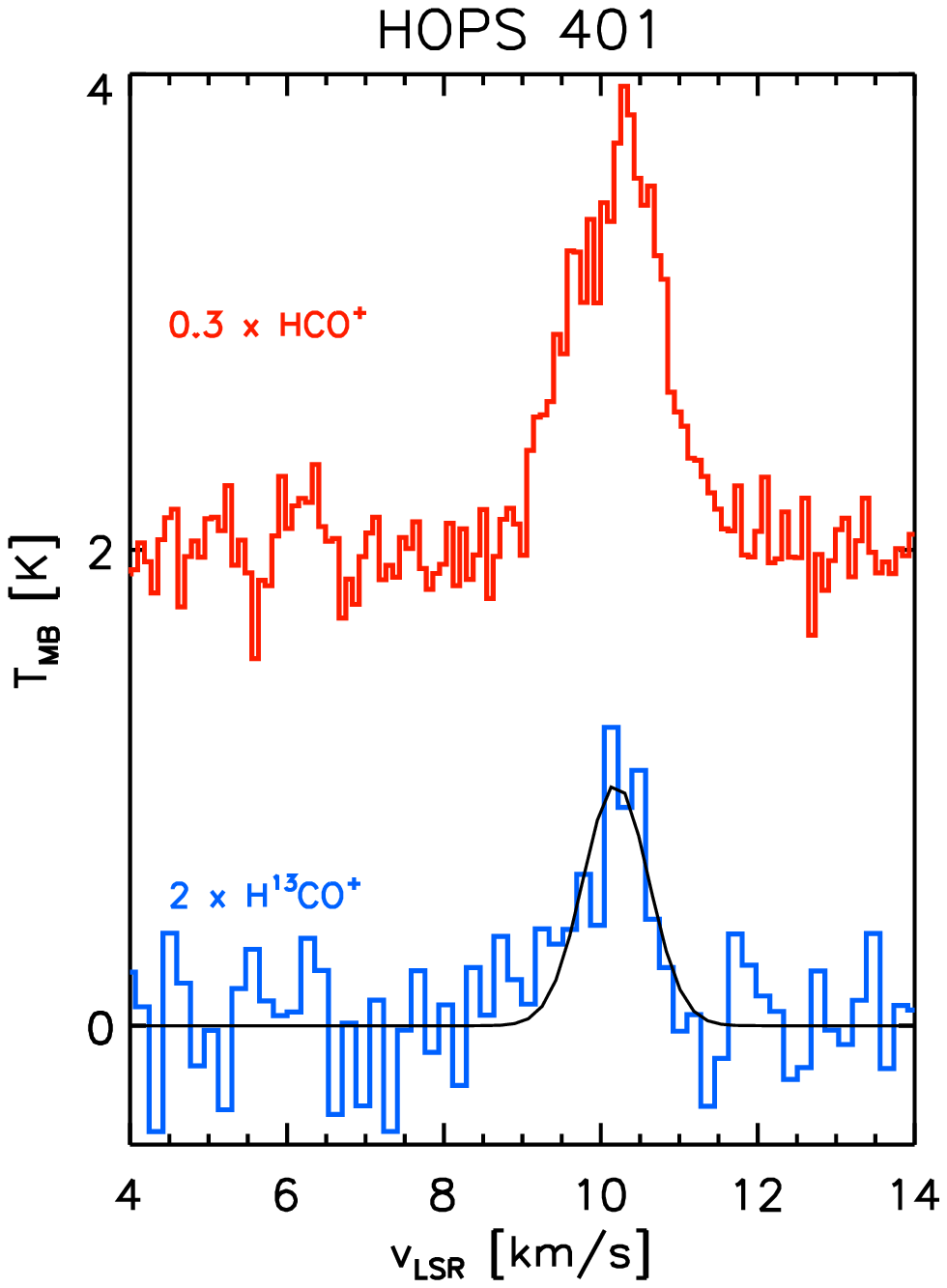}
\includegraphics[width=4.7cm]{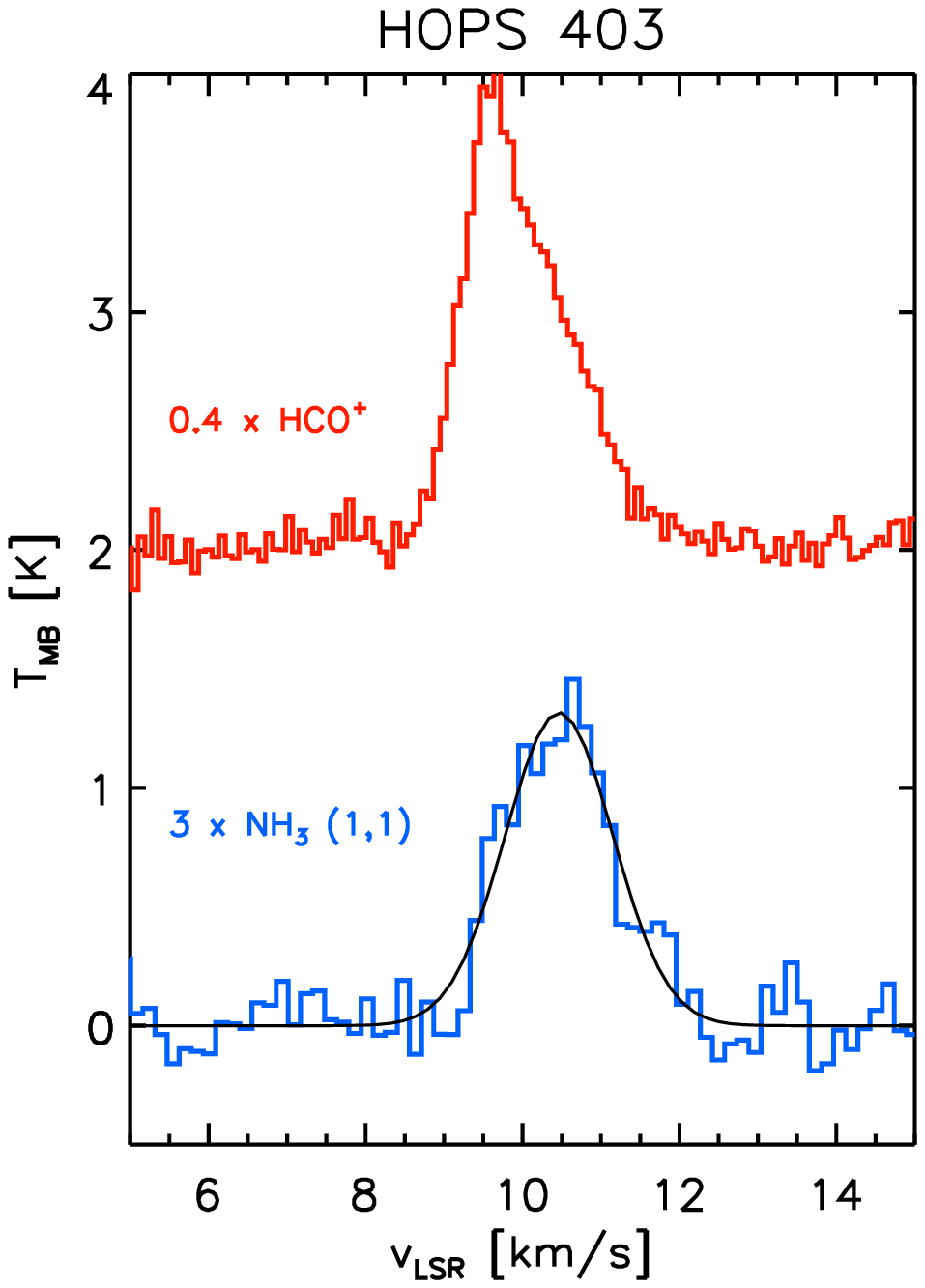}
\includegraphics[width=4.7cm]{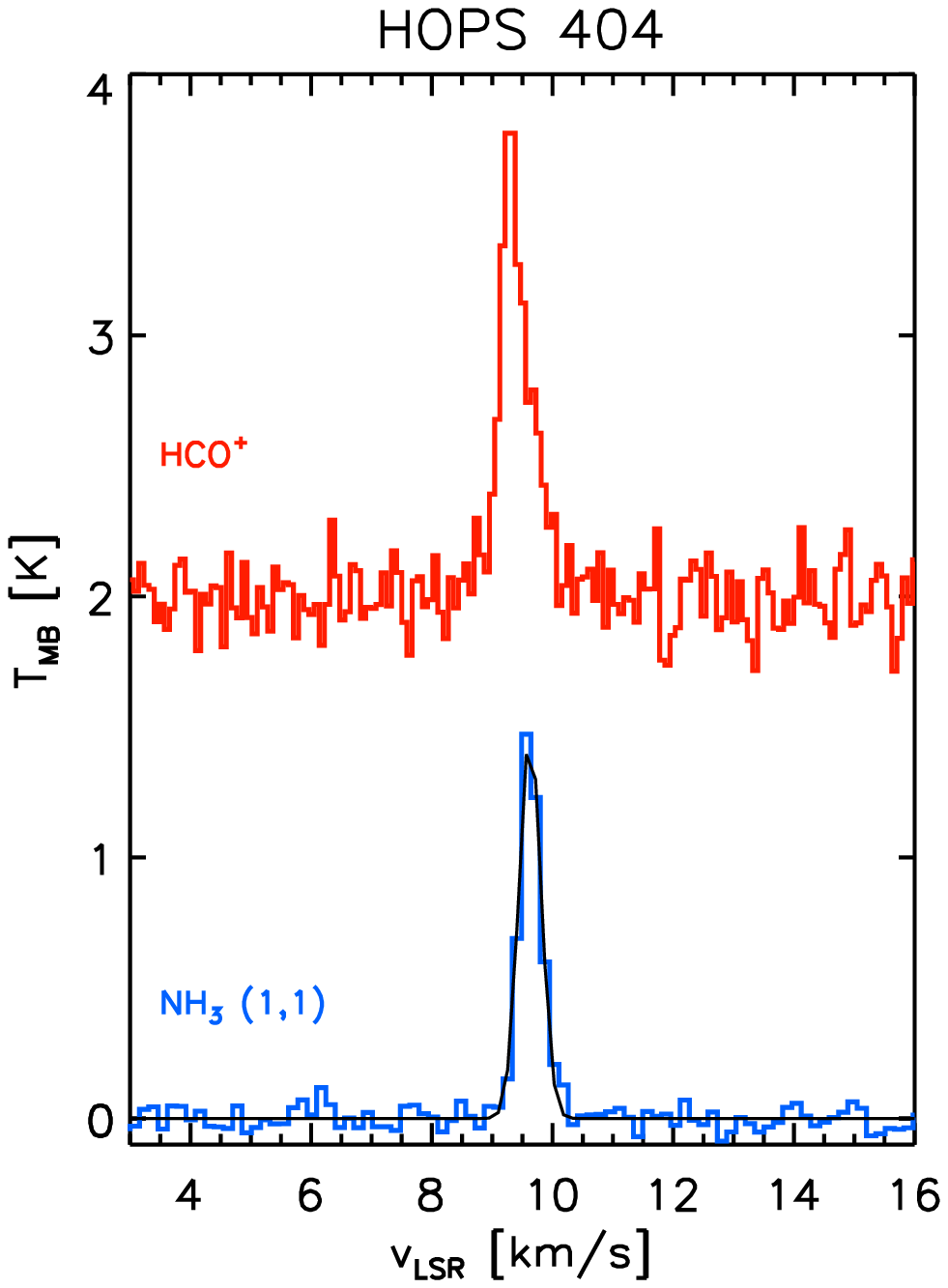}
\includegraphics[width=4.7cm]{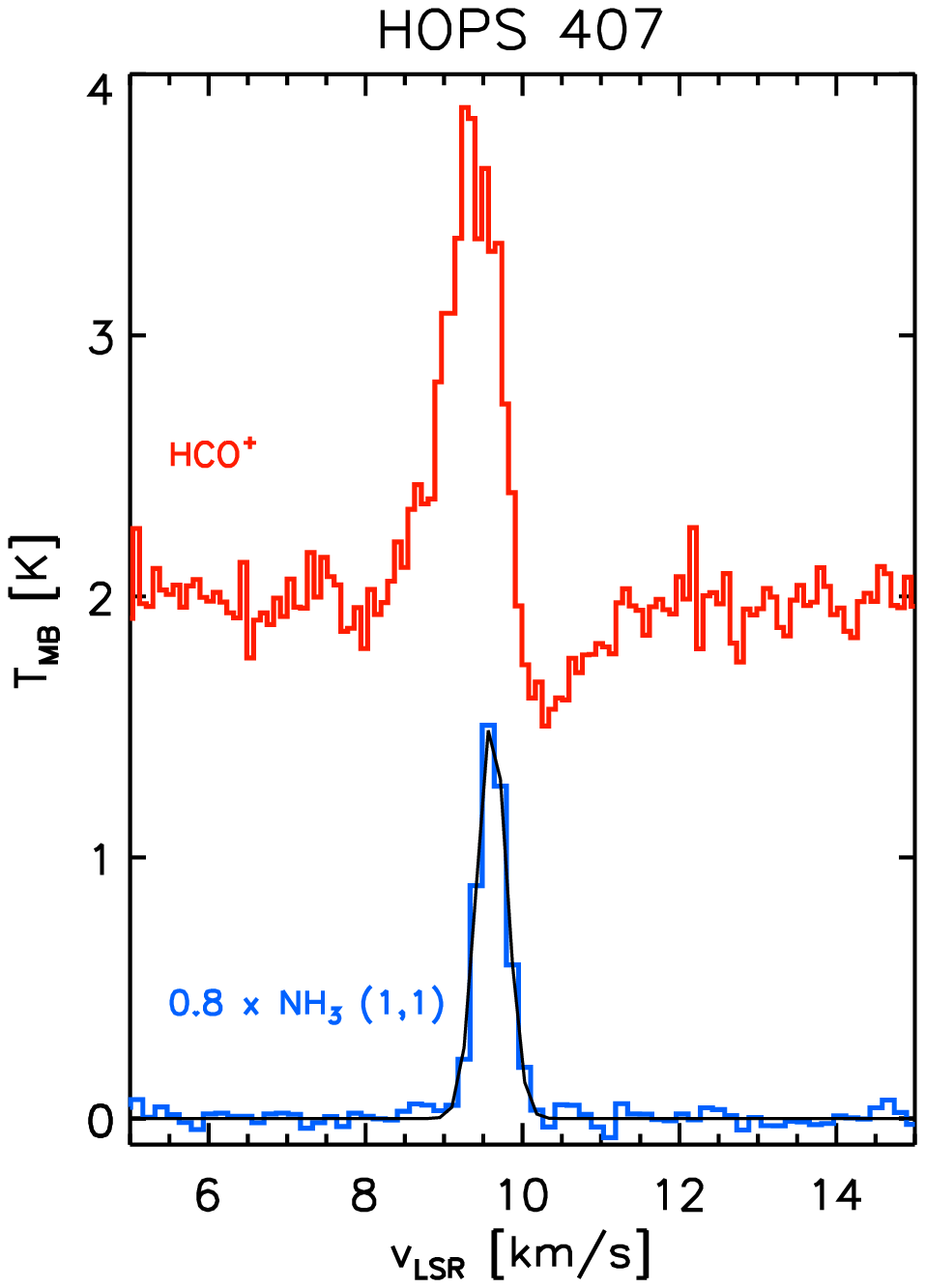}
\caption{HCO$^+$ (3-2), H$^{13}$CO$^+$ (3-2), and NH$_3$ (1,1) line profiles observed toward the PBRs to search for infall signatures. The Gaussian fits are also shown in case of the H$^{13}$CO$^+$ (3-2) and NH$_3$ (1,1) lines.}
\label{HCOp_HNC_NH3_spec}
\end{center}
\end{figure*}

\clearpage

\section{Intensity maps}
\label{sect_channelmaps}

\begin{figure*}[!h]
\centering
\includegraphics[height=5.2cm, trim=0cm 0cm 0cm 14.5cm,clip=true]{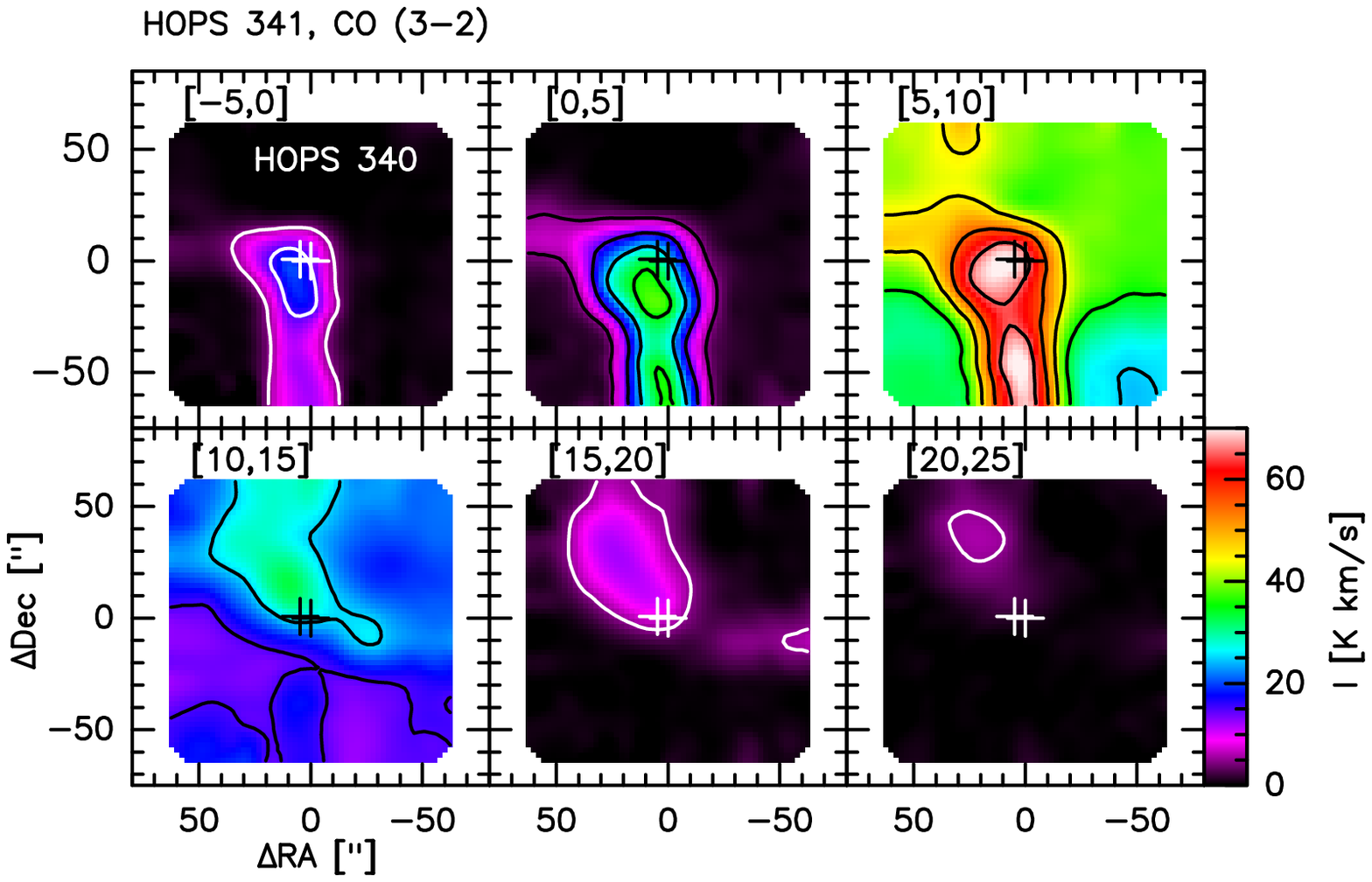}
\hskip+0.3cm
\includegraphics[height=5.2cm, trim=0cm 0cm 0cm 14.5cm,clip=true]{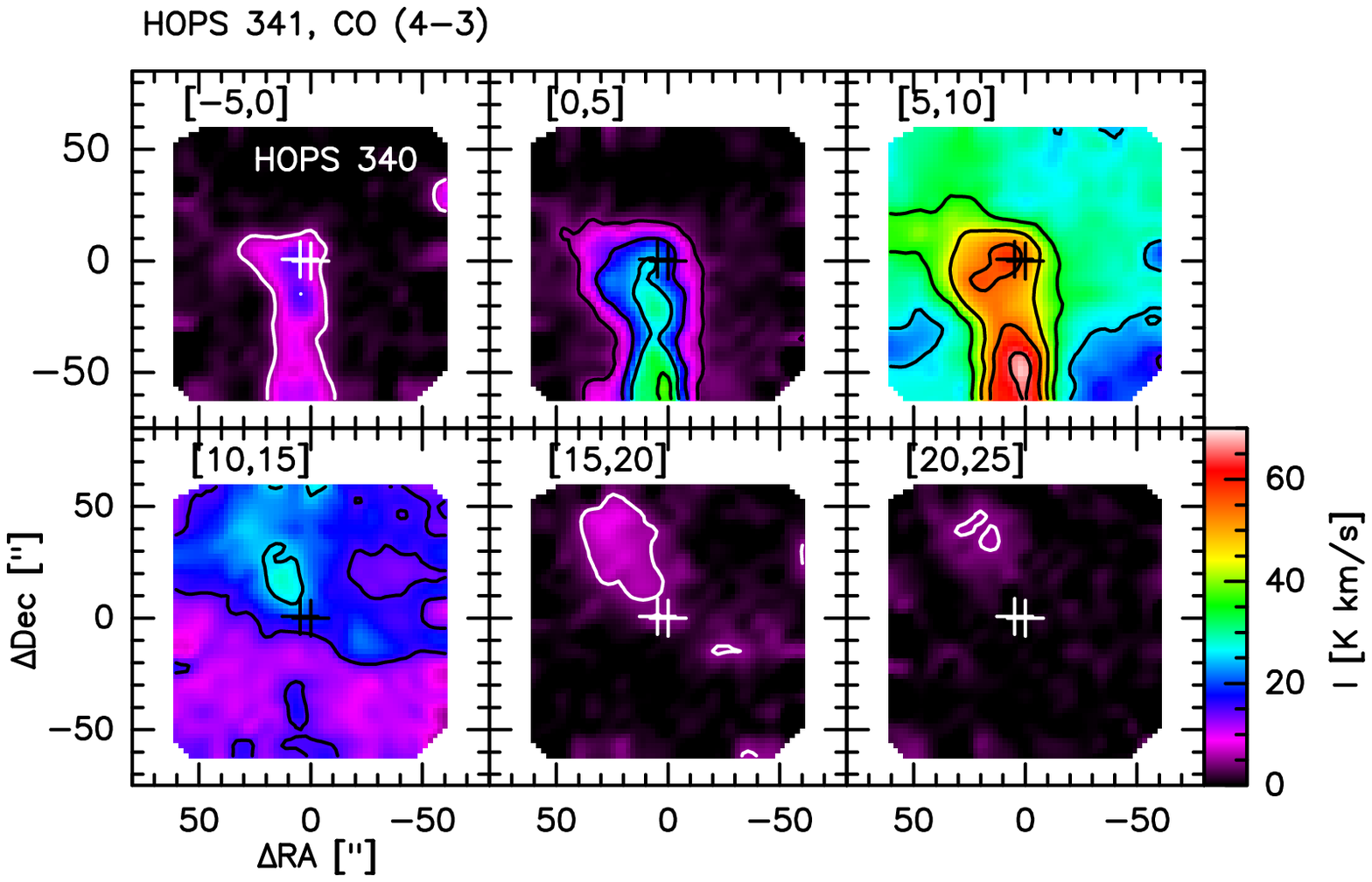}
\includegraphics[height=5.2cm, trim=0cm 0cm 4cm 14.5cm,clip=true]{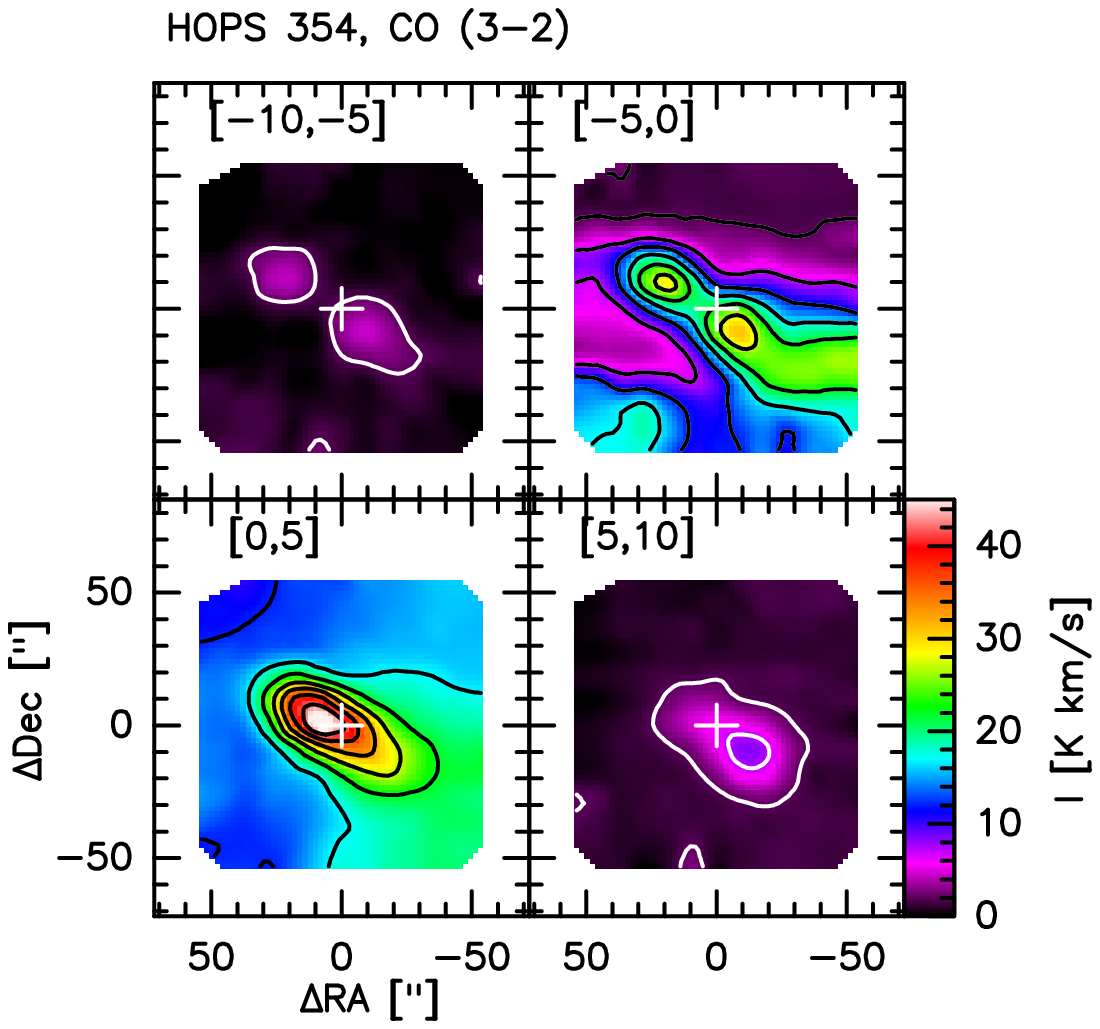}
\hskip+0.3cm
\includegraphics[height=5.2cm, trim=0cm 0cm 4cm 14.5cm,clip=true]{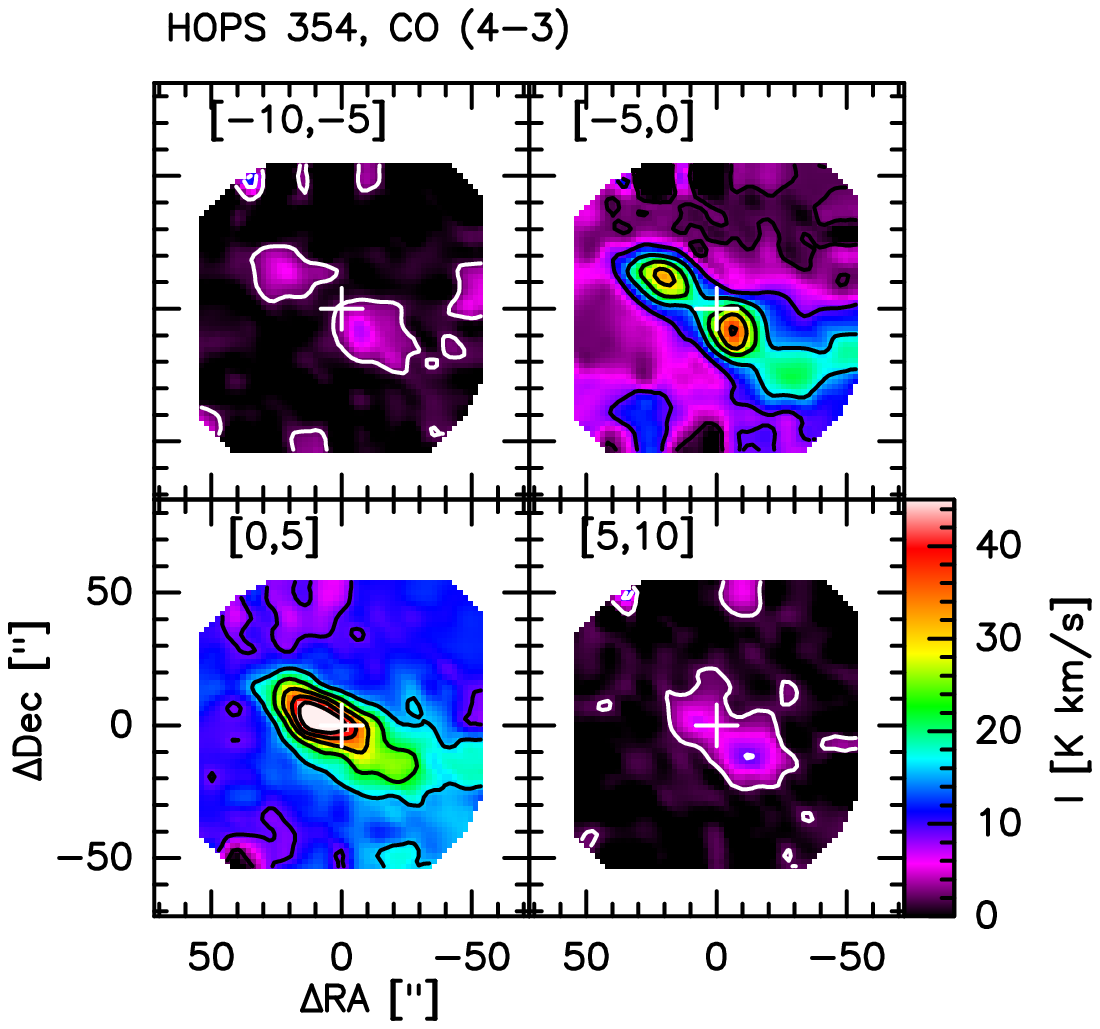}
\includegraphics[height=5.2cm, trim=0cm 0cm 0cm 14.5cm,clip=true]{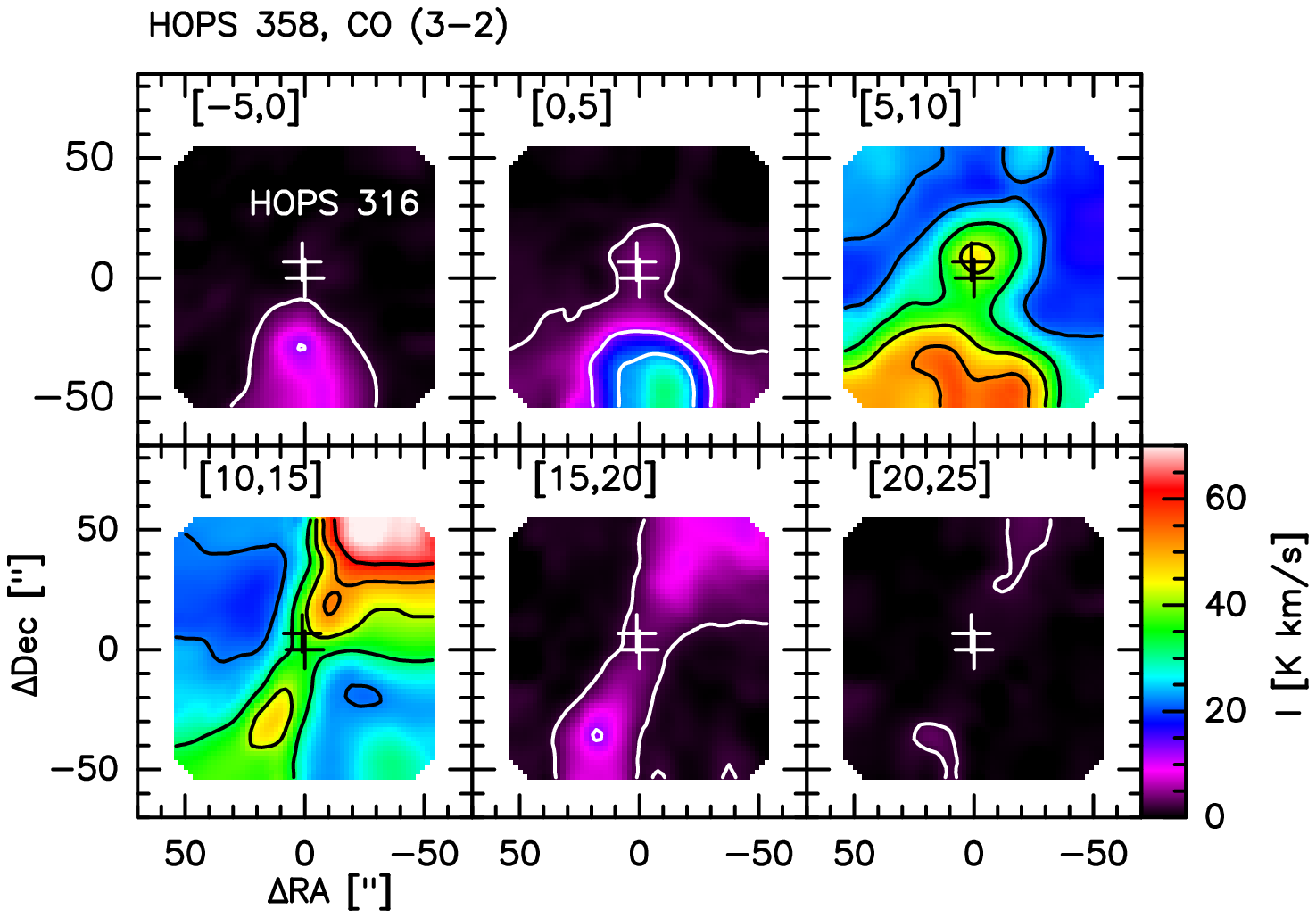}
\hskip+0.3cm
\includegraphics[height=5.2cm, trim=0cm 0cm 0cm 14.5cm,clip=true]{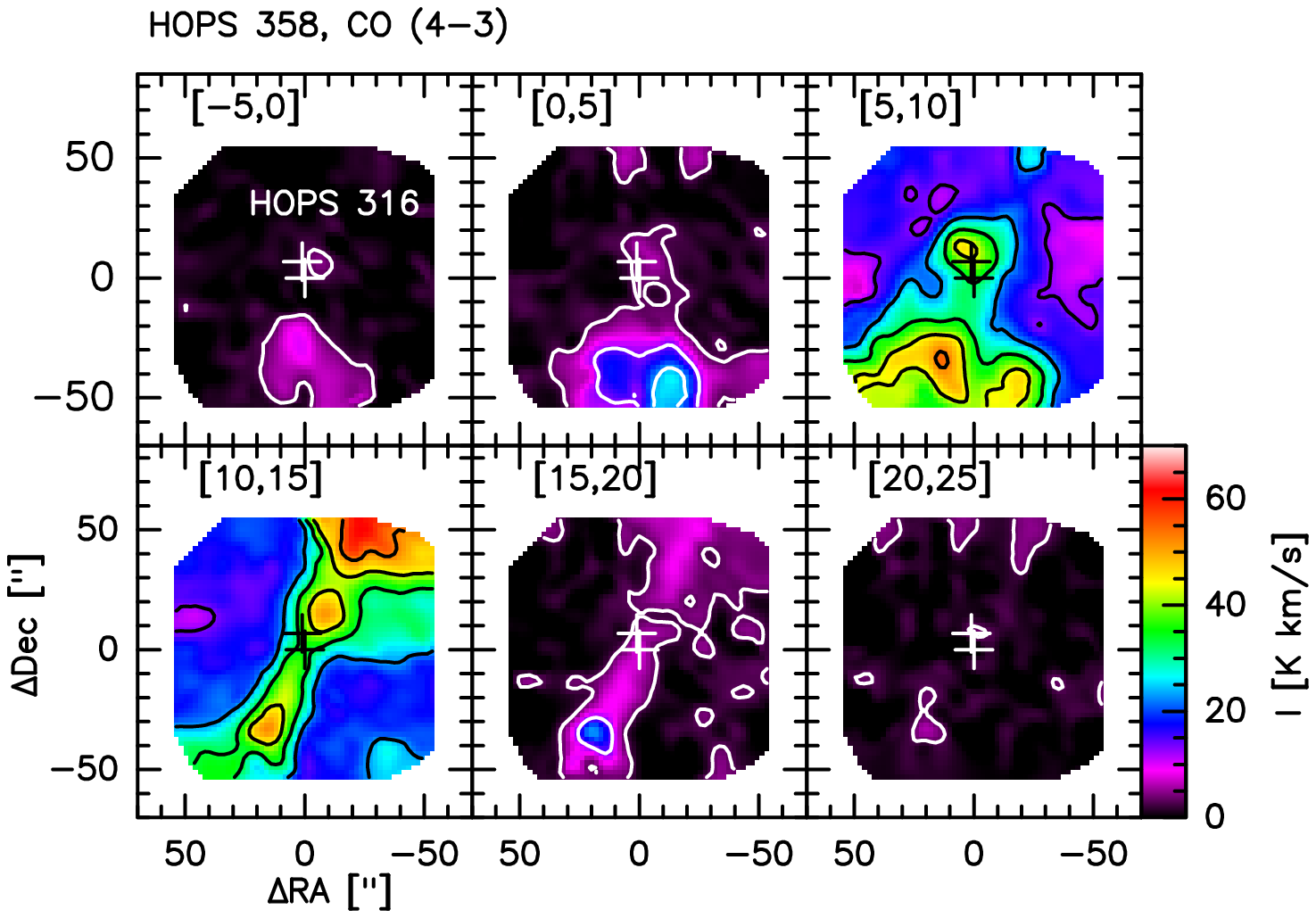}
\includegraphics[height=5.2cm, trim=0cm 0cm 4cm 14.5cm,clip=true]{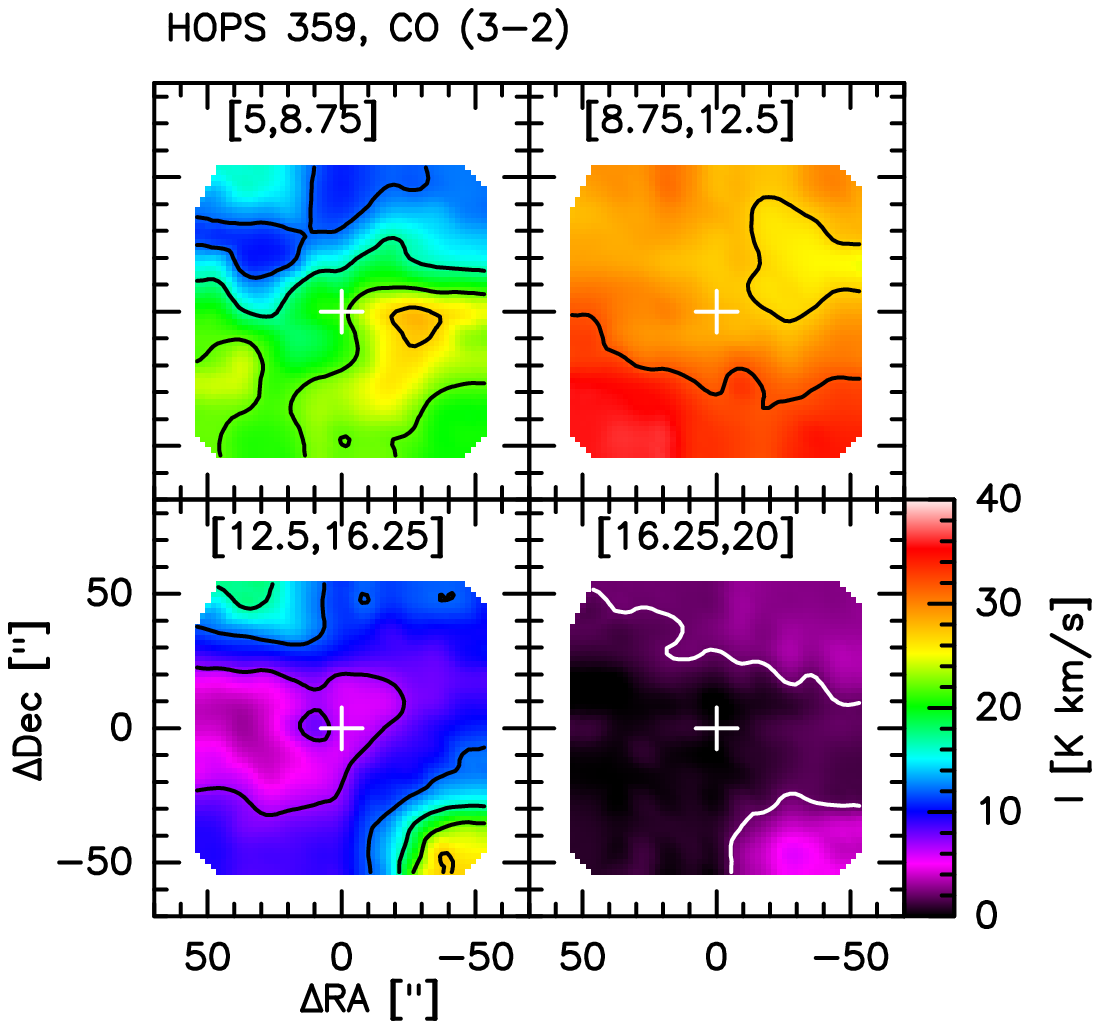}
\hskip+0.3cm
\includegraphics[height=5.2cm, trim=0cm 0cm 4cm 14.5cm,clip=true]{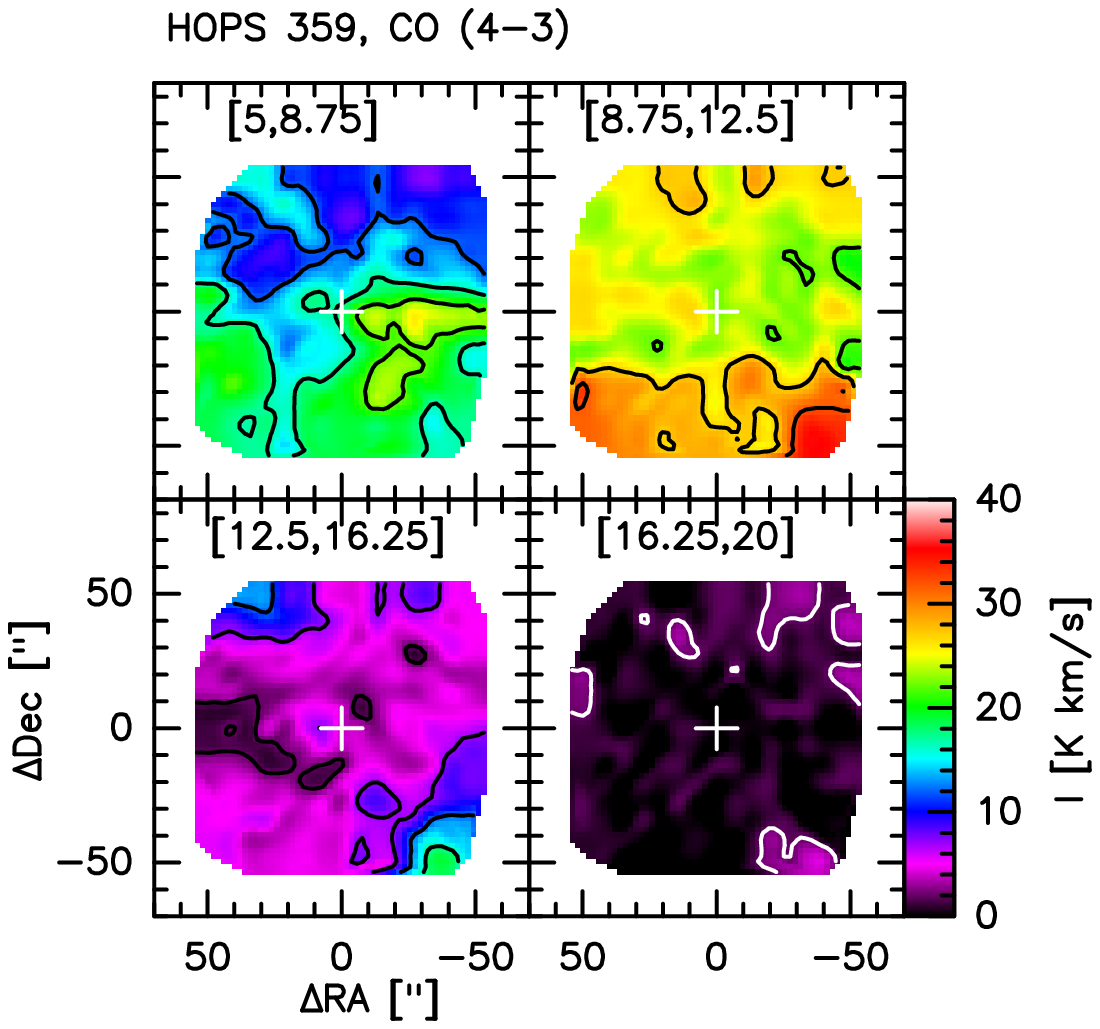}
\label{channelmaps_hops341_32}
\caption{Intensity maps centered on HOPS 341, HOPS 354, HOPS 358, and HOPS 359 integrated over the velocity ranges shown in the upper-left corner of each subfigure.
Contour levels: between 5 and 70 K km s$^{-1}$ at intervals of 10 K km s$^{-1}$ for HOPS 341 CO $J$=3-2 and CO $J$=4-3, between 2 and 45 K km s$^{-1}$ at intervals of 5 K km s$^{-1}$ for HOPS 354 CO $J$=3-2, between 2 and 45 K km s$^{-1}$ at intervals of 7 K km s$^{-1}$ for HOPS 354 CO $J$=4-3, between 2 and 70 K km s$^{-1}$ at intervals of 10 K km s$^{-1}$ for HOPS 358 CO $J$=3-2, between 3 and 70 K km s$^{-1}$ at intervals of 10 K km s$^{-1}$ for HOPS 358 CO $J$=4-3, between 2 and 40 K km s$^{-1}$ at intervals of 5 K km s$^{-1}$ for HOPS 359 CO $J$=3-2 and $J$=4-3.
}
\end{figure*}

\begin{figure*}[!h]
\centering
\includegraphics[height=5.5cm, trim=0cm 0cm 0cm 14.5cm,clip=true]{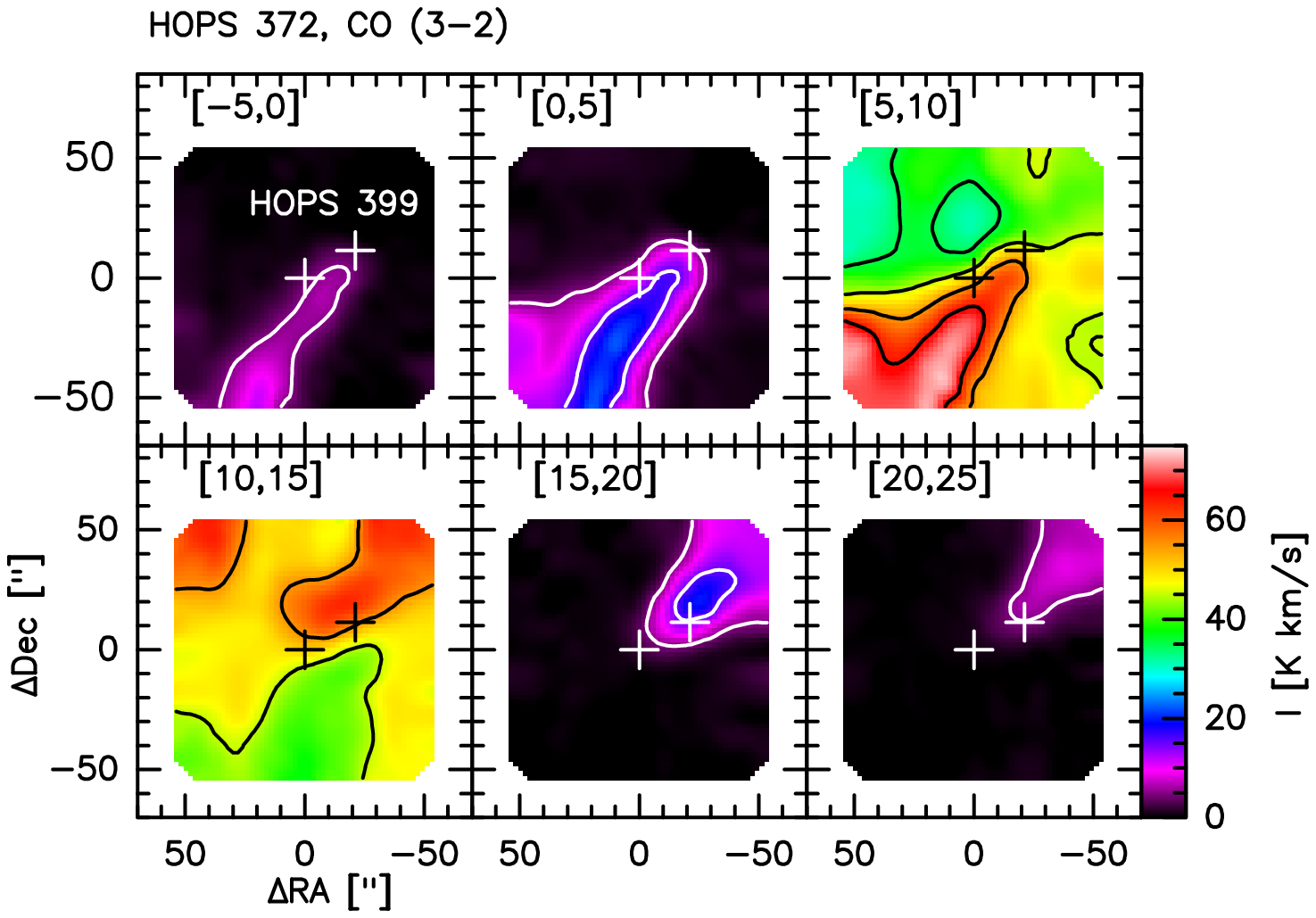}
\hskip+0.3cm
\includegraphics[height=5.5cm, trim=0cm 0cm 0cm 14.5cm,clip=true]{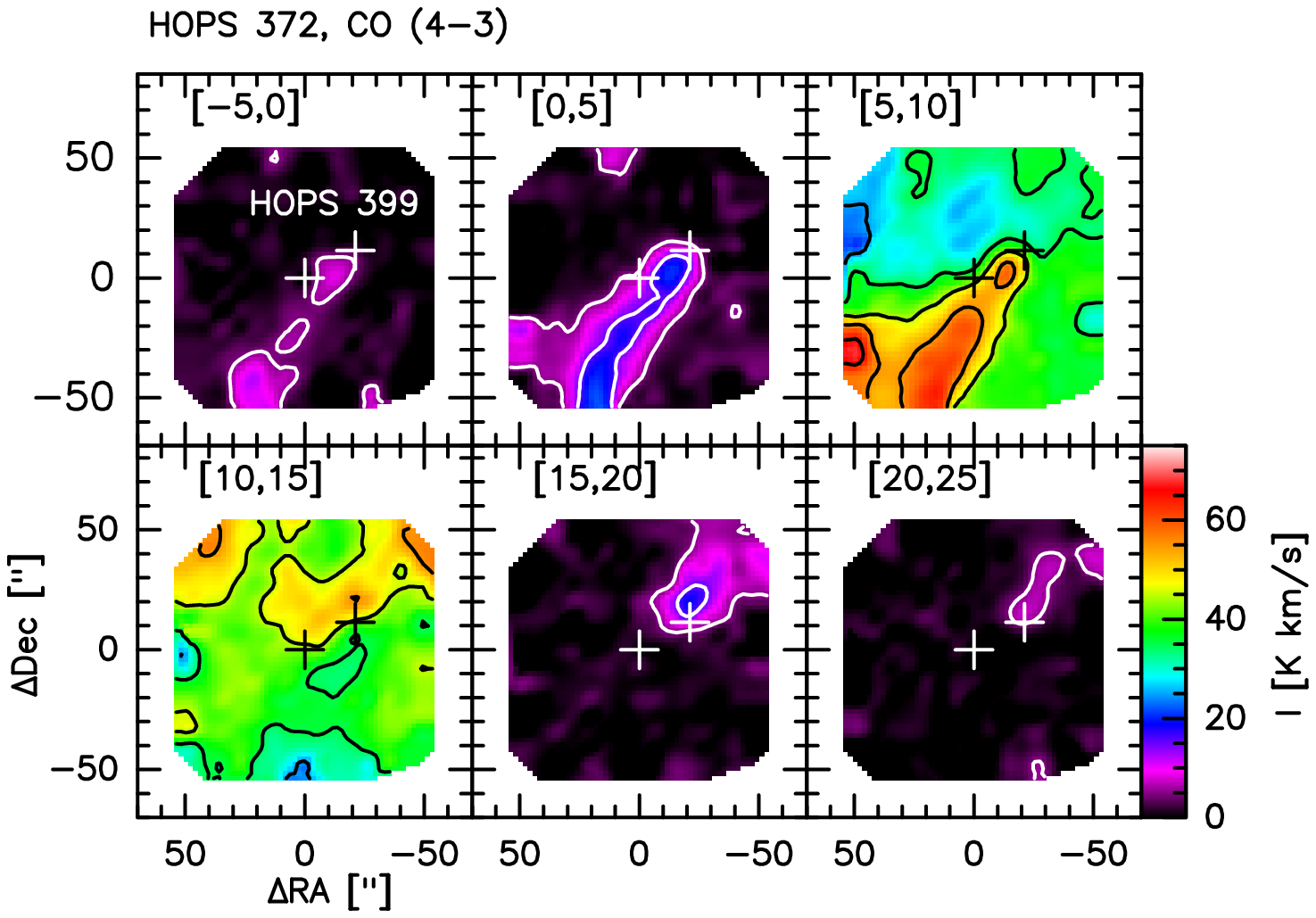}
\includegraphics[height=5.5cm, trim=0cm 0cm 4cm 14.5cm,clip=true]{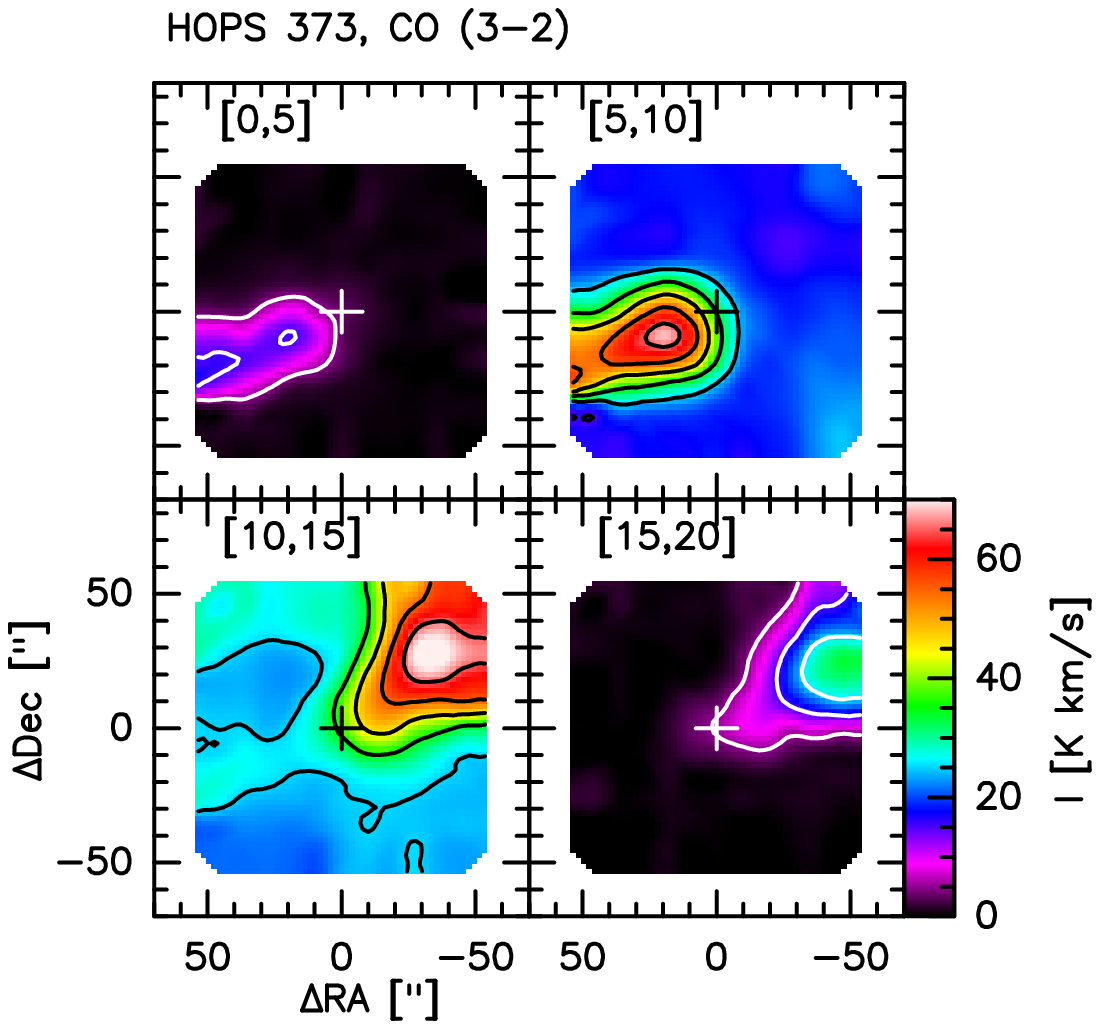}
\hskip+0.3cm
\includegraphics[height=5.5cm, trim=0cm 0cm 4cm 14.5cm,clip=true]{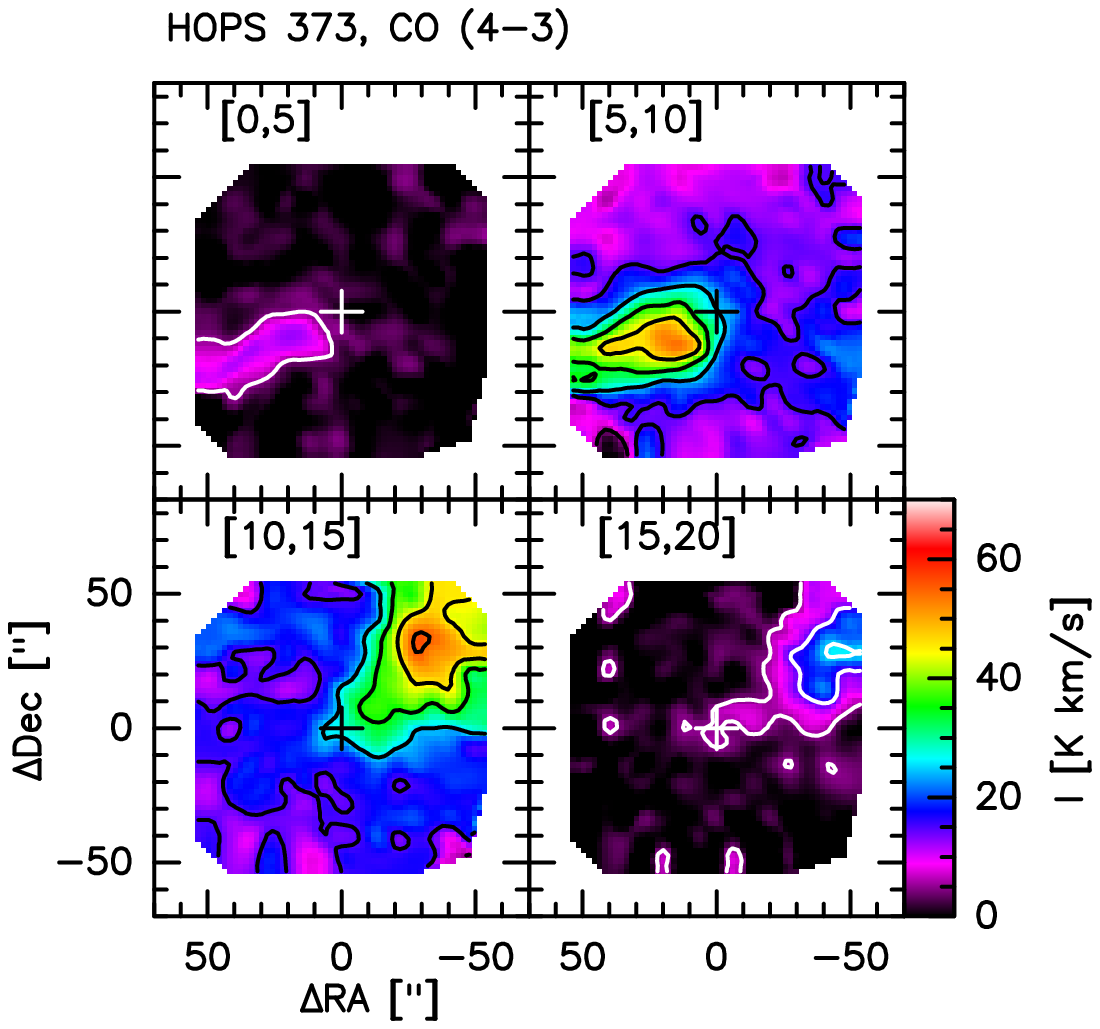}
\includegraphics[height=5.5cm, trim=0cm 0cm 4cm 14.5cm,clip=true]{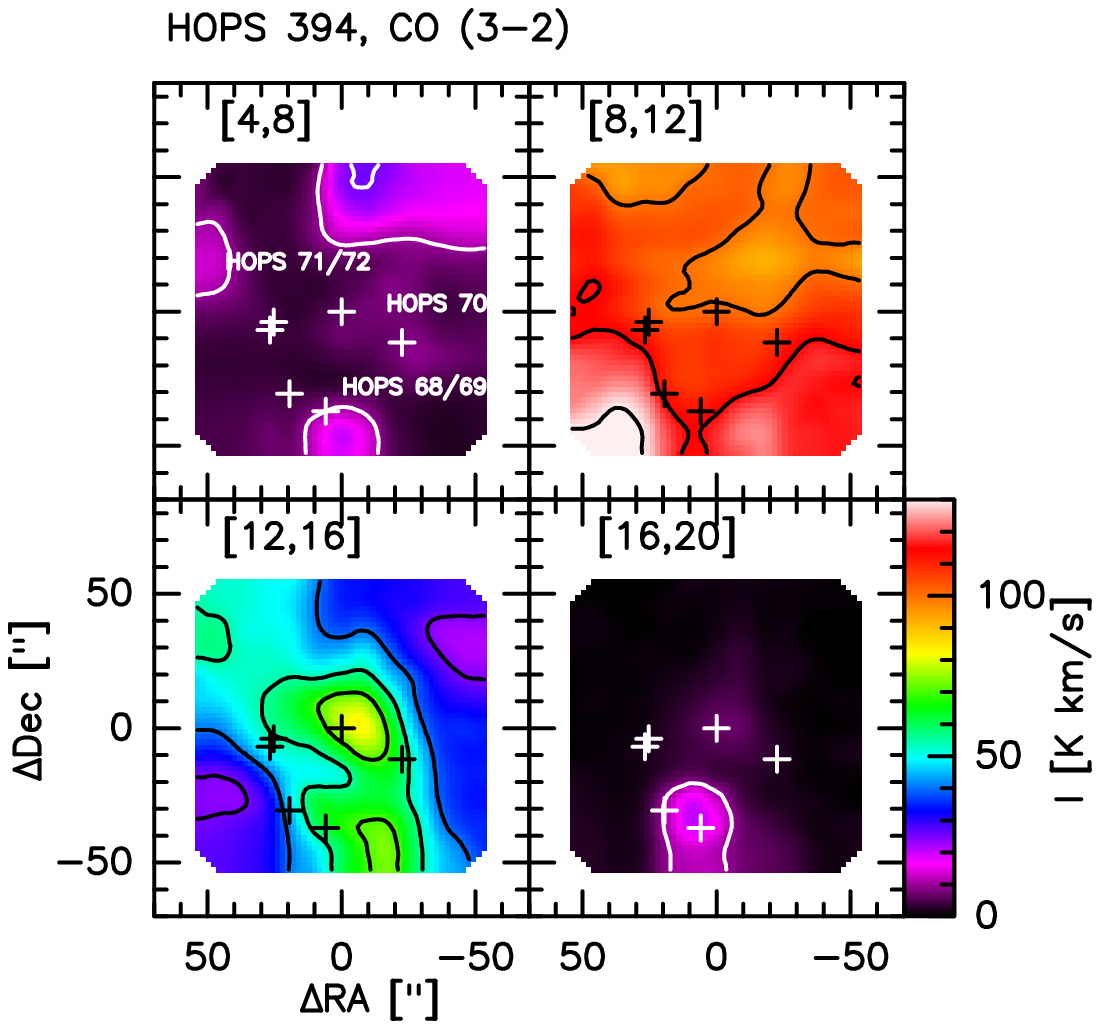}
\hskip+0.3cm
\includegraphics[height=5.5cm, trim=0cm 0cm 4cm 14.5cm,clip=true]{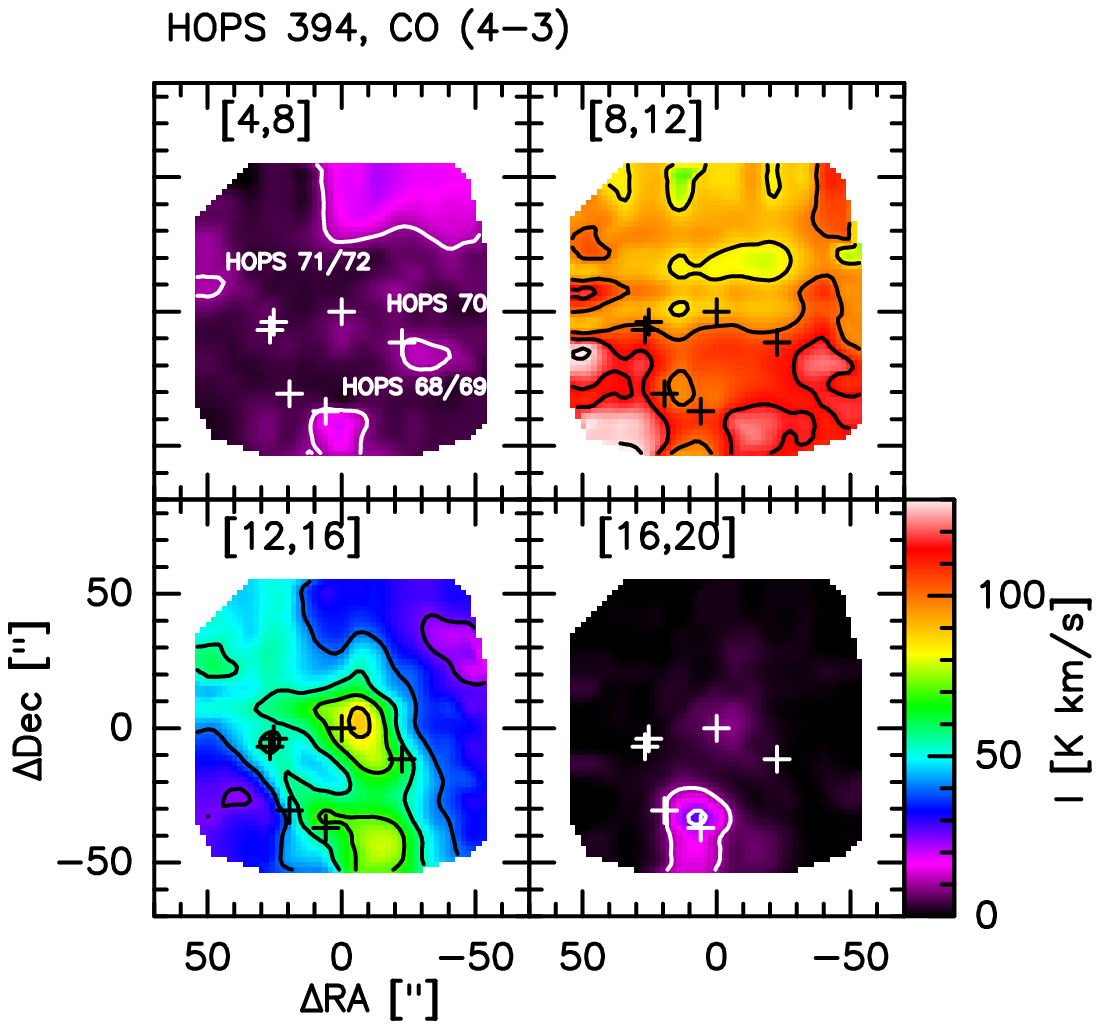}
\includegraphics[height=5.5cm, trim=0cm 0cm 4cm 14.5cm,clip=true]{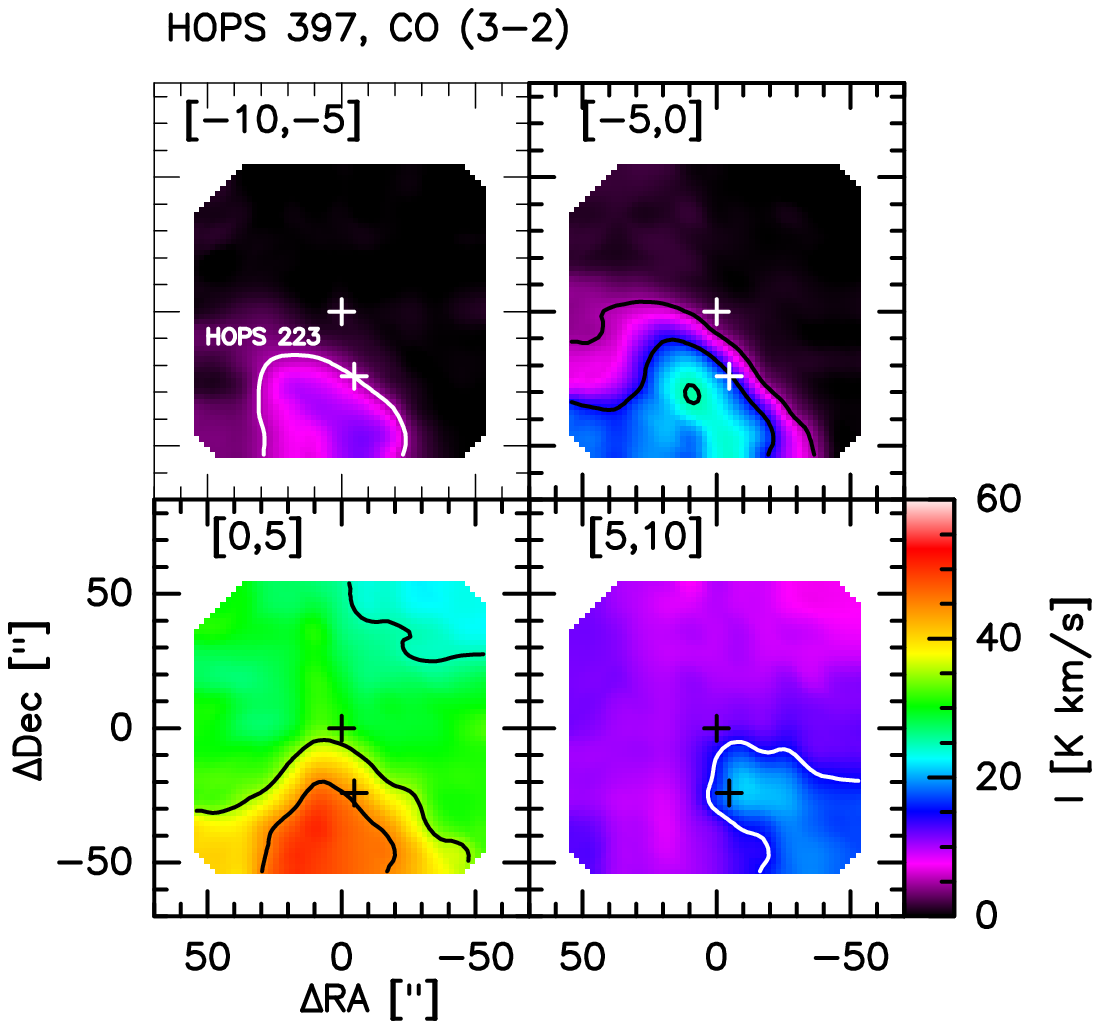}
\hskip+0.3cm
\includegraphics[height=5.5cm, trim=0cm 0cm 4cm 14.5cm,clip=true]{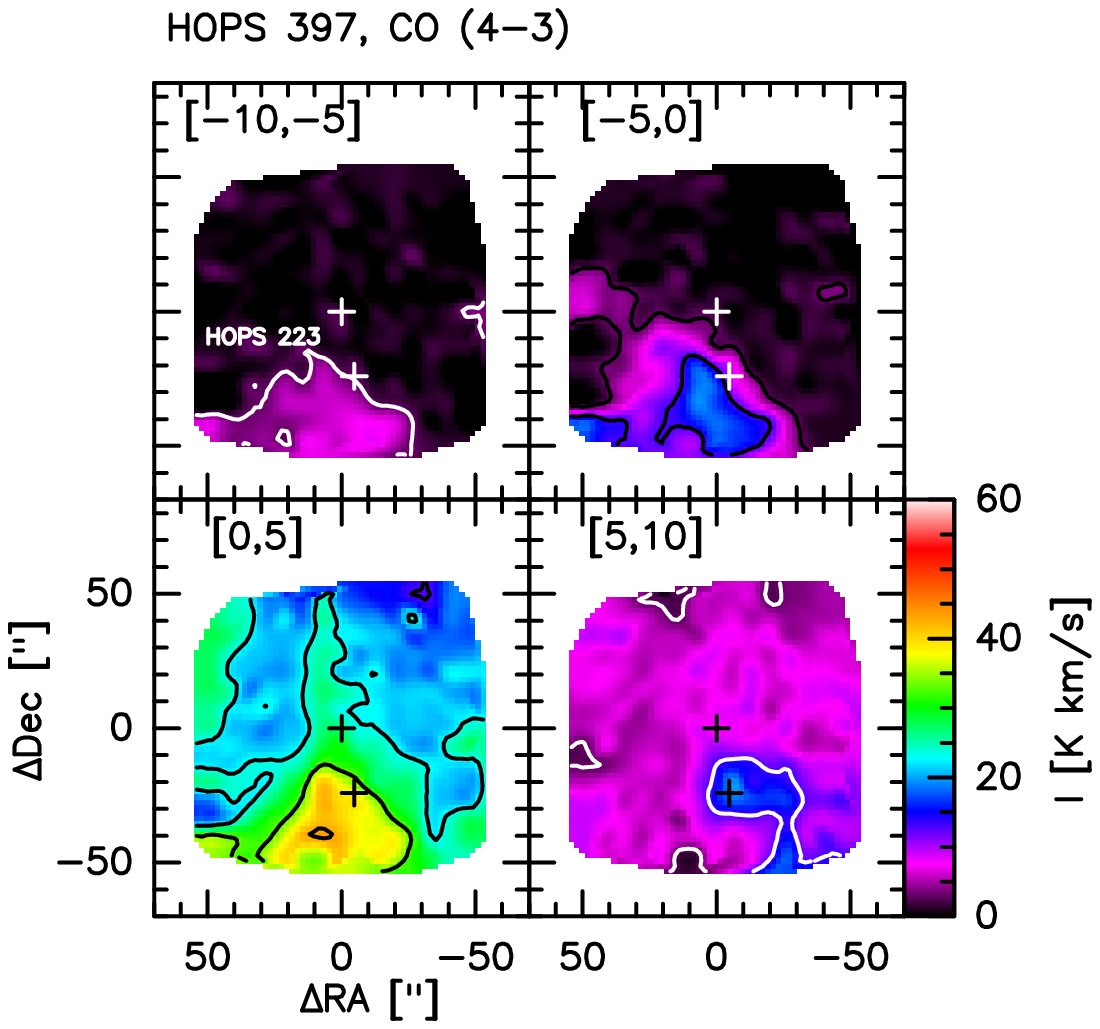}
\label{channelmaps_hops372_32}
\caption{Intensity maps centered on HOPS 372, HOPS 373, HOPS 394, and HOPS 397 integrated over the velocity ranges shown in the upper-left corner of each subfigure.
Contour levels: between 5 and 75 K km s$^{-1}$ at intervals of 10 K km s$^{-1}$ for both CO $J$=3-2 and $J$=4-3 for HOPS 372 and HOPS 373, between 10 and 130 at intervals of 15 K km s$^{-1}$ for HOPS 394 CO $J$=3-2 and $J$=4-3, between 5 and 60 K km s$^{-1}$ at intervals of 10 K km s$^{-1}$ for HOPS 397 CO $J$=3-2, between 3 and 60 K km s$^{-1}$ at intervals of 10 K km s$^{-1}$ for HOPS 397 CO $J$=4-3.
}
\end{figure*}

\begin{figure*}[!h]
\centering
\includegraphics[height=5.5cm, trim=0cm 0cm 4cm 14.5cm,clip=true]{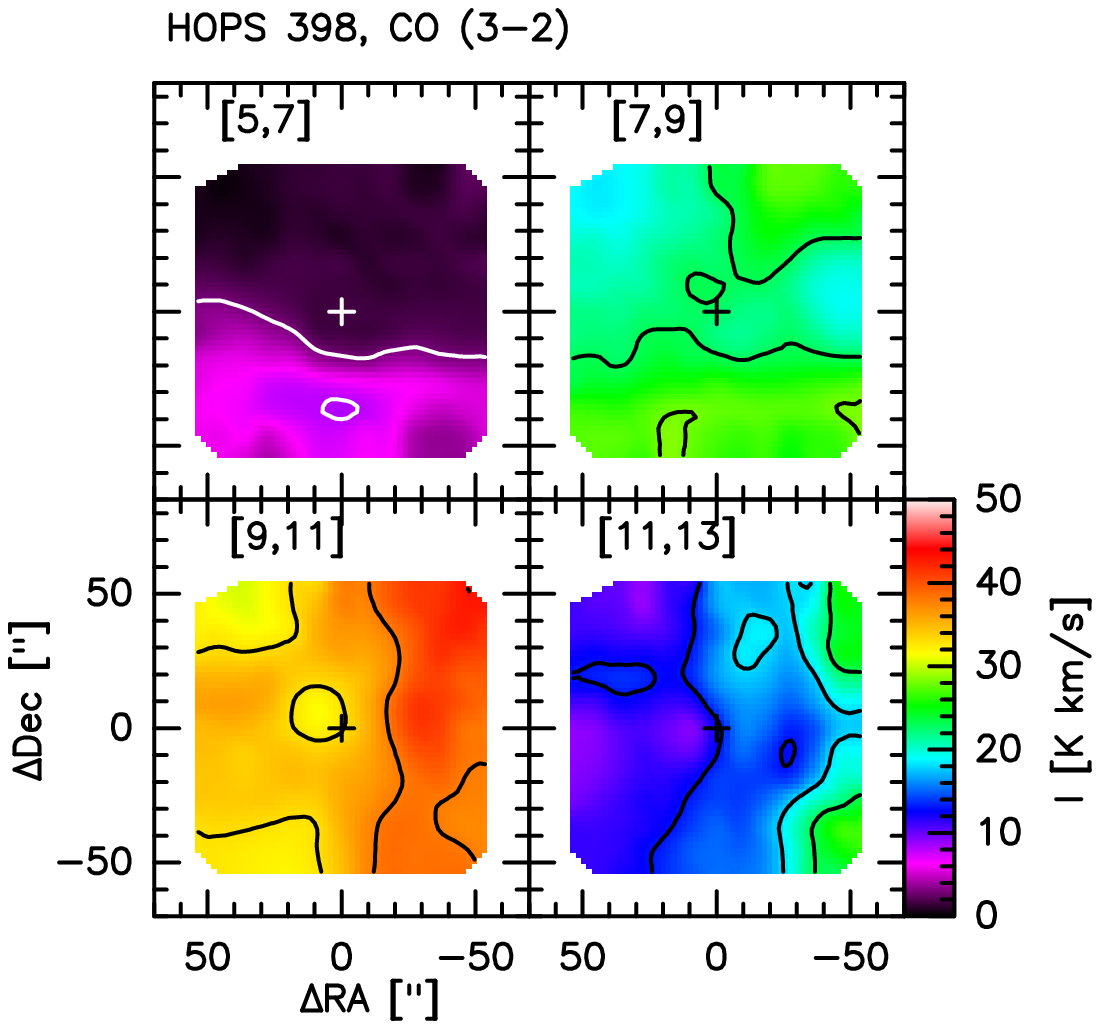}
\hskip+0.3cm
\includegraphics[height=5.5cm, trim=0cm 0cm 4cm 14.5cm,clip=true]{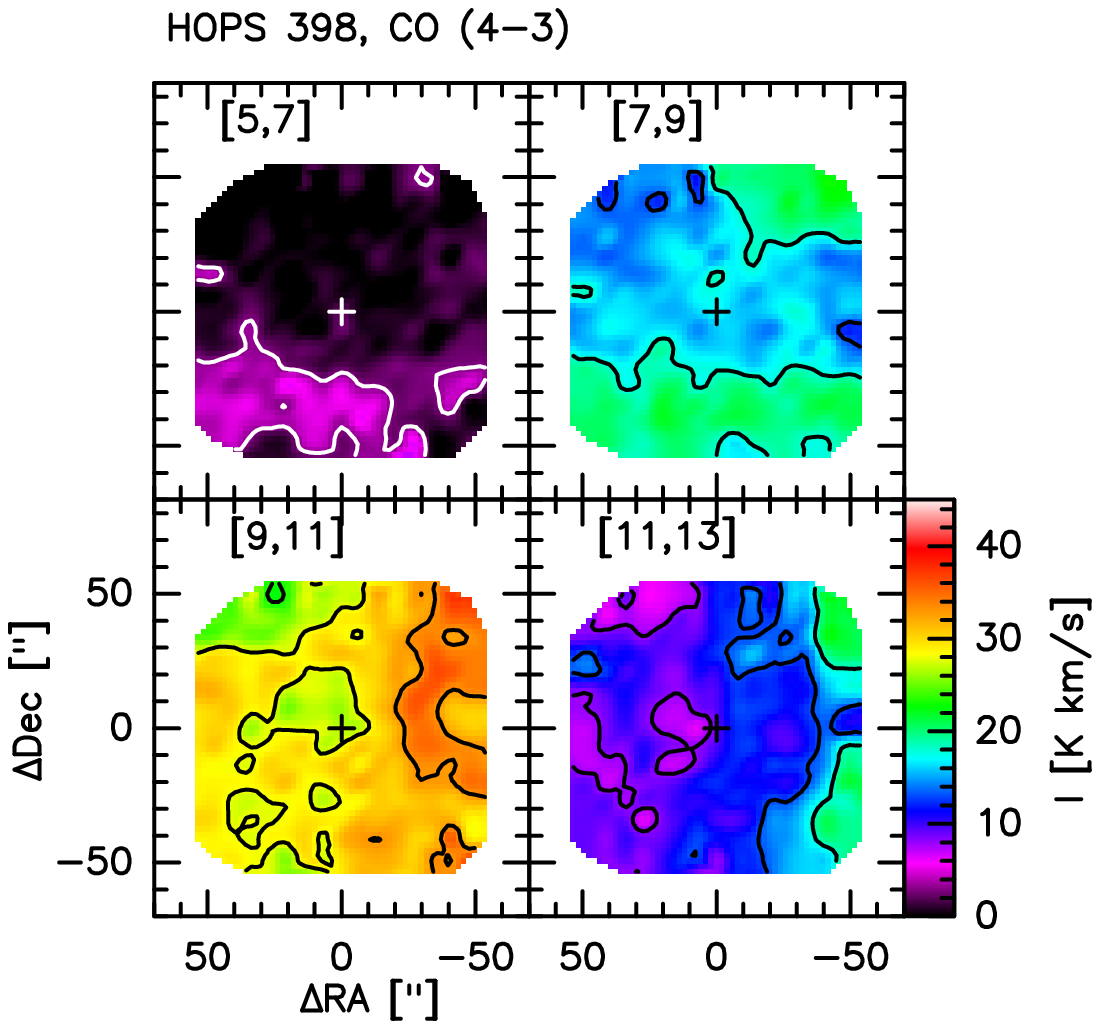}
\includegraphics[height=5.5cm, trim=0cm 0cm 0cm 14.5cm,clip=true]{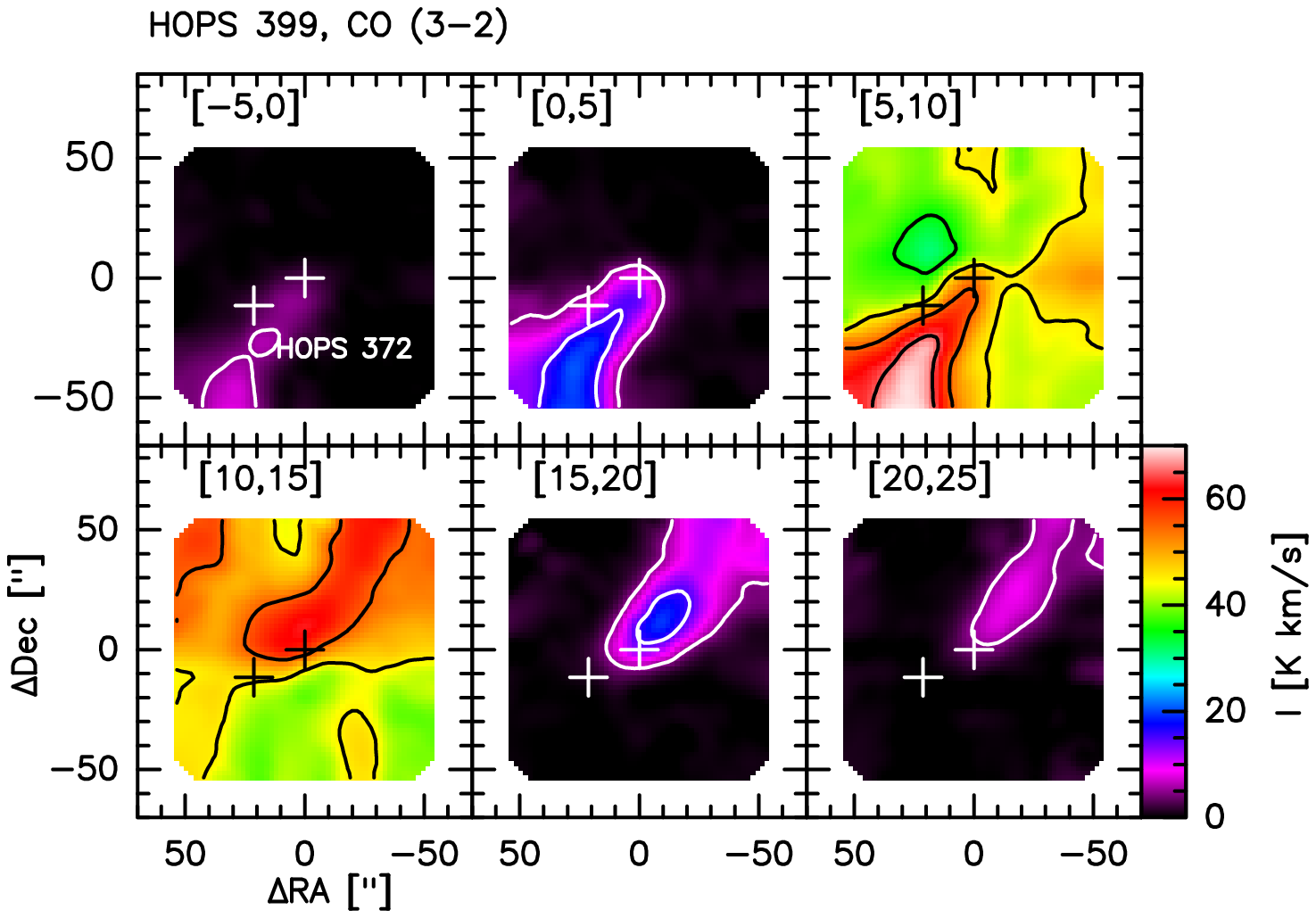}
\hskip+0.3cm
\includegraphics[height=5.5cm, trim=0cm 0cm 0cm 14.5cm,clip=true]{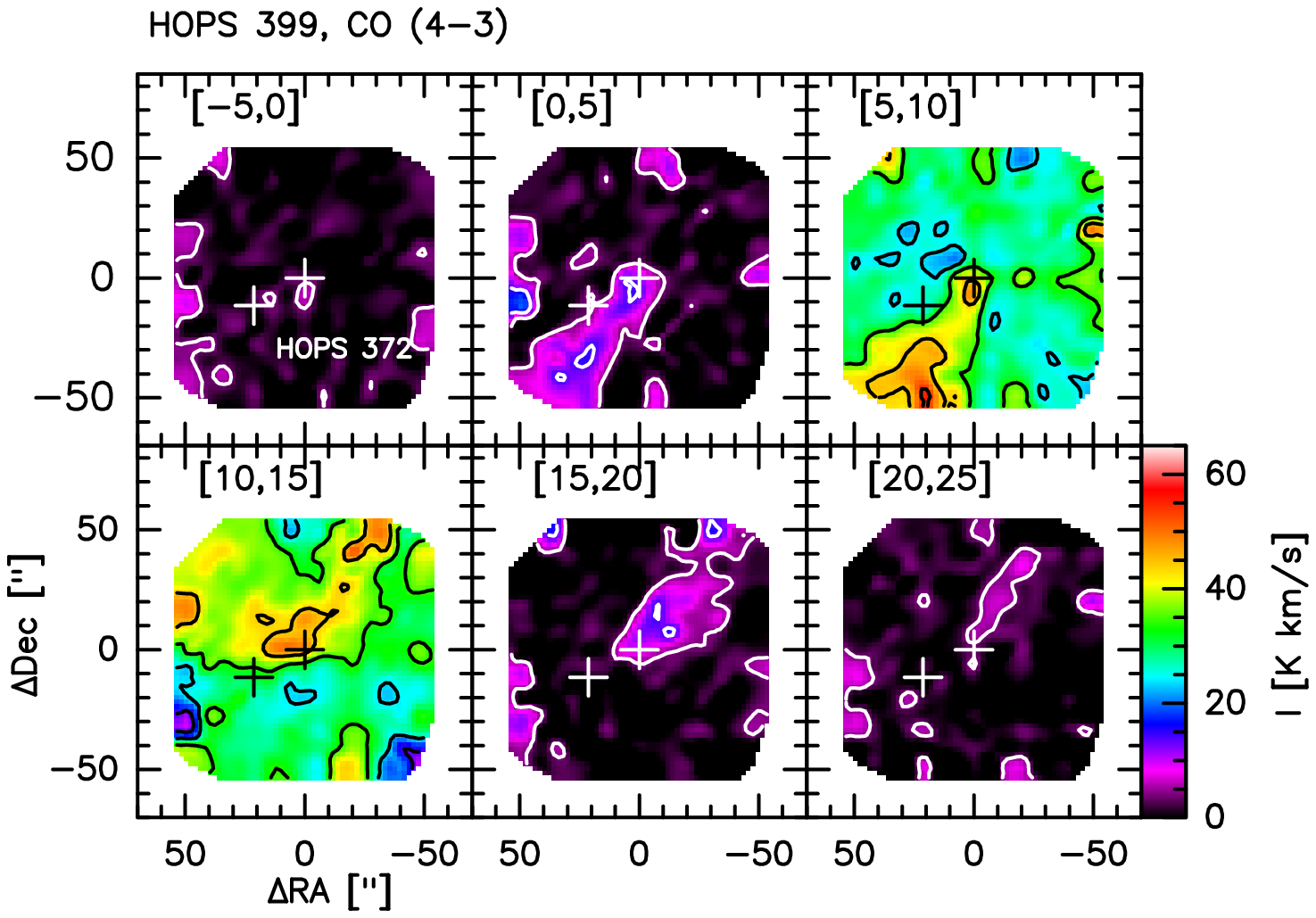}
\includegraphics[height=5.5cm, trim=0cm 0cm 4cm 14.5cm,clip=true]{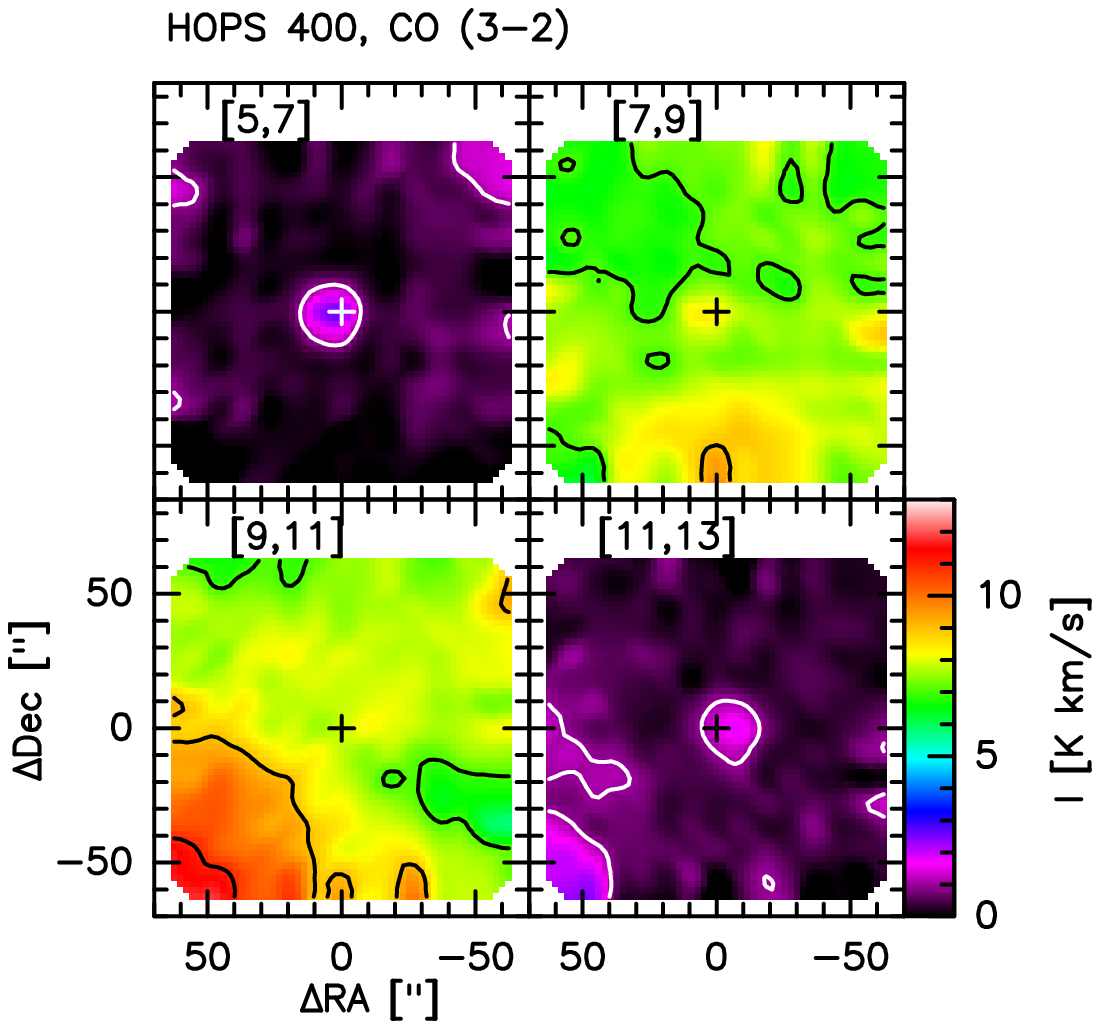}
\hskip+0.3cm
\includegraphics[height=5.5cm, trim=0cm 0cm 4cm 14.5cm,clip=true]{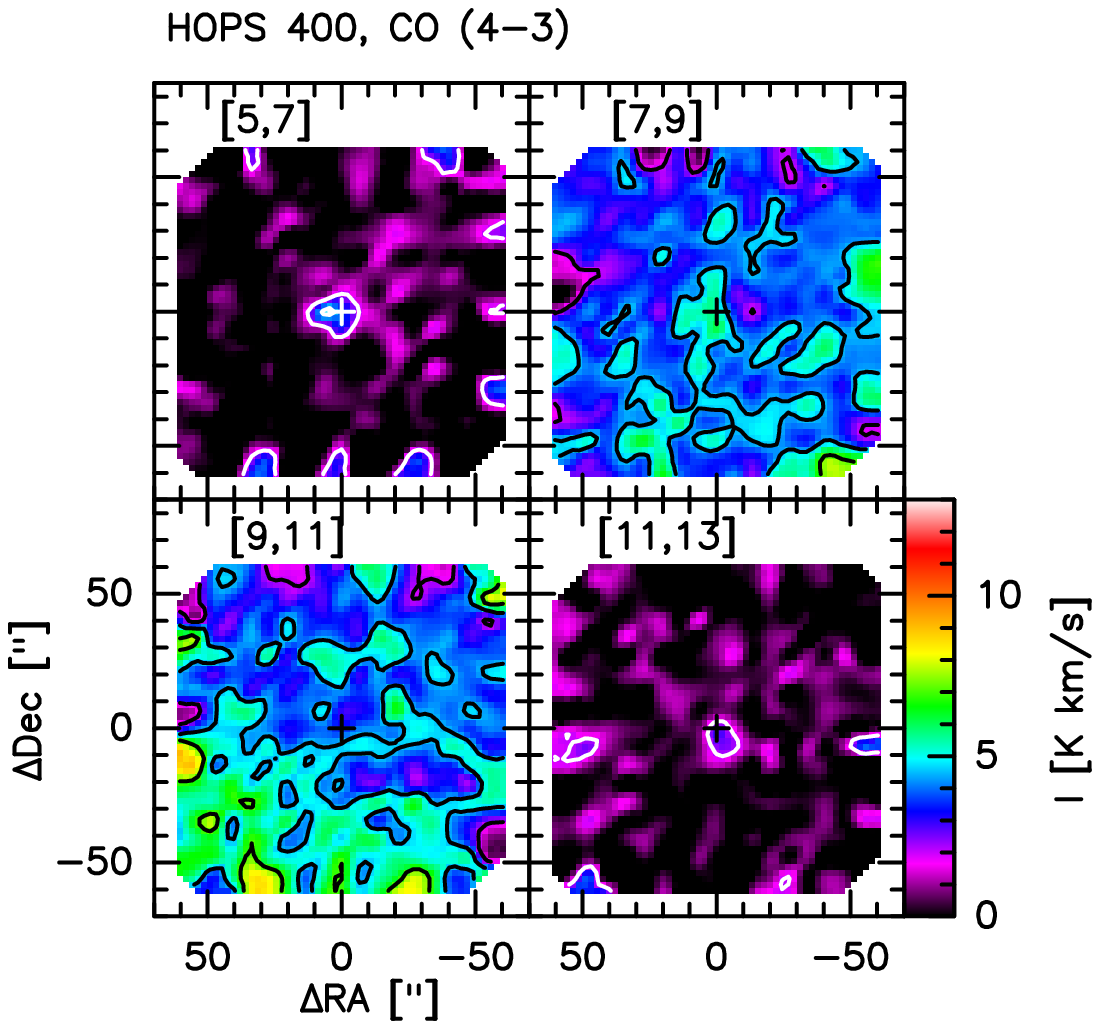}
\includegraphics[height=5.5cm, trim=0cm 0cm 4cm 14.5cm,clip=true]{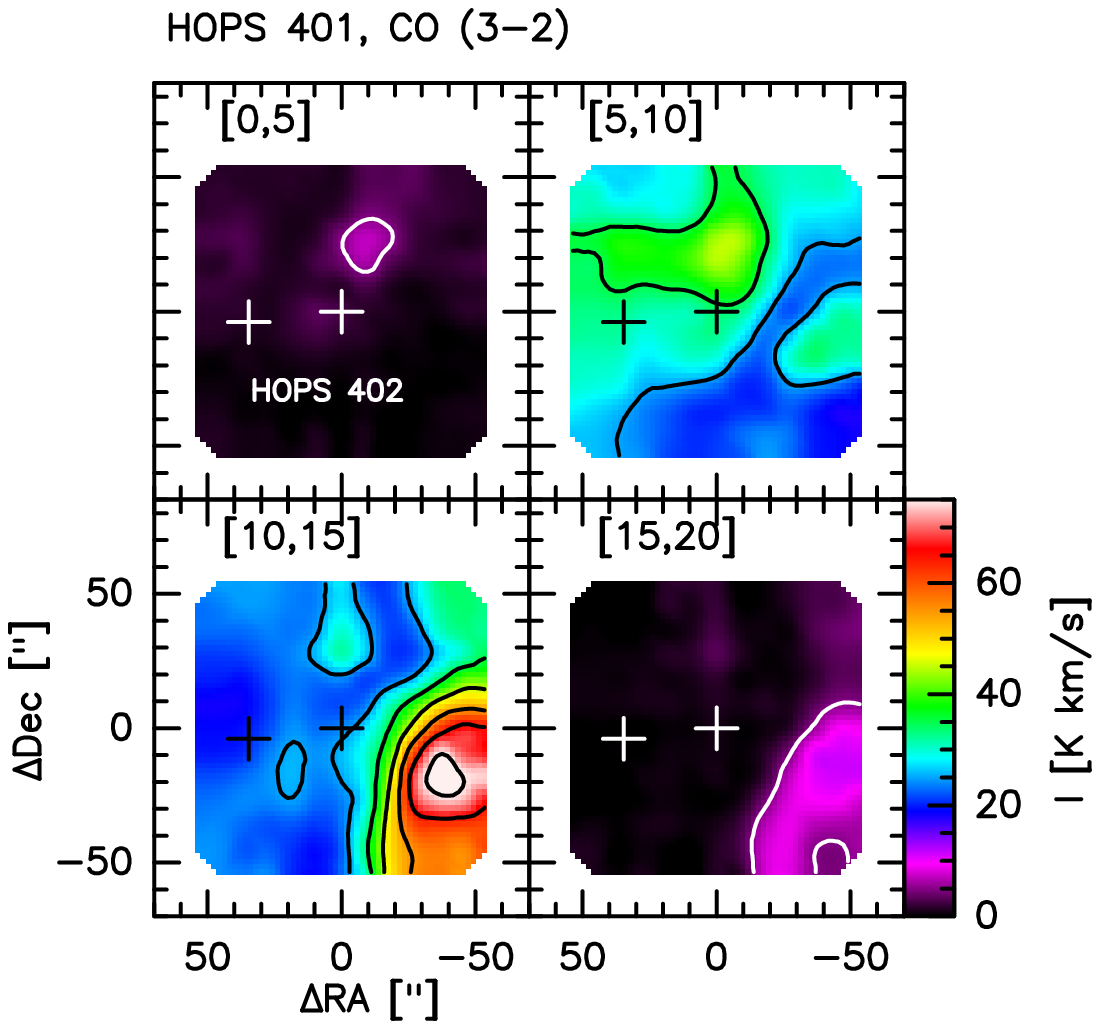}
\hskip+0.3cm
\includegraphics[height=5.5cm, trim=0cm 0cm 4cm 14.5cm,clip=true]{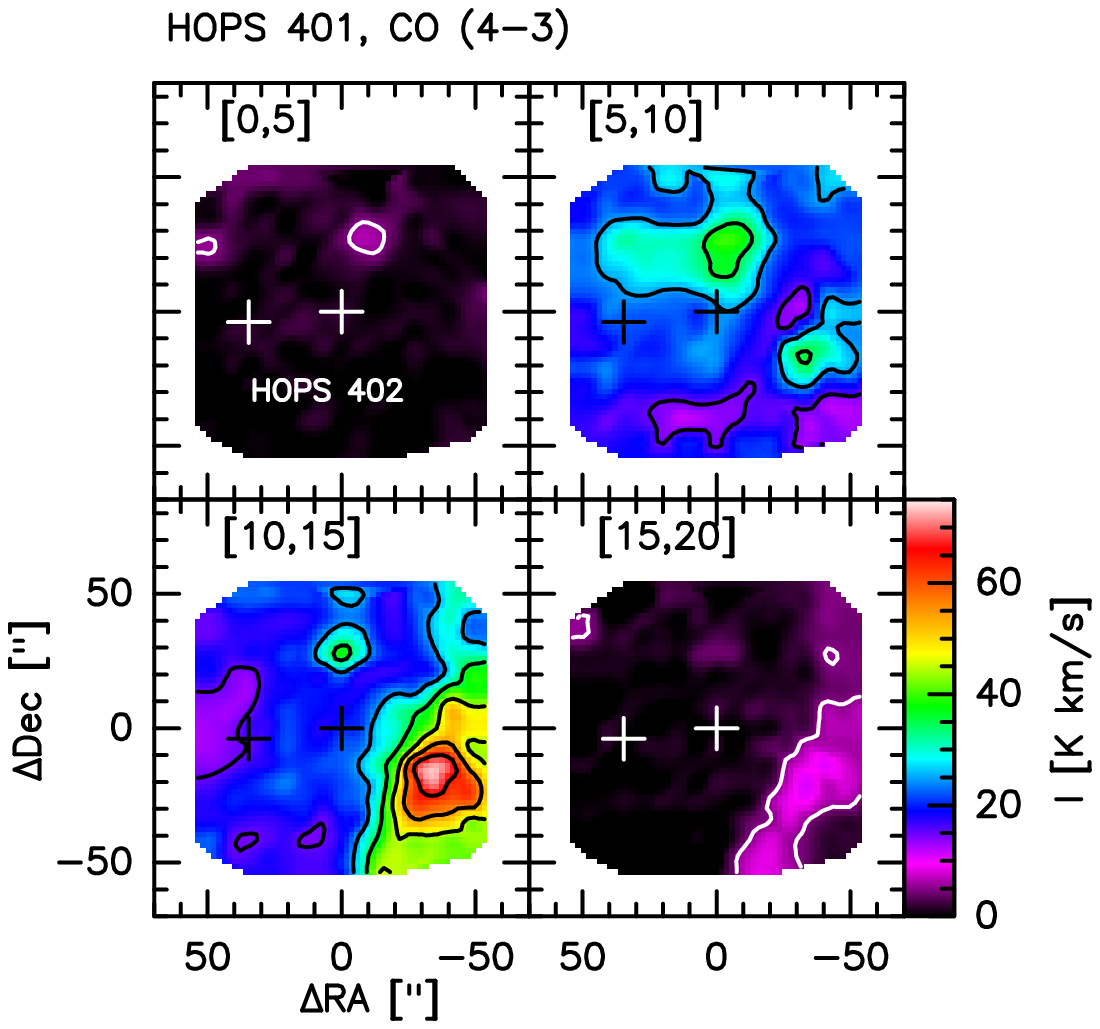}
\label{channelmaps_hops398_32}
\caption{Intensity maps centered on HOPS 398, HOPS 399, HOPS 400, and HOPS 401 integrated over the velocity ranges shown in the upper-left corner of each subfigure.
Contour levels: between 3 and 50 K km s$^{-1}$ at intervals of 5 K km s$^{-1}$ for HOPS 398 CO $J$=3-2 and $J$=4-3, between 5 and 70 K km s$^{-1}$ at intervals of 10 K km s$^{-1}$ for HOPS 399 CO $J$=3-2, between 4 and 70 K km s$^{-1}$ at intervals of 10 K km s$^{-1}$ for HOPS 399 CO $J$=4-3, between 1 and 13 K km s$^{-1}$ at intervals of 2 K km s$^{-1}$ for HOPS 400 CO $J$=3-2, between 2 and 13 K km s$^{-1}$ at intervals of 2.5 K km s$^{-1}$ for HOPS 400 CO $J$=4-3, between 5 and 75 K km s$^{-1}$ at intervals of 10 K km s$^{-1}$ for HOPS 401 CO $J$=3-2 and $J$=4-3.
}
\end{figure*}

\begin{figure*}[!h]
\centering
\includegraphics[height=5.5cm, trim=0cm 0cm 4cm 14.5cm,clip=true]{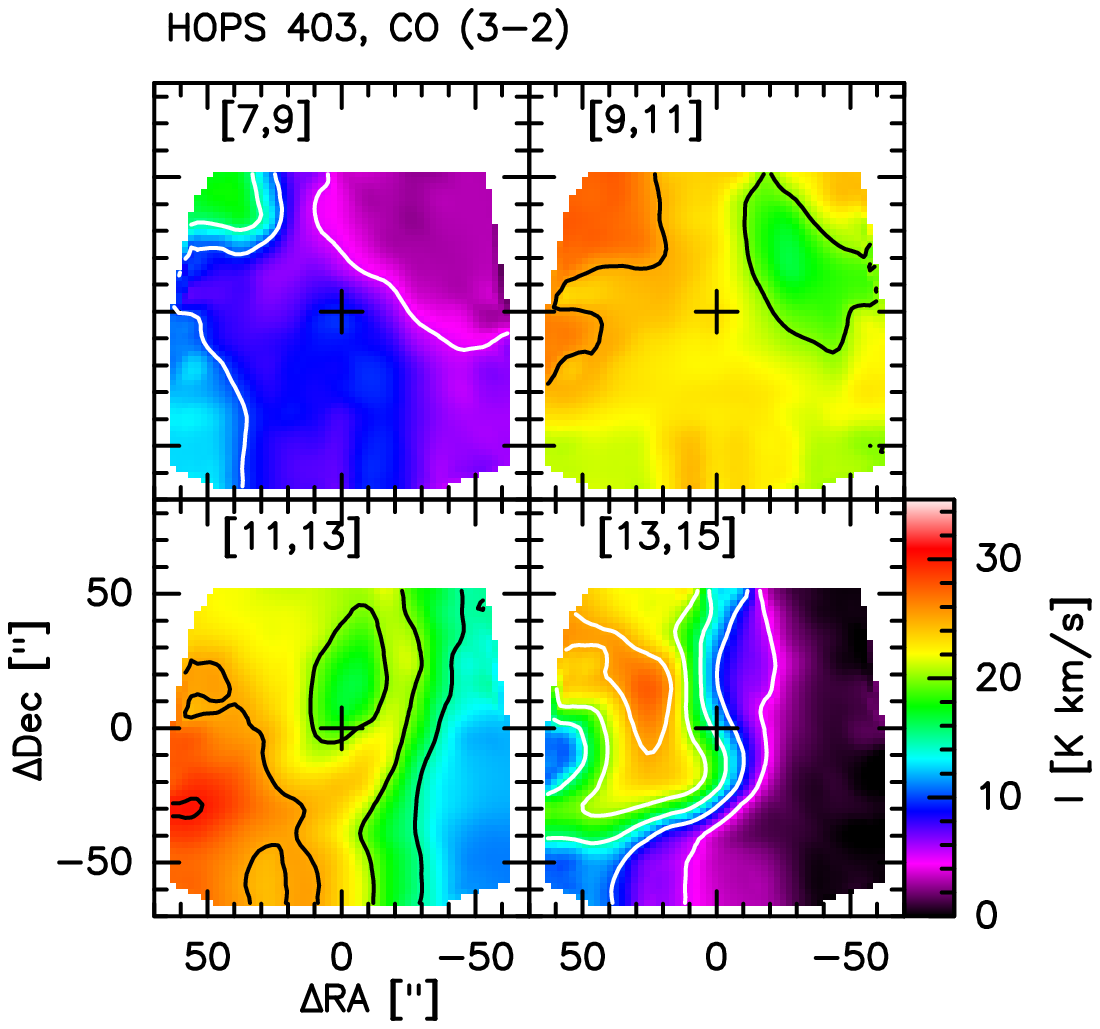}
\hskip+0.3cm
\includegraphics[height=5.5cm, trim=0cm 0cm 4cm 14.5cm,clip=true]{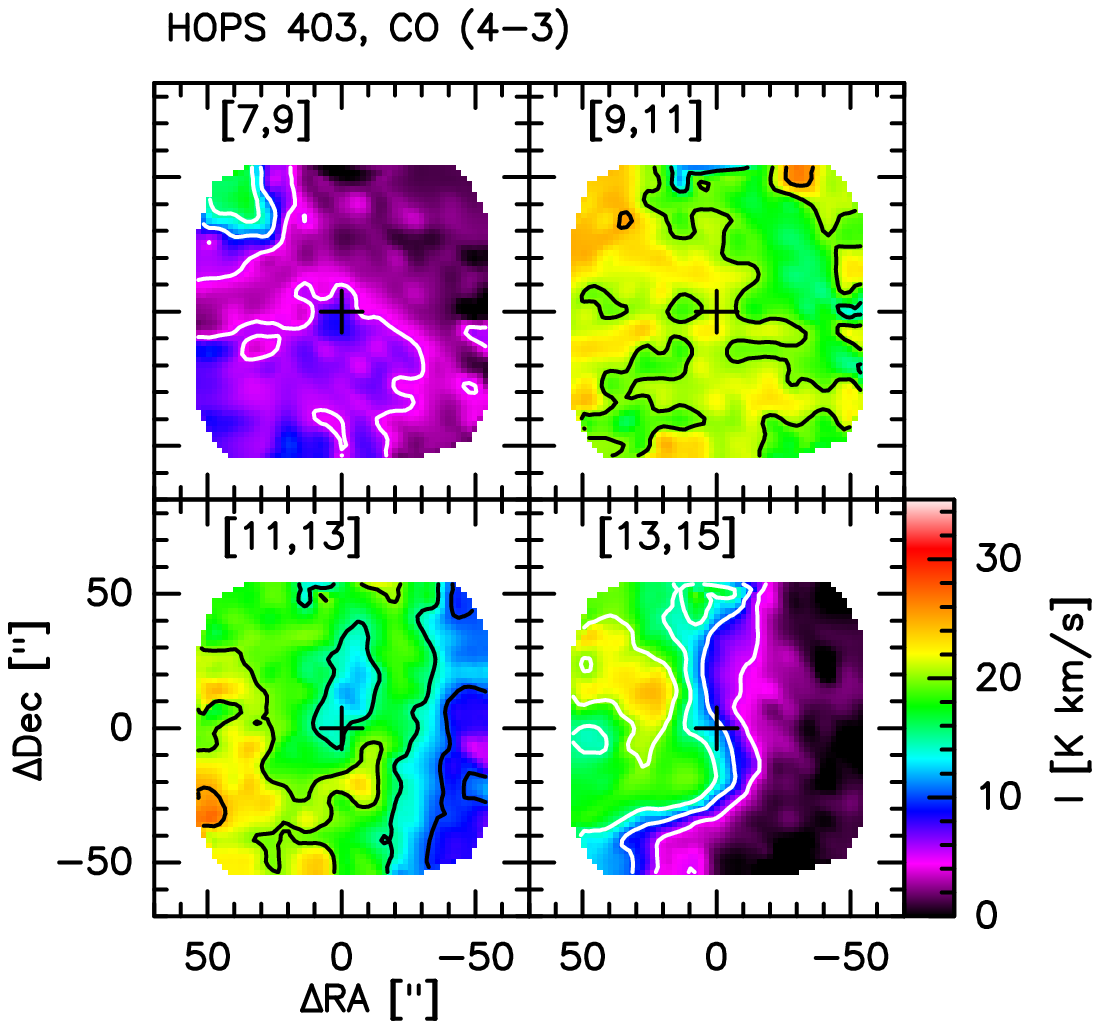}
\includegraphics[height=5.5cm, trim=0cm 0cm 4cm 14.5cm,clip=true]{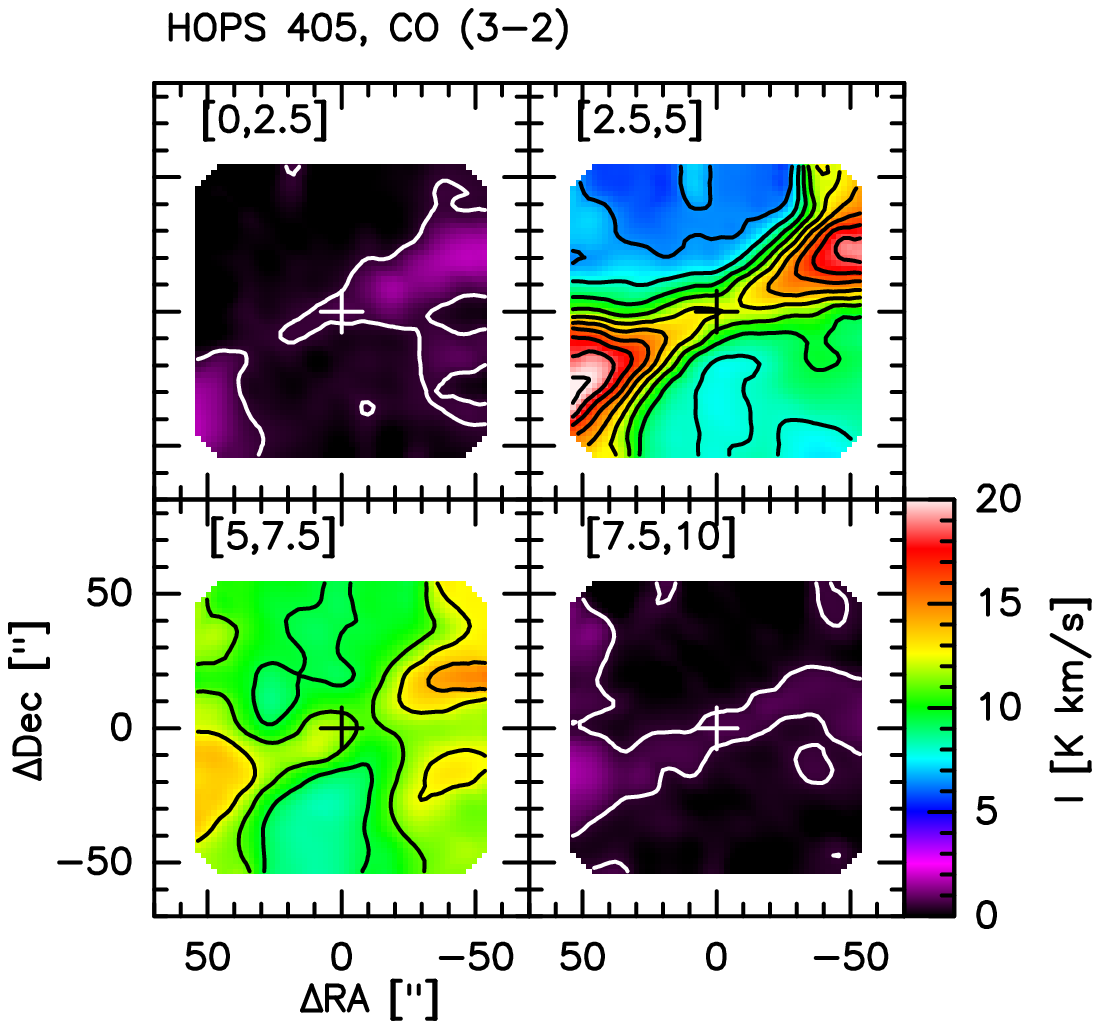}
\hskip+0.3cm
\includegraphics[height=5.5cm, trim=0cm 0cm 4cm 14.5cm,clip=true]{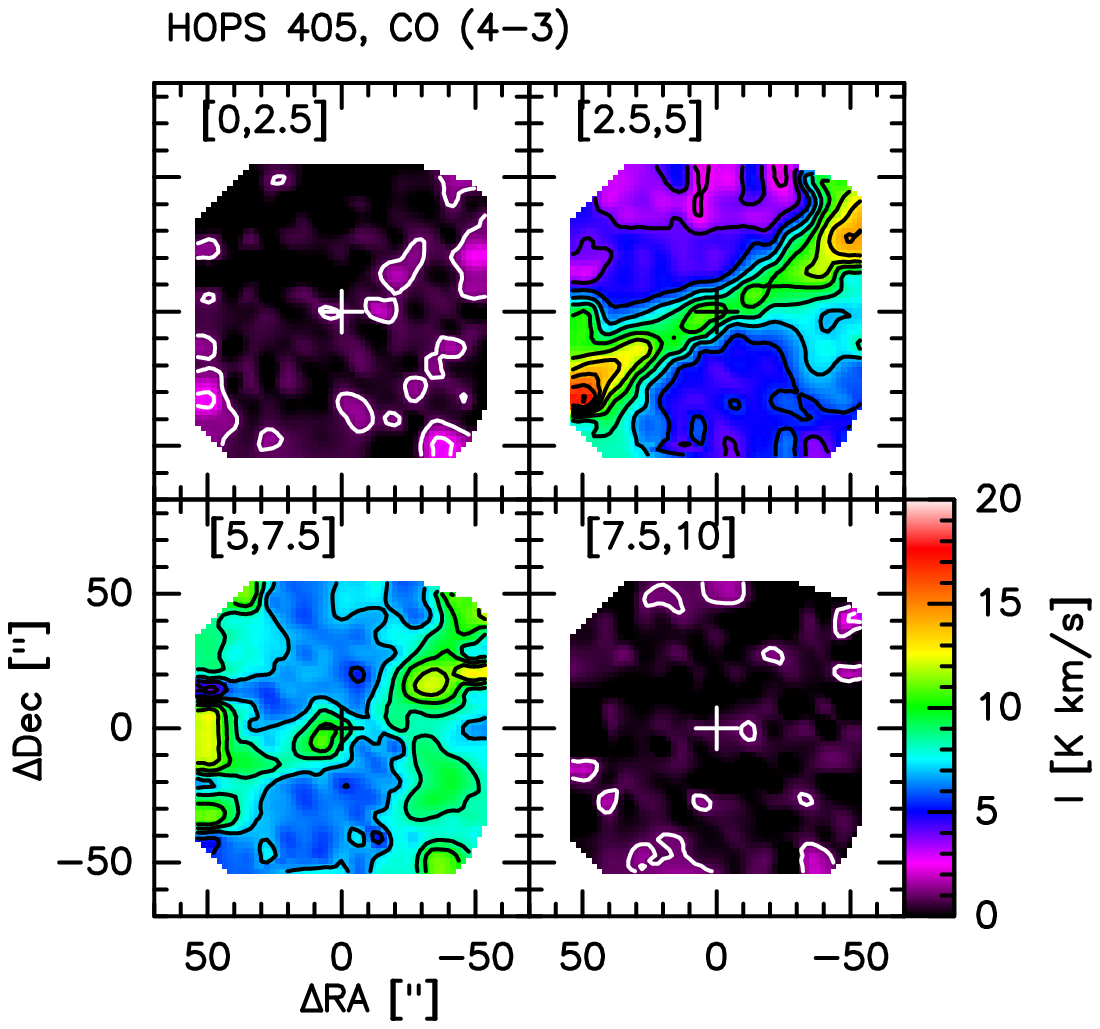}
\includegraphics[height=5.5cm, trim=0cm 0cm 4cm 14.5cm,clip=true]{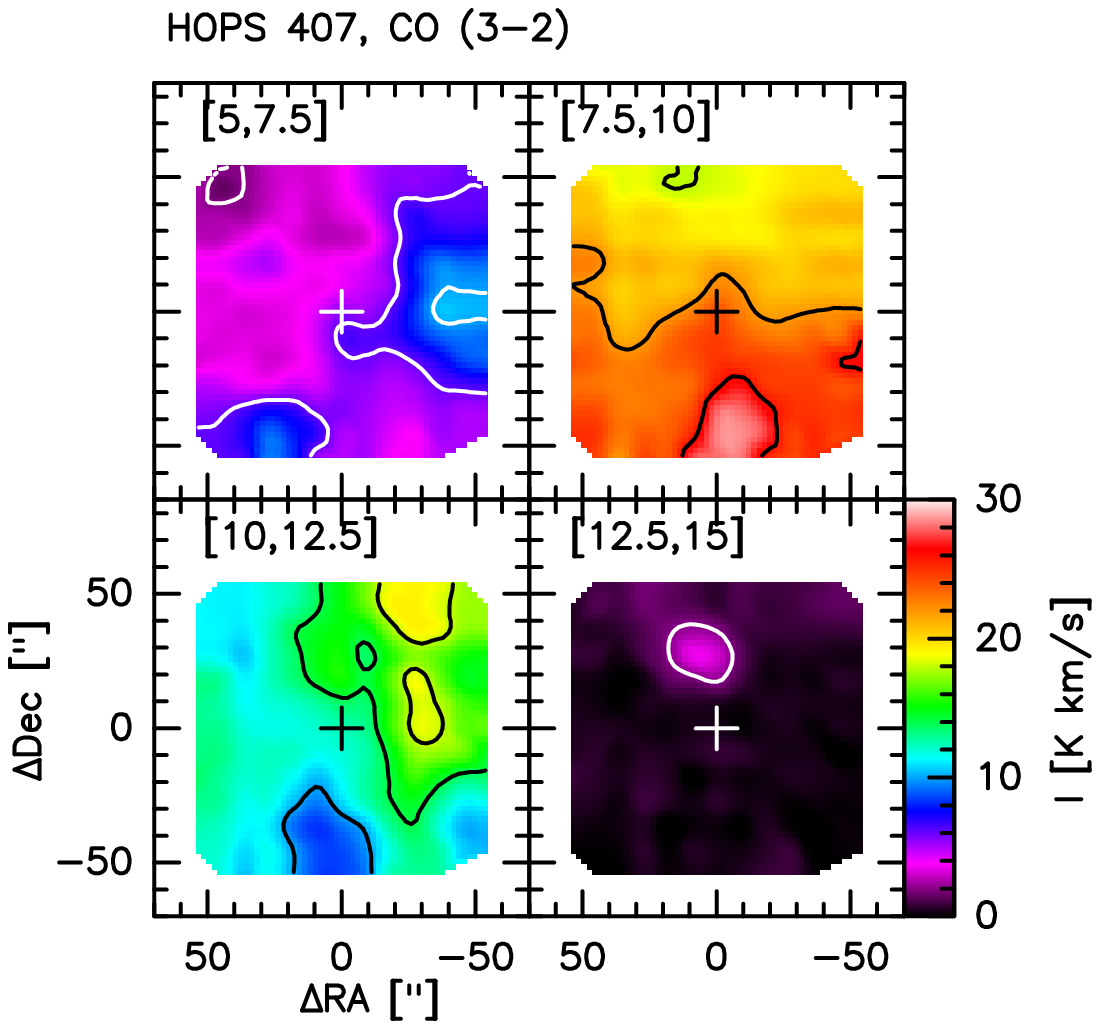}
\hskip+0.3cm
\includegraphics[height=5.5cm, trim=0cm 0cm 4cm 14.5cm,clip=true]{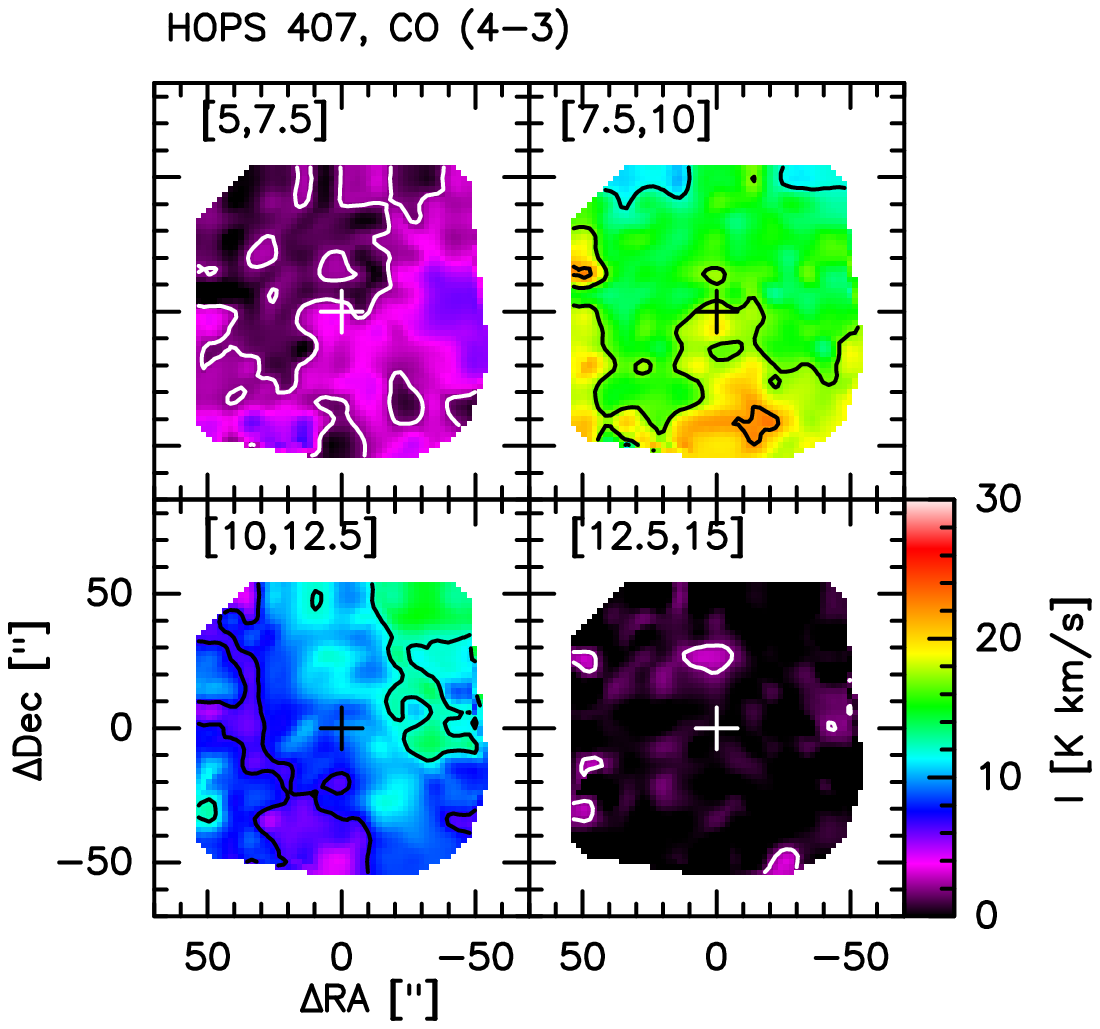}
\includegraphics[height=5.5cm, trim=0cm 0cm 4cm 14.5cm,clip=true]{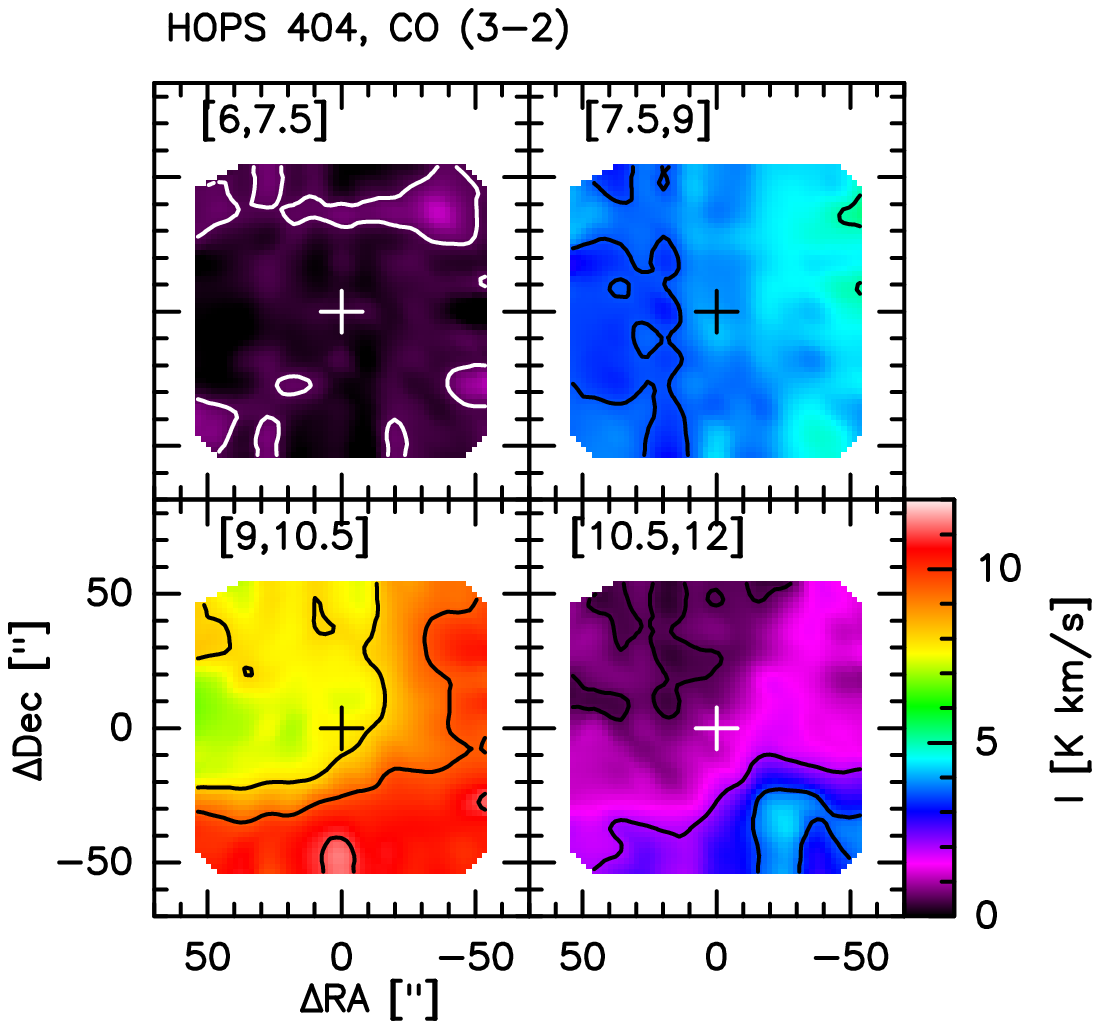}
\label{channelmaps_hops403_32}
\caption{Intensity maps centered on HOPS 403, HOPS 405, HOPS 407, and HOPS 404 integrated over the velocity ranges shown in the upper-left corner of each subfigure.
Contour levels: between 5 and 35 K km s$^{-1}$ at intervals of 5 K km s$^{-1}$ for HOPS 403 $J$=3-2 and $J$=4-3, between 0.5 and 20 K km s$^{-1}$ at intervals of 1.5 K km s$^{-1}$ for HOPS 405 CO $J$=3-2, between 1 and 20 K km s$^{-1}$ at intervals of 1.5 K km s$^{-1}$ for HOPS 405 $J$=4-3, between 2 and 30 K km s$^{-1}$ at intervals of 4 K km s$^{-1}$ for HOPS 407 CO $J$=3-2, between 2 and 30 K km s$^{-1}$ at intervals of 5 K km s$^{-1}$ for HOPS 407 CO $J$=4-3, between 0.5 and 12 K km s$^{-1}$ at intervals of 1.5 K km s$^{-1}$ for HOPS 404 CO $J$=3-2.
}
\end{figure*}

\clearpage

\section{Outflow maps}
\label{sec:int_maps}

\begin{figure*}[!h]

\includegraphics[width=5.9cm, trim=0.5cm 1cm 0.4cm 8cm,clip=true]{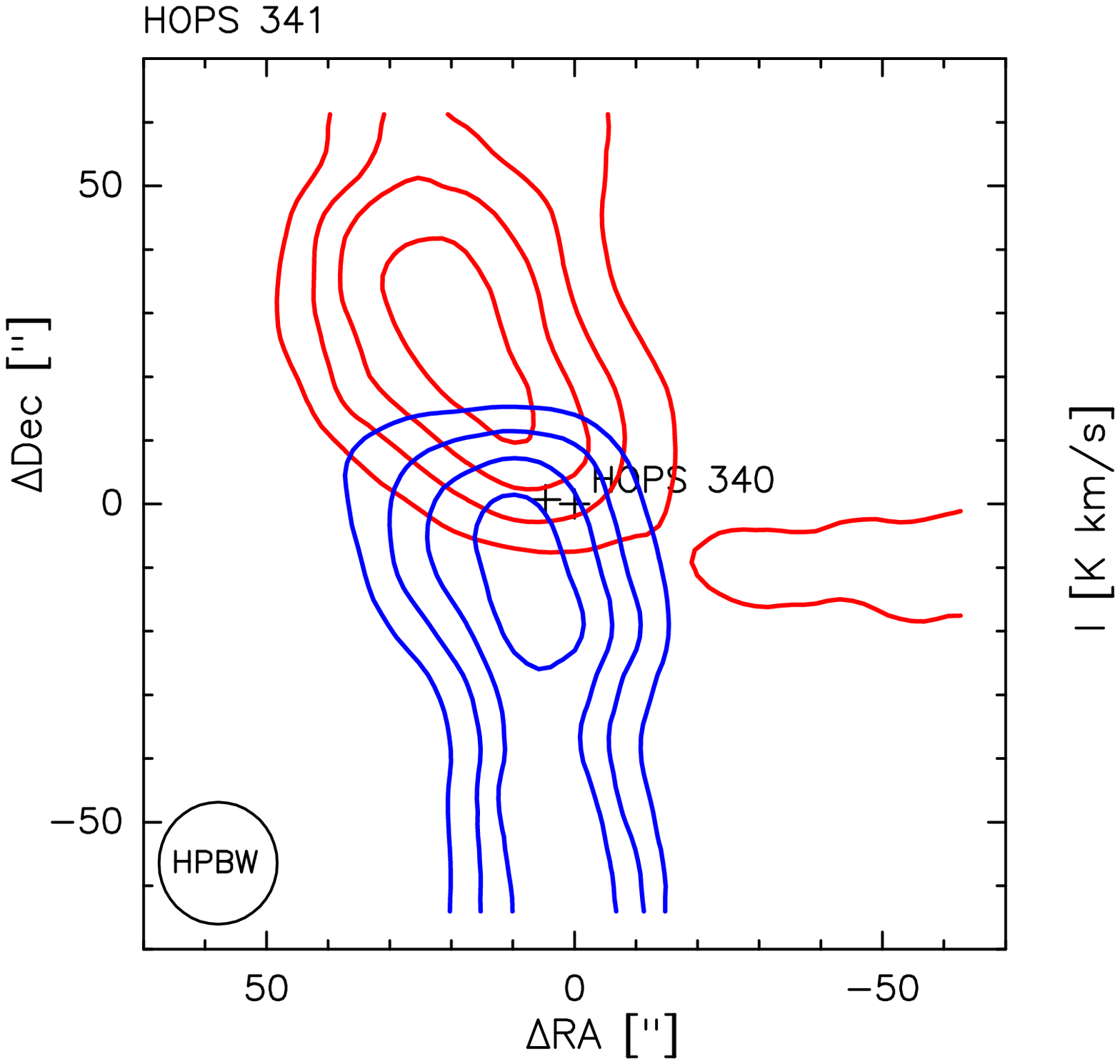}
\includegraphics[width=5.9cm, trim=0.5cm 1cm 0.4cm 8cm,clip=true]{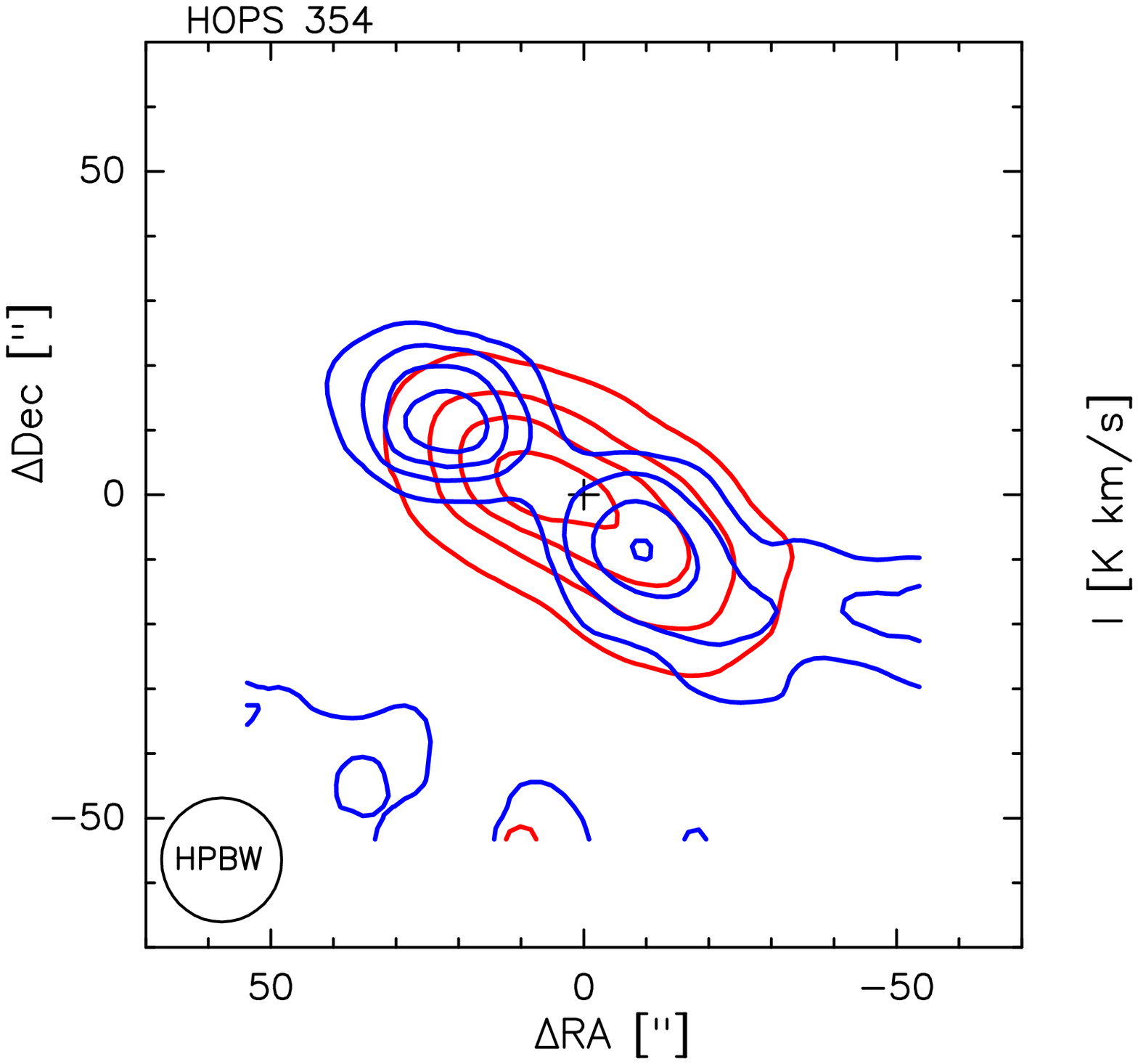}
\includegraphics[width=5.9cm, trim=0.5cm 1cm 0.4cm 8cm,clip=true]{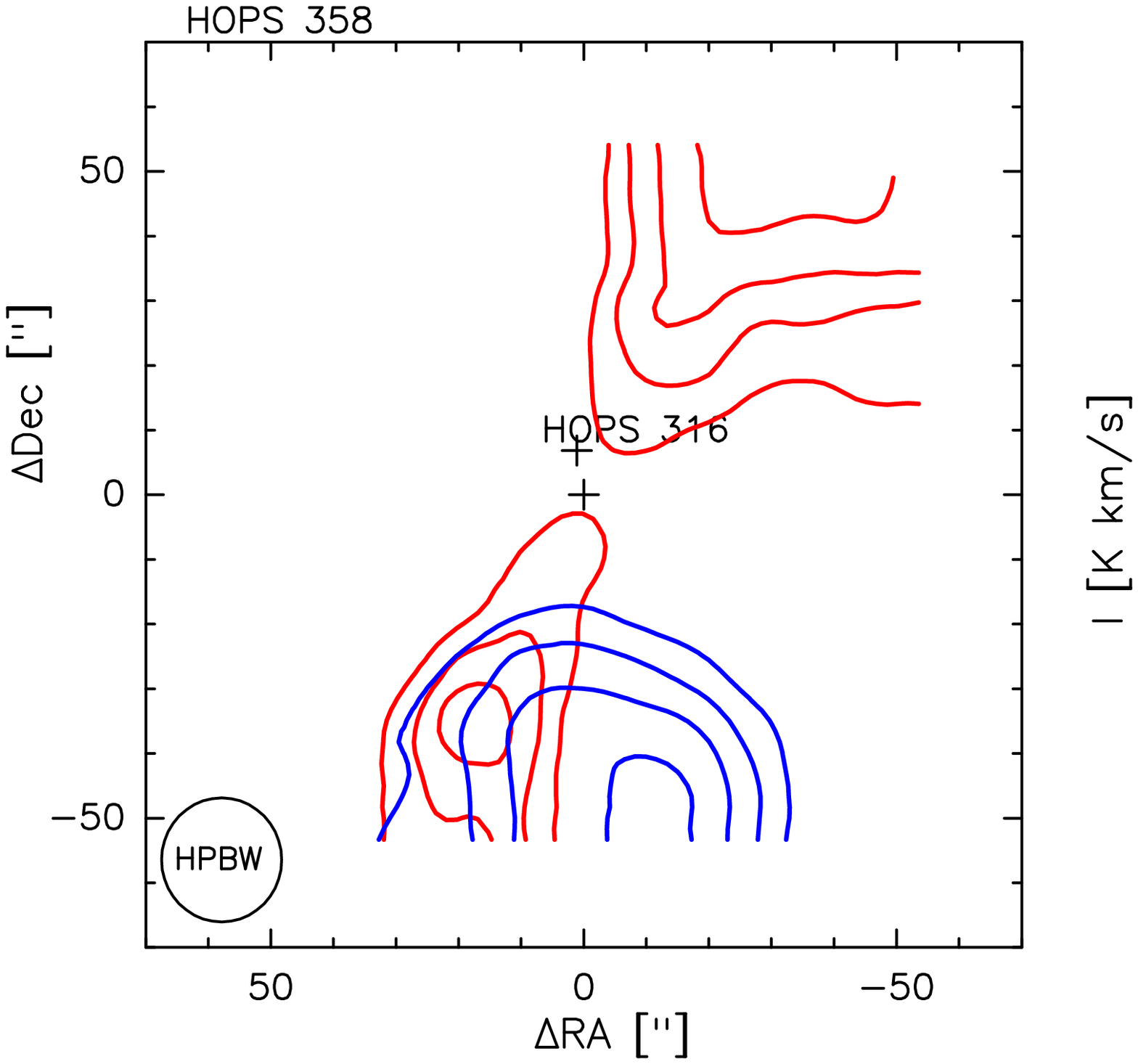}
\includegraphics[width=5.9cm, trim=0.5cm 1cm 0.4cm 8cm,clip=true]{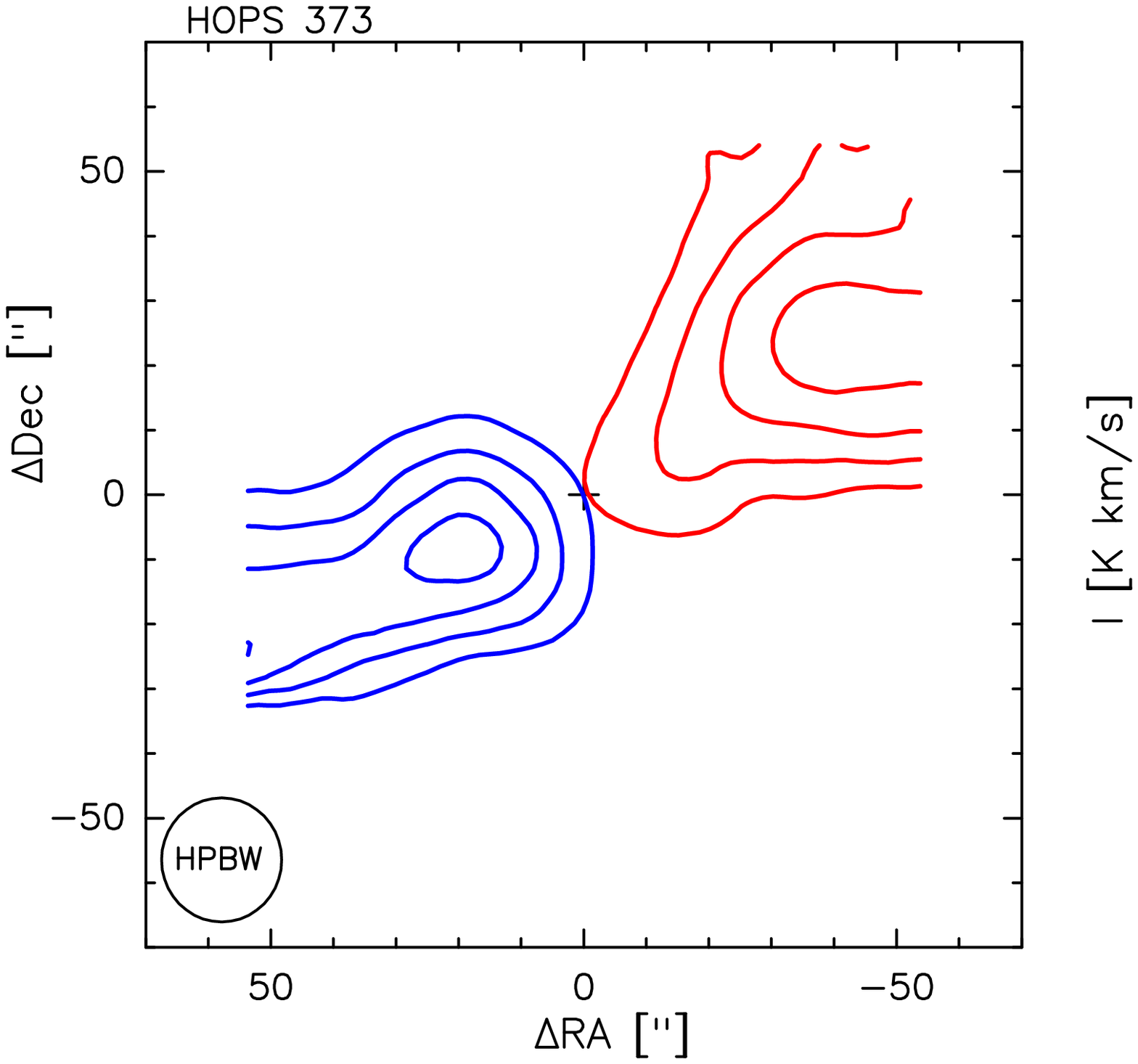}
\includegraphics[width=5.9cm, trim=0.5cm 1cm 0.4cm 8cm,clip=true]{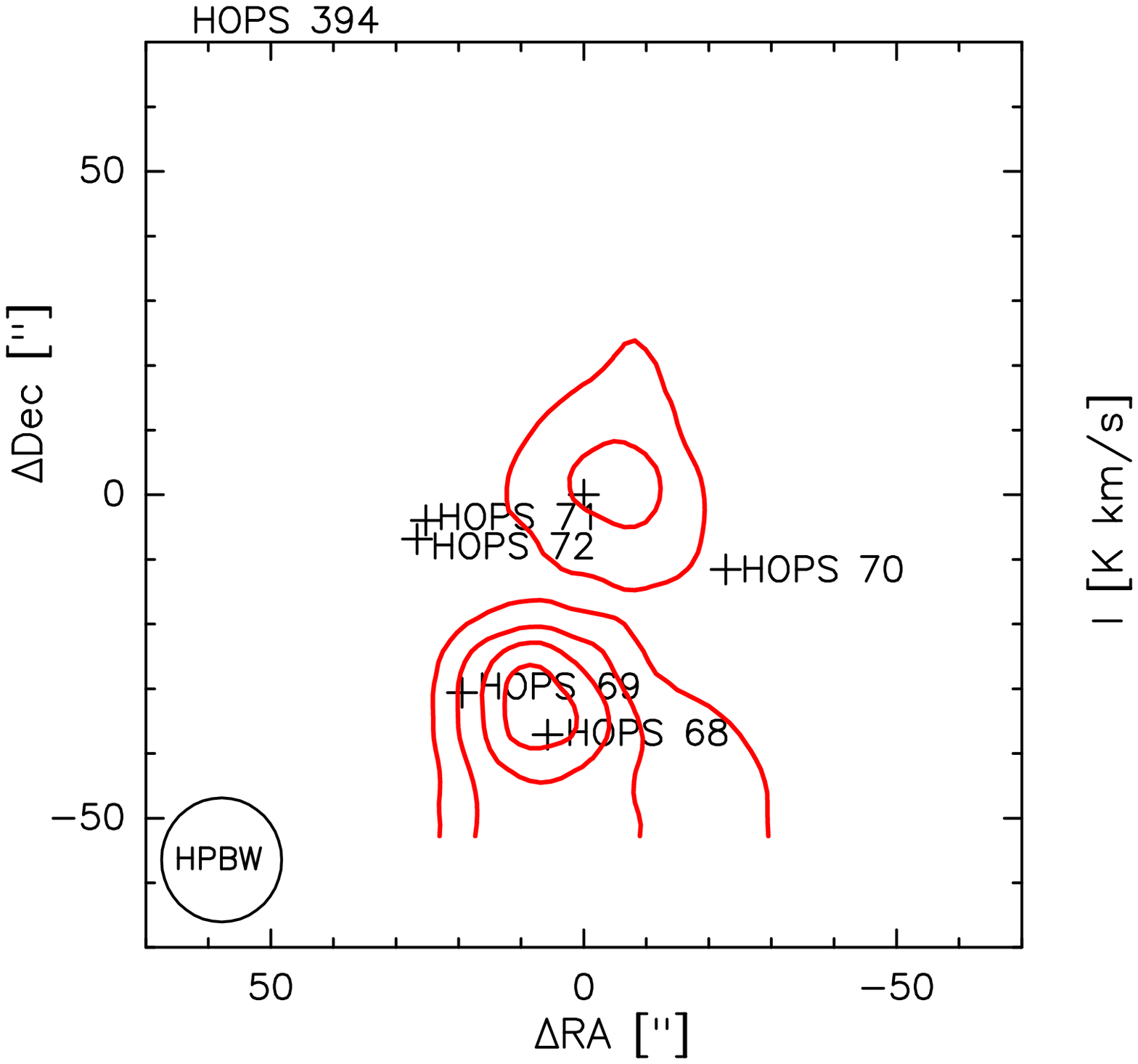}
\includegraphics[width=5.9cm, trim=0.5cm 1cm 0.4cm 8cm,clip=true]{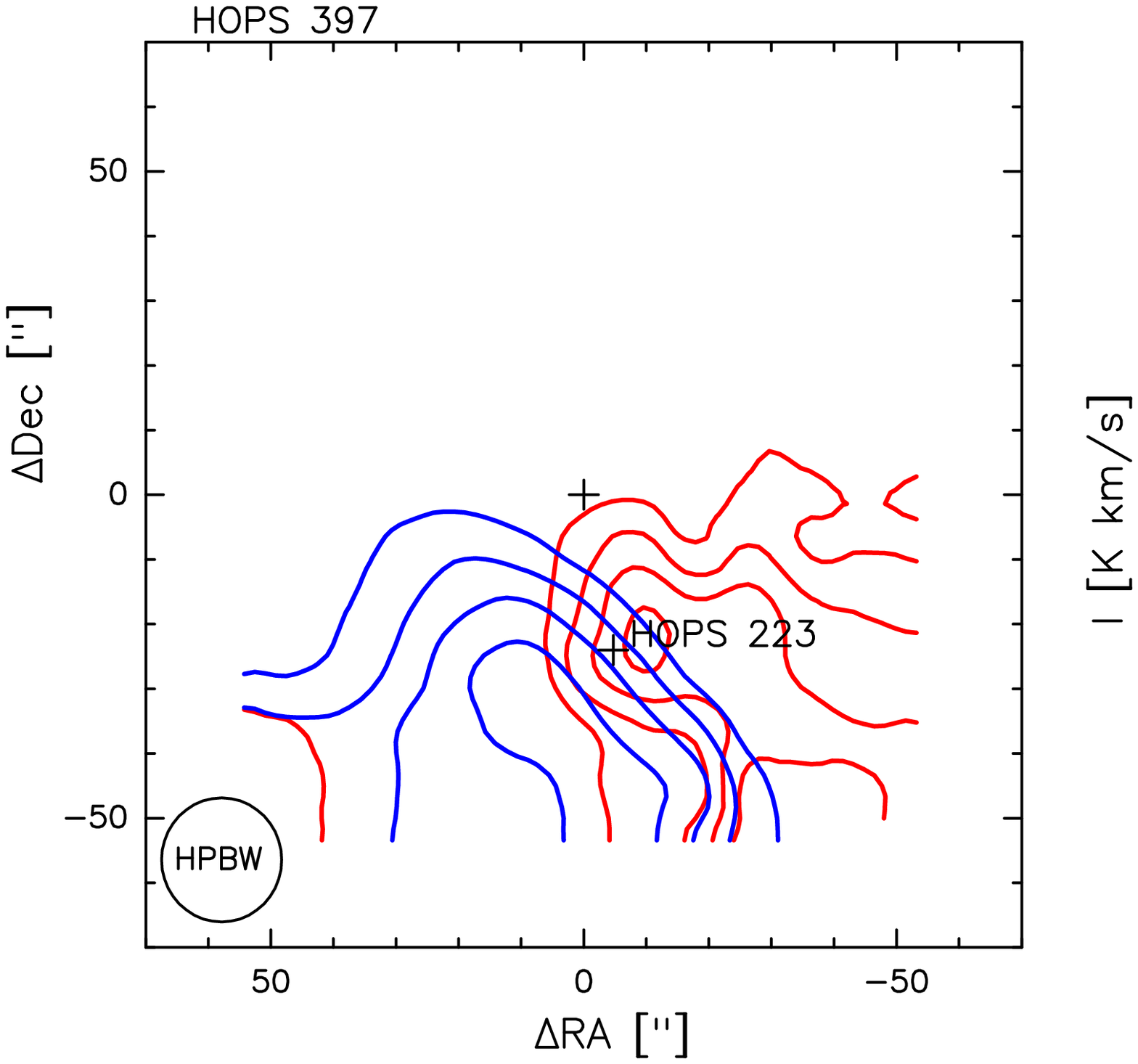}
\includegraphics[width=5.9cm, trim=0.5cm 1cm 0.4cm 8cm,clip=true]{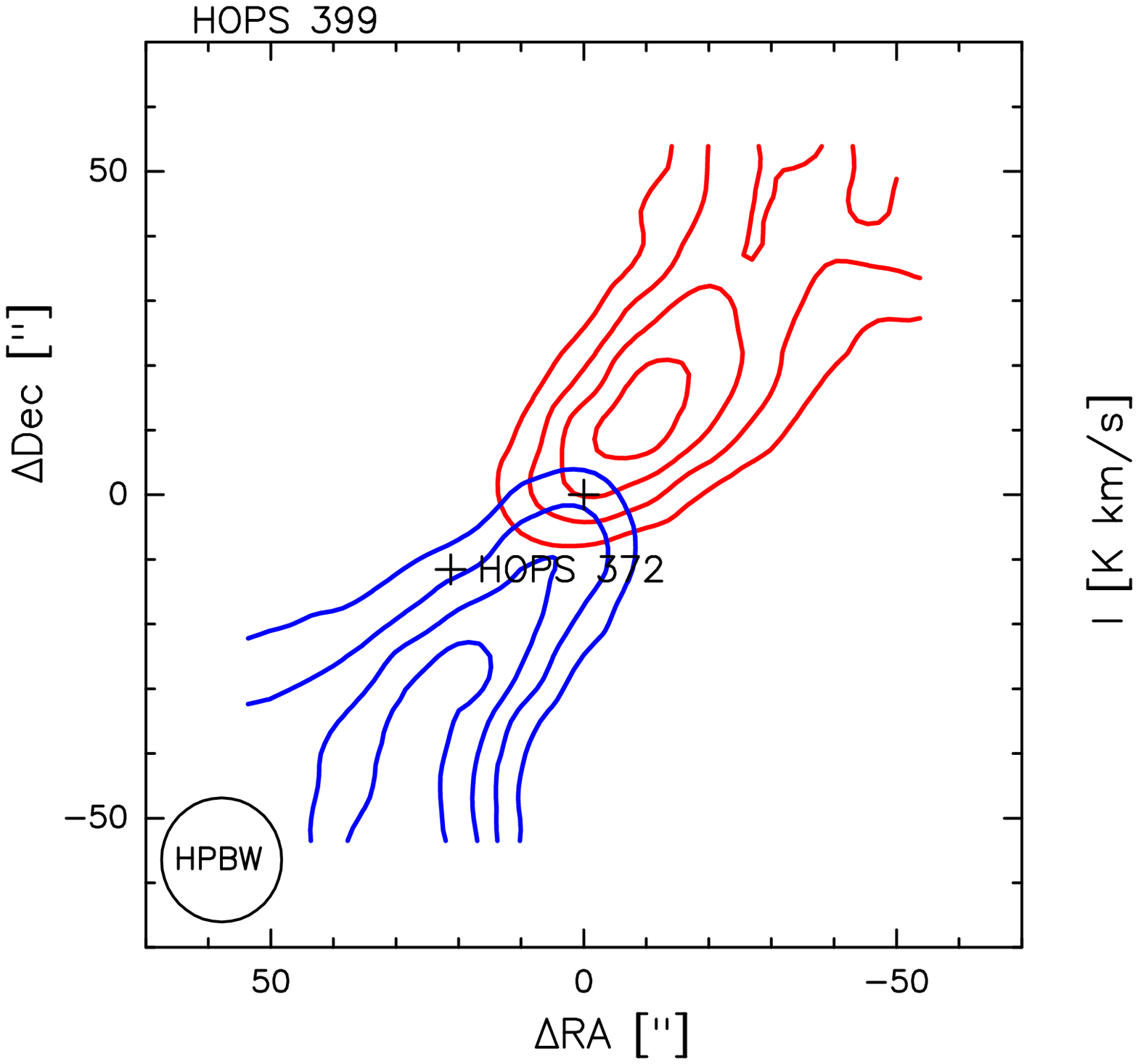}
\includegraphics[width=5.9cm, trim=0.5cm 1cm 0.4cm 8cm,clip=true]{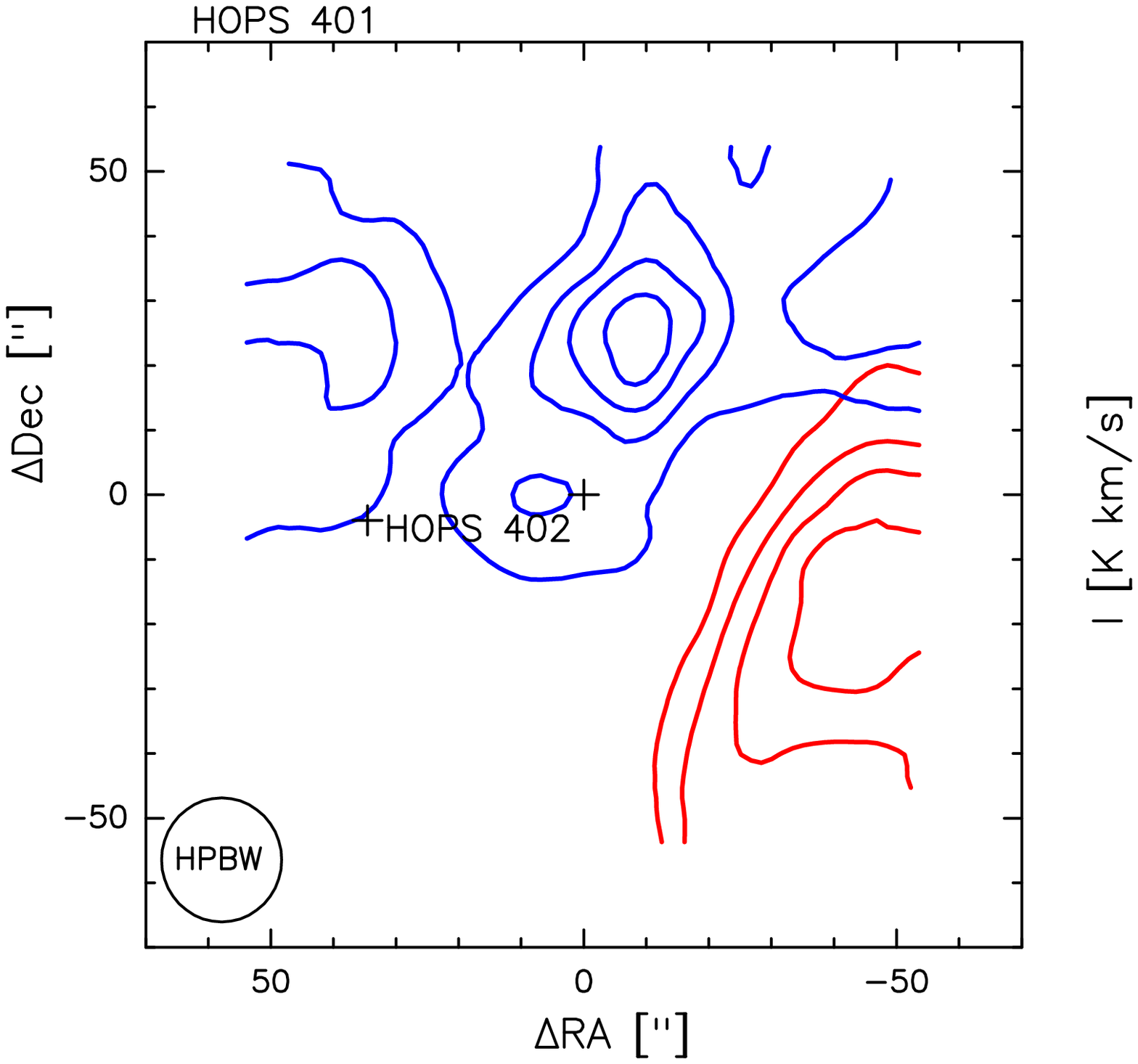}
\caption{CO $J$=3-2 integrated intensities of the blue- and red-shifted outflows (blue and red contours) for the sources with clear outflow detections. The integration limits are listed in Table \ref{table:outflowmass}.
The contours are 30\%, 50\%, 70\%, and 90\% of the maximum values of the intensities corresponding to the red- and blue-shifted outflow lobes in $T_{\rm{MB}}$ units toward the different maps, which are: 23.8 K km/s (red) and 68.0 K km/s (blue) toward HOPS 341, 30.5 K km/s (red) and 12.4 K km/s (blue) toward HOPS 354, 27.0 K km/s (red) and 58.4 K km/s (blue) toward HOPS 358, 68.9 K km/s (red) and 54.9 K km/s (blue) for HOPS 373, 36.9 K km/s (red) for HOPS 394, 13.2 K km/s (red) and 41.4 K km/s (blue) for HOPS 397, 40.0 K km/s (red) and 44.8 K km/s (blue) for HOPS 399, and 38.4 K km/s (red) and 12.8 K km/s (blue) for HOPS 401.
}
\label{fig:int_co32}
\end{figure*}

\begin{figure*}[!h]
\includegraphics[width=5.9cm, trim=0.5cm 1cm 0.4cm 8cm,clip=true]{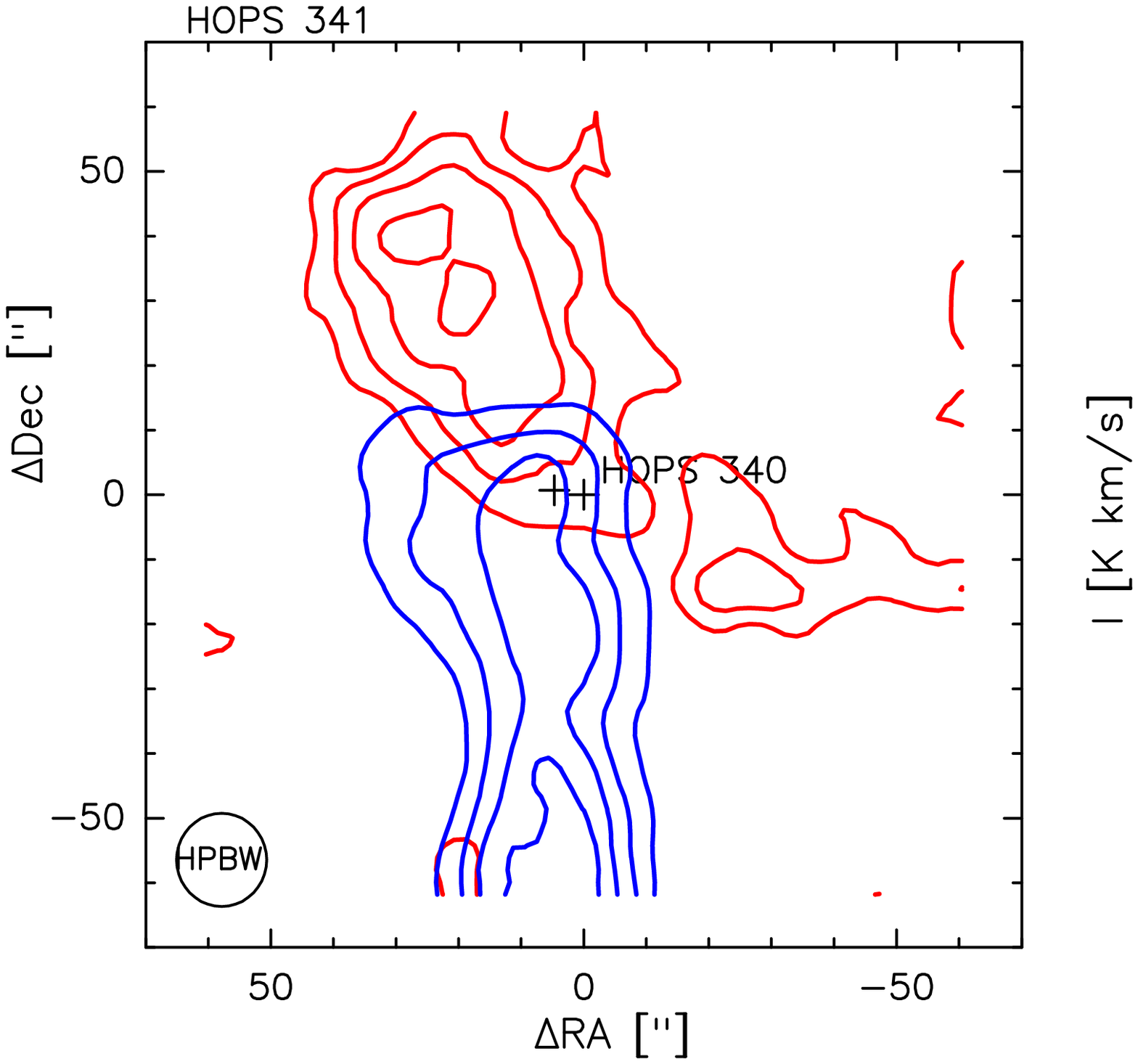}
\includegraphics[width=5.9cm, trim=0.5cm 1cm 0.4cm 8cm,clip=true]{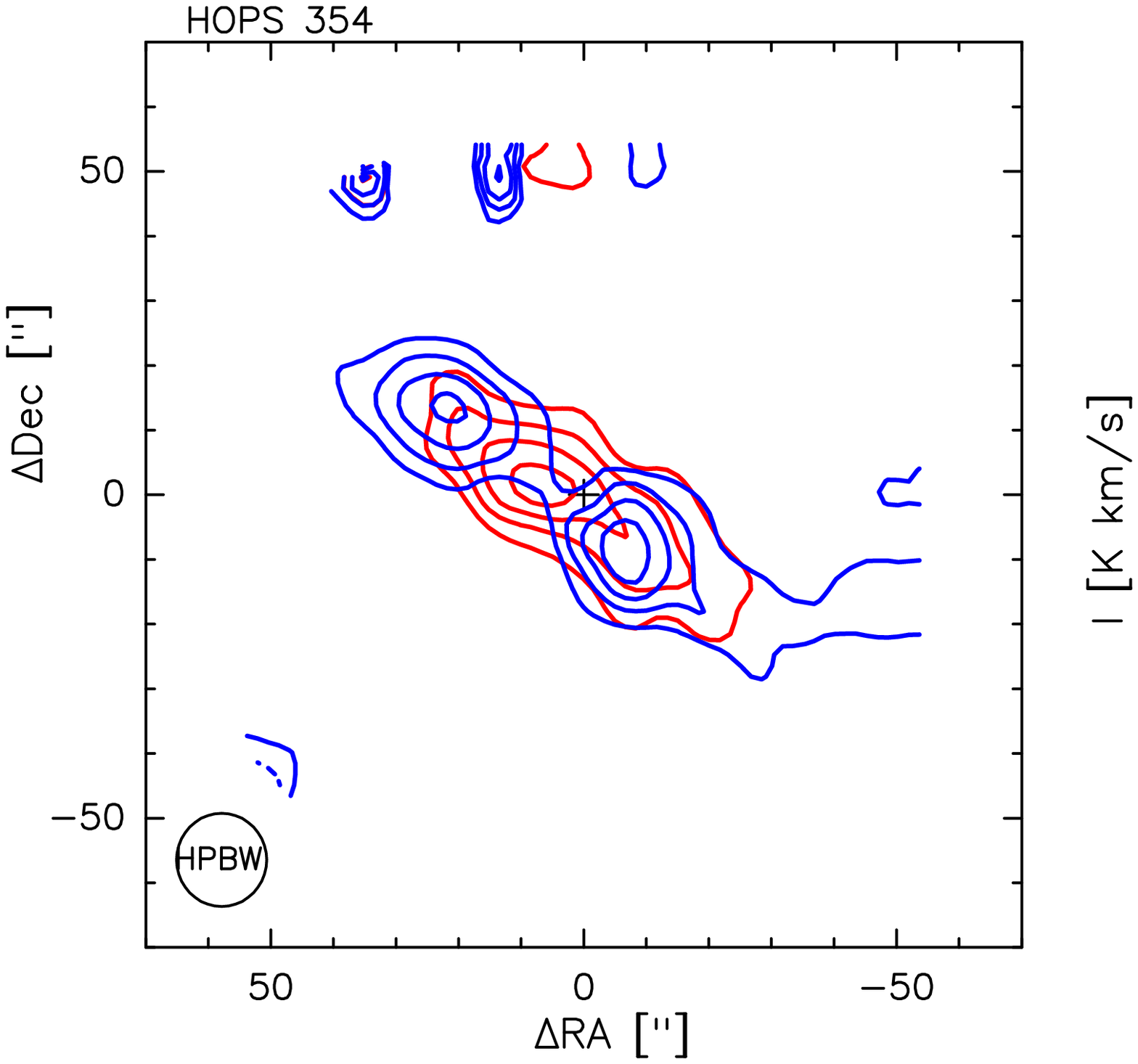}
\includegraphics[width=5.9cm, trim=0.5cm 1cm 0.4cm 8cm,clip=true]{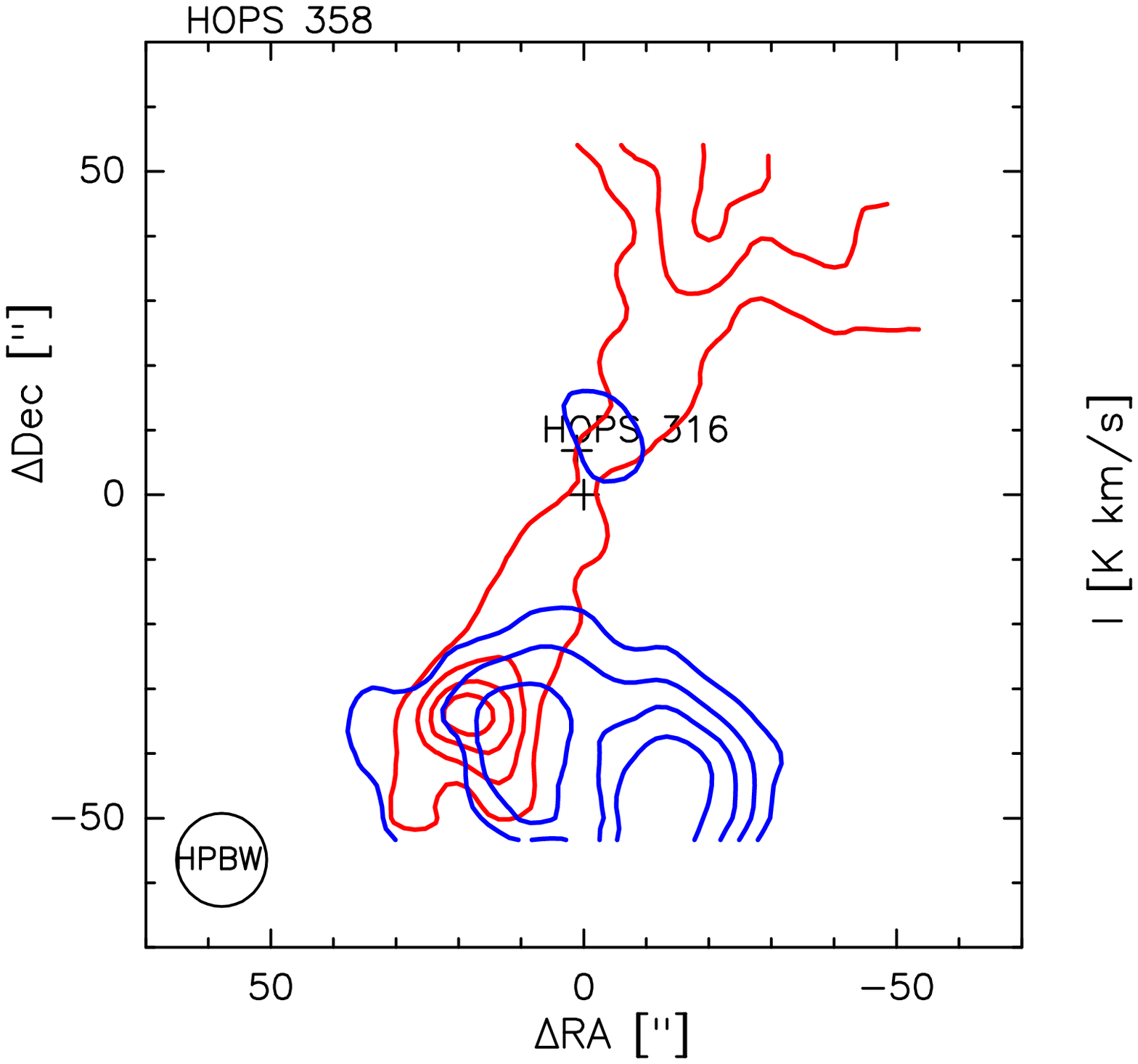}
\includegraphics[width=5.9cm, trim=0.5cm 1cm 0.4cm 8cm,clip=true]{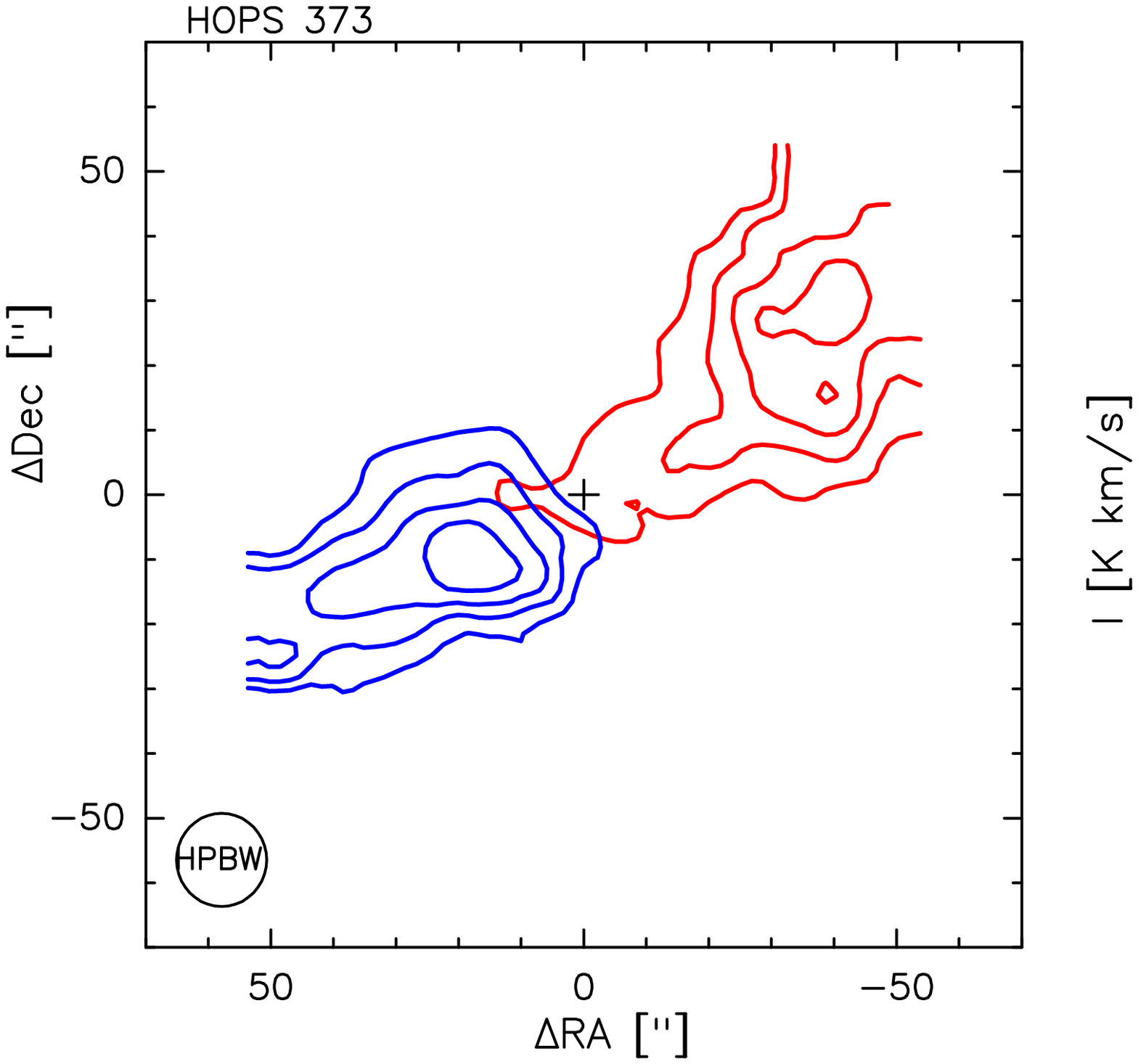}
\includegraphics[width=5.9cm, trim=0.5cm 1cm 0.4cm 8cm,clip=true]{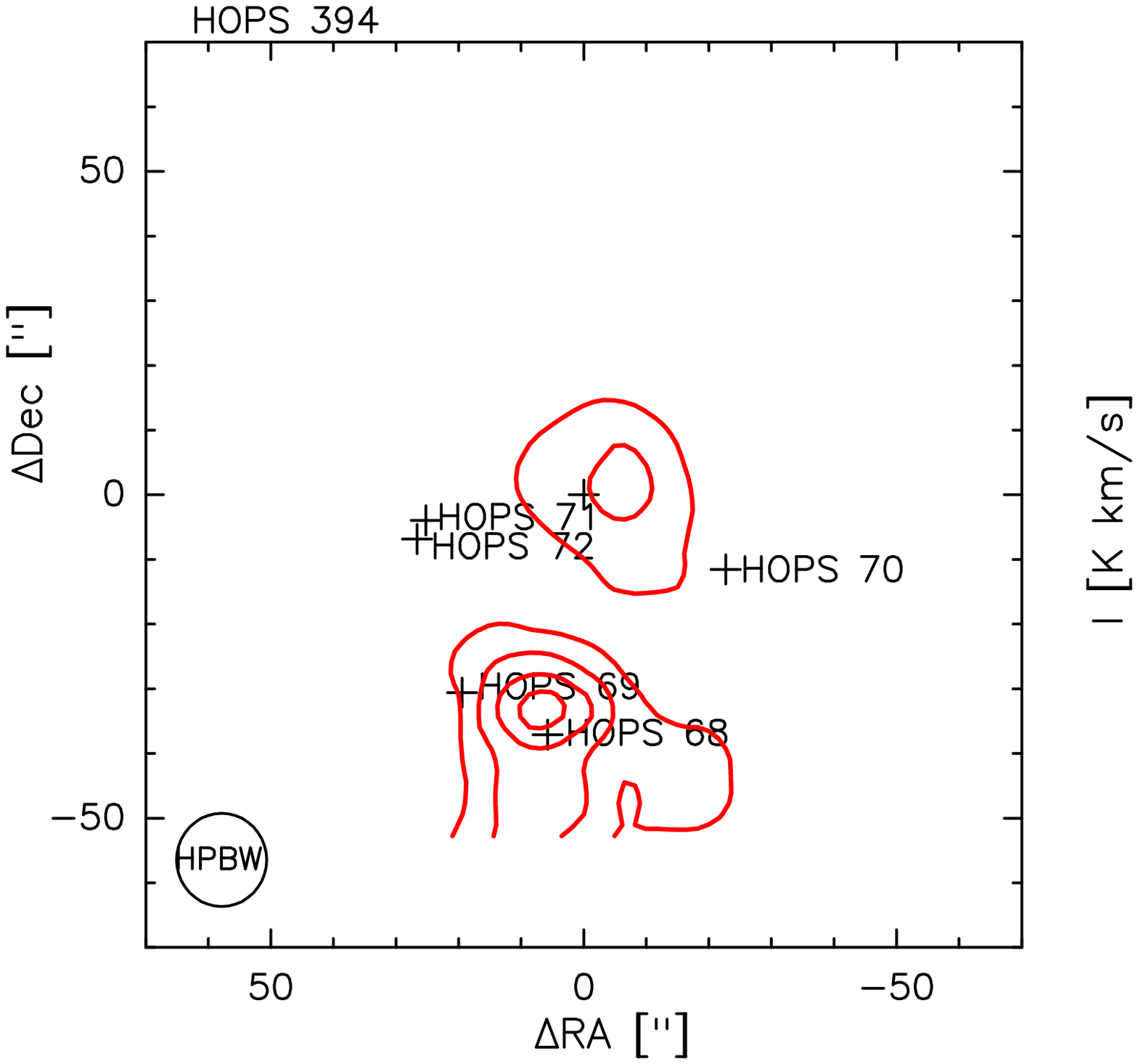}
\includegraphics[width=5.9cm, trim=0.5cm 1cm 0.4cm 8cm,clip=true]{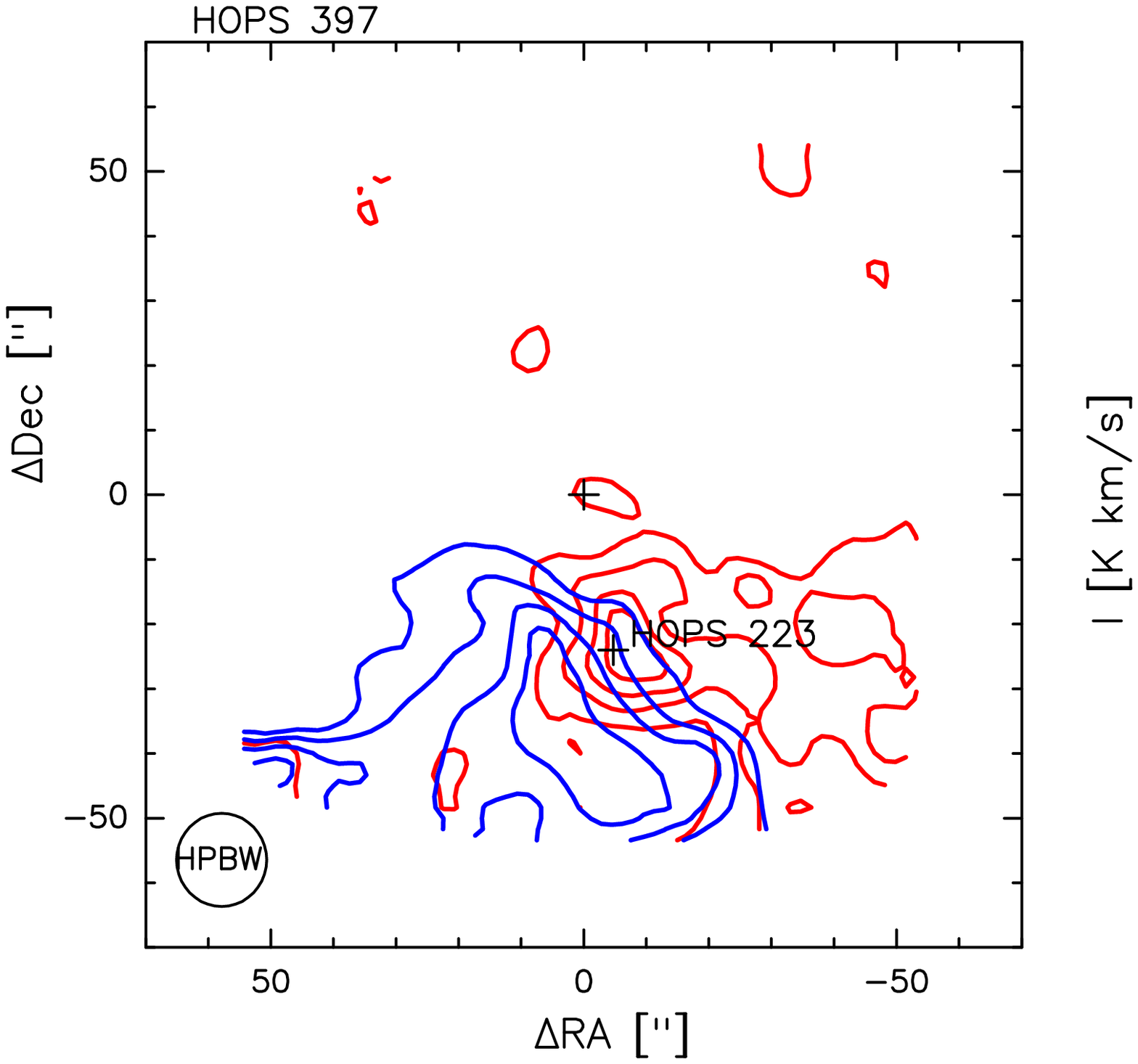}
\includegraphics[width=5.9cm, trim=0.5cm 1cm 0.4cm 8cm,clip=true]{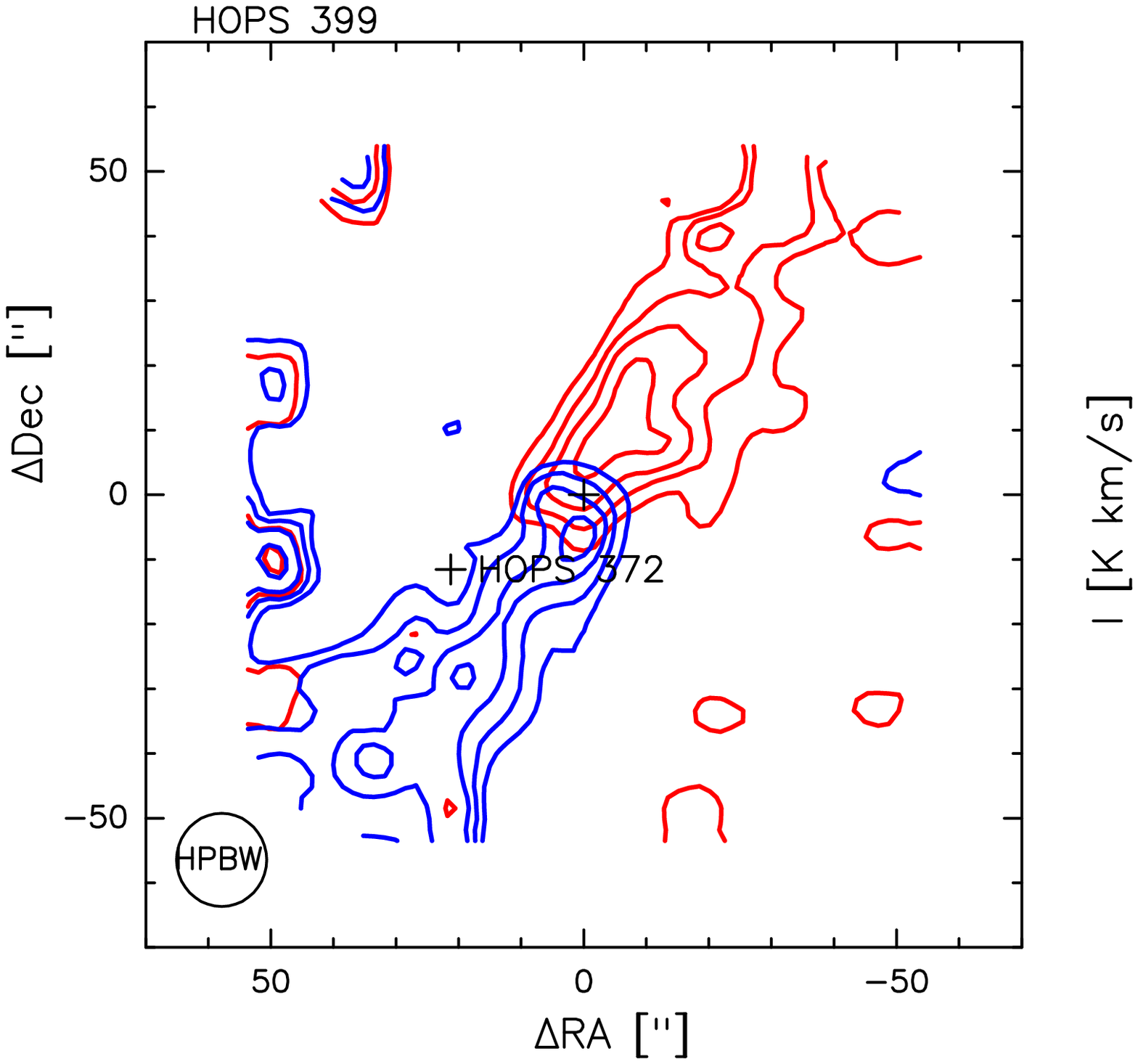}
\includegraphics[width=5.9cm, trim=0.5cm 1cm 0.4cm 8cm,clip=true]{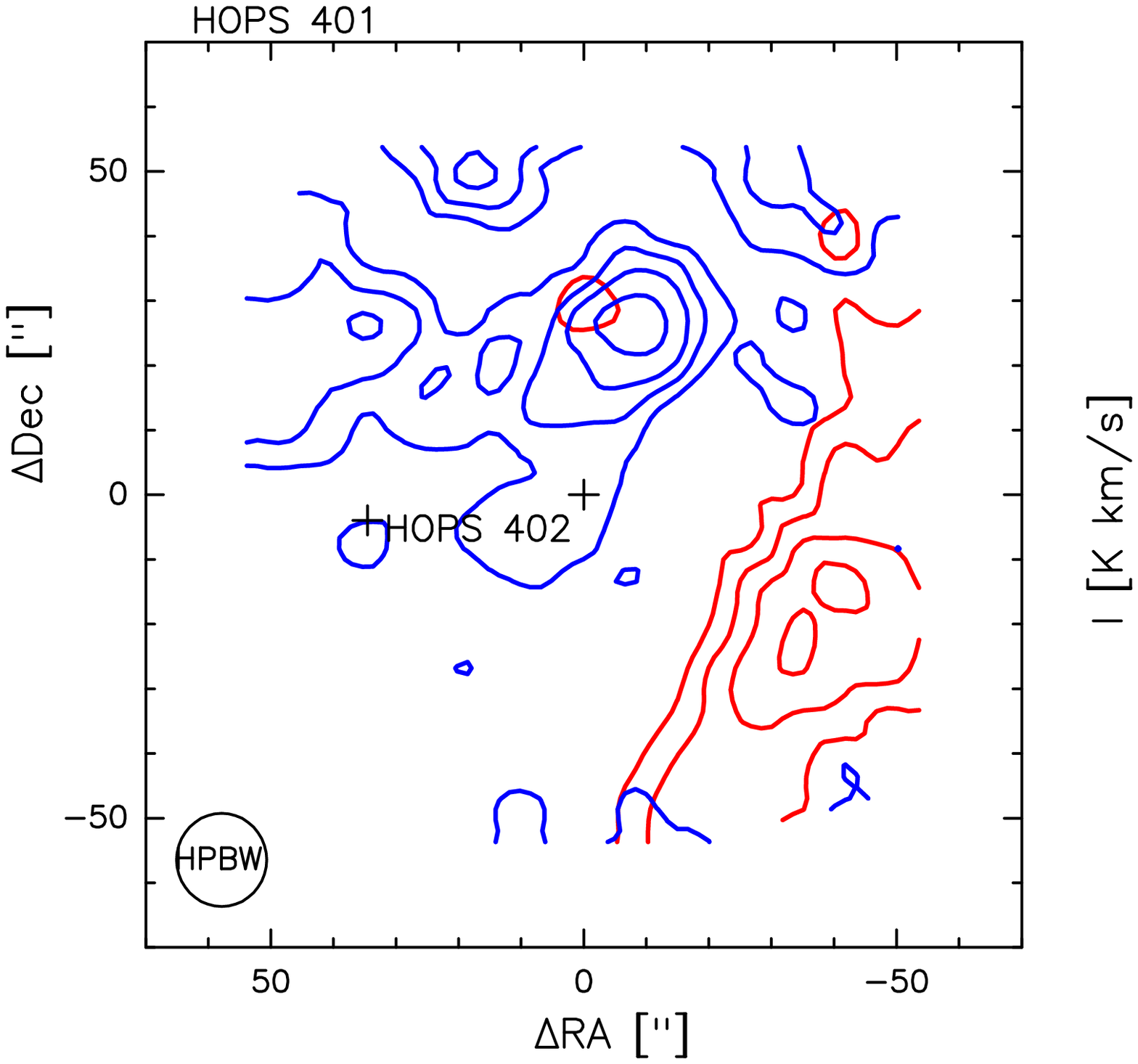}
\caption{CO $J$=4-3 integrated intensities of the blue- and red-shifted outflows (blue and red contours) for the sources with clear outflow detections. The integration limits are listed in Table \ref{table:outflowmass}.
The contours are 30\%, 50\%, 70\%, and 90\% of the maximum values of the intensities corresponding to the red- and blue-shifted outflow lobes in $T_{\rm{MB}}$ units toward the different maps, which are: 19.8 K km/s (red) and 63.6 K km/s (blue) toward HOPS 341, 31.2 K km/s (red) and 21.5 K km/s (blue) toward HOPS 354, 48.8 K km/s (red) and 46.7 K km/s (blue) toward HOPS 358, 42.4 K km/s (red) and 39.8 K km/s (blue) for HOPS 373, 56.0 K km/s (red) for HOPS 394, 14.3 K km/s (red) and 32.5 K km/s (blue) for HOPS 397, 29.5 K km/s (red) and 34.1 K km/s (blue) for HOPS 399, and 34.7 K km/s (red) and 13.8 K km/s (blue) for HOPS 401.
}
\label{fig:int_co43}
\end{figure*}

\end{appendix}

\end{document}